\documentclass[showpacs,preprintnumbers,amsmath,amssymb]{revtex4}
\usepackage{graphicx}
\usepackage{dcolumn}
\usepackage{bm}
\usepackage{epsfig}

\newcommand{\dhd}{{\textstyle d}
\lower.03ex\hbox{\kern-0.40em$^{\scriptstyle-}$}\kern-0.08em{}}  

\newcommand{\half}{{1\over 2}}
\newcommand{\bu}{{\bullet}}

\newcommand{\baru}{{\bar u}}

\newcommand{\barz}{{\bar z}}

\newcommand{\calo}{{\cal O}}  
\newcommand{\calr}{{\cal R}}  
 
\newcommand{\calu}{{\cal U}} 
\newcommand{\calv}{{\cal V}} 
\newcommand{\calz}{{\cal Z}}

\newcommand{\halo}{\hat{\cal O}}  

\begin{document}

\preprint{JLAB-THY-10-1160}

\title{High-energy amplitudes in the next-to-leading order}

\author{Ian Balitsky }
\affiliation{
Physics Dept., ODU, Norfolk VA 23529, \\
and \\
Theory Group, Jlab, 12000 Jefferson Ave, Newport News, VA 23606
}
\email{balitsky@jlab.org}

\date{\today}

\begin{abstract}
I review the calculation of the next-to-leading order behavior of high-energy
amplitudes in ${\cal N}=4$ SYM and QCD using the operator expansion in Wilson lines.

\end{abstract}



\maketitle

\tableofcontents

\section{\label{sec:in}Introduction }
The standard way to analyze the high-energy behavior of pQCD amplitudes is the direct summation of 
Feynman diagrams. In the Regge limit $s>>t,m^2$ (where $m$ is a characteristic mass or virtuality of colliding particles) there is an   
extra parameter in addition to $\alpha_s$, namely $\alpha_s\ln{s\over m^2}$. At pre-asymptotic energies we may have a window where the
leading logarithmic approximation (LLA) 
\begin{equation}
\alpha_s\ll 1, ~~~~\alpha_s\ln{s\over m^2}\sim 1
\label{lla}
\end{equation}
is valid. Following the pioneering work by Lipatov \cite{lip76}, the amplitude in this region of energies was shown 
to be determined by the BFKL pomeron \cite{bfkl}  which leads to the behavior of QCD cross sections of the type
\begin{equation}
\sigma(s)~=~\int\! d\nu ~\Big({s\over m^2}\Big)^{\omega(\nu)}F(\nu)~\sim~\Big({s\over m^2}\Big)^{4{\alpha_s\over\pi}N_c\ln 2}
\label{flob2}
\end{equation}
where $N_c$ is a number of colors ($N_c=3$ for QCD),  $\omega(\nu)={\alpha_s\over \pi}N_c[2\psi(1)-\psi(\half+i\nu)-\psi(\half-i\nu)]$ is the BFKL intercept, 
and $F(\nu)$ is the ``pomeron residue'' which depends on the process.

The power behavior of the cross
section  (\ref{flob2}) violates  the Froissart bound $\sigma\leq \ln^2s$, and therefore the BFKL
pomeron describes only  the pre-asymptotic behavior at intermediate energies
when the cross sections are small in comparison to the geometric cross section 
$2\pi R^2$. In order to find the exact window of the BFKL applicability one needs to 
calculate the first correction to the BFKL amplitude. 

The importance of  corrections to LLA result in high-energy QCD is twofold.
As I mentioned above, to get the region of application of a leading-order result 
one needs to find the next-to-leading order (NLO) corrections. 
However, in  the case of the BFKL amplitude (\ref{flob2}) there is another reason why NLO corrections are essential.  
Unlike for example the DGLAP evolution, the argument of the coupling constant in Eq. (\ref{flob2}) is left undetermined in 
the LLA, and usually it is set by hand to be of order of characteristic transverse momenta. Careful analysis of this argument is very 
important  from both theoretical and experimental points of view, 
and the starting point of the analysis of the argument of $\alpha_s$ in Eq.  (\ref{flob2}) is the calculation of the NLO BFKL correction.

The  NLO correction to the BFKL pomeron intercept $\omega(\nu)$ was found by two groups  - Lipatov and Fadin \cite{nlobfkl1} and
Ciafaloni and Camichi \cite{nlobfkl2}  - after almost ten years of calculations. The result has the form
\begin{eqnarray}
&&\hspace{-2mm} 
\omega(\nu)~=~{\alpha_s\over \pi}N_c\Big[\chi(\nu)+{\alpha_sN_c\over 4\pi}\delta(\nu)
\Big],
\nonumber\\
&&\hspace{-1mm}
\delta(\nu)~=~-{b\over 2N_c}\chi^2(\nu)+
6\zeta(3)+\Big[{67\over 9}-
{\pi^2\over 3}-{10n_f\over 9N_c}\Big]\chi(\nu)+\chi''(\nu)
-~2\Phi(\nu)-2\Phi(-\nu)
\label{eigen1}
\end{eqnarray}
where $b={11\over 3}N_c-{2\over 3}n_f$,  $\chi(\nu)=-2C-\psi(\half +i\nu)-\psi(\half -i\nu)$ ($C=-\psi(1)$ is Euler's constant), and 
\begin{eqnarray}
&&\hspace{-1mm}
\Phi(\nu)~=~-\int_0^1\!{dt\over 1+t}~t^{-\half+i\nu}
\Big[{\pi^2\over 6}+2{\rm Li}_2(t)\Big].
\label{fi}
\end{eqnarray}
The term proportional to $b$ depends on the choice of the argument of coupling constant.
The choice of this term in the r.h.s. of Eq. (\ref{eigen1}) corresponds to $\alpha(q_1q_2)$ where $q_i$ are momenta
of scattered BFKL gluons, see the discussion in Ref. \cite{nlobfkl1}.  The pomeron intercept is related to anomalous dimensions of
 the twist-2 gluon operators - it determines the asymptotics of anomalous dimensions $\gamma_j$ as $j-1=\omega\rightarrow 0$.
 It is worth noting that the result of explicit calculation of 3-loop anomalous dimensions  at the 3-loop level \cite{3loops} 
 agrees with Eq. (\ref{eigen1}).  
 As we see from Eq. (\ref{eigen1}) the full NLO description of the amplitude implies the knowledge of the ``pomeron residue'' $F(\nu)$
 at the NLO level. A classical example is the scattering of virtual photons in QCD where $F(\nu)$ is proportional to product of two
 amplitudes of emission of two t-channel gluons by the upper and the lower virtual photon. In the leading order, these ``impact factors''  
are known for a long time since they differ from the corresponding QED result only by trivial color factors. However, to the best of 
my knowledge, there is no complete analytical calculation of the NLO photon impact factor in the literature (for a combination of 
numerical and analytical results, see Ref. \cite{bart1}). I will present the analytical result for the photon impact factor in the coordinate space 
which is compact and conformally invariant; however,  the Fourier transformation to the momentum space is not performed yet. 
The calculations use different approach to high-energy scattering based on the operator expansion in Wilson lines \cite{npb96}  rather than
on the direct summation of Feynman diagrams. 
 
 A general feature of high-energy scattering is that a fast particle moves along its straight-line classical trajectory and the only 
 quantum effect is the eikonal phase factor acquired along this propagation path. In QCD, for the fast quark or gluon scattering 
 off some target, this eikonal phase factor is a Wilson line - the infinite gauge link ordered along the straight line collinear to particle's velocity $n^\mu$:
\begin{equation}
U^Y(x_\perp)={\rm Pexp}\Big\{-ig\int_{-\infty}^\infty\!\!  du ~n_\mu
~A^\mu(un+x_\perp)\Big\},~~~~
\label{defU}
\end{equation}
Here $A_\mu$ is the gluon field of the target, $x_\perp$ is the transverse
position of the particle which remains unchanged throughout the collision, and the 
index $Y$ labels the rapidity of the particle. Repeating the above argument for the target (moving fast in the spectator's frame) we see that 
particles with very different rapidities perceive each other as Wilson lines and
therefore these Wilson-line operators form
the convenient effective degrees of freedom in high-energy QCD (for a review, see ref. \cite{mobzor}).

As an example of high-energy process let us consider the deep inelastic scattering from a hadron at small 
$x_B=Q^2/(2p\cdot q)$.  The virtual photon decomposes into a pair of fast quarks 
 moving along straight lines separated by some transverse distance.
The propagation of this quark-antiquark pair reduces  to the 
``propagator of the color dipole''  $U(x_\perp)U^\dagger(y_\perp)$ - two Wilson lines ordered along the direction collinear to quarks' velocity. 
The structure function of a hadron is proportional to a matrix element of this color dipole operator
\begin{equation}
\hat{\cal U}^Y(x_\perp,y_\perp)=1-{1\over N_c}
{\rm Tr}\{\hat{U}^Y(x_\perp)\hat{U}^{\dagger Y}(y_\perp)\}
\label{fla1}
\end{equation}
switched between the target's states ($N_c=3$ for QCD).  The gluon parton density is 
approximately
\begin{equation}
x_BG(x_B,\mu^2=Q^2)~
\simeq ~\left.\langle p|~\hat{\cal U}^Y(x_\perp,0)|p\rangle
\right|_{x_\perp^2=Q^{-2}}
\label{fla2}
\end{equation}
where $Y=\ln{1\over x_B}$. (As usual, we denote operators by ``hat'').
The energy dependence of the structure function is translated then into the dependence of the color dipole on on the rapidity $Y$.
There are two ways to restrict the rapidity of Wilson lines: one can consider Wilson lines with the support line collinear
 to the velocity of the fast-moving particle or
one can take the light-like Wilson line and cut the rapidity integrals ``by hand'' (see Eq. (\ref{cutoff}) below).
While the former method appears to be more natural, 
it is technically simpler to get the final results with the latter method of ``rigid cutoff'' in the longitudinal direction. 

Eq. (\ref{fla2}) means that the  small-x behavior of the structure functions is  governed by the 
rapidity evolution of color dipoles \cite{mu94,nnn}. 
At relatively high energies and for sufficiently small dipoles we can use the leading logarithmic approximation (\ref{lla})
($ \alpha_s\ll 1,~ \alpha_s\ln x_B\sim 1$) and get the non-linear BK evolution equation for the color
dipoles \cite{npb96,yura}:
\begin{eqnarray}
&&\hspace{-1mm}
{d\over dY}~\hat{\cal U}^Y(z_1,z_2)~=~
{\alpha_sN_c\over 2\pi^2}\!\int\!d^2z_3~ {z_{12}^2\over z_{13}^2z_{23}^2}
[\hat{\cal U}^Y(z_1,z_3)+\hat{\cal U}^Y(z_3,z_2))-\hat{\cal U}^Y(z_1,z_3)-\hat{\cal U}^Y(z_1,z_3)\hat{\cal U}^Y(z_3,z_2)]
\label{bk}
\end{eqnarray}
where $Y=\ln{1\over x_B}$ and $z_{12}\equiv z_1-z_2$ etc. 
The first three terms correspond to the linear BFKL evolution \cite{bfkl} and describe the parton emission 
while the last term is responsible for the parton annihilation. For sufficiently low $x_B$ the parton emission 
balances the parton annihilation so the partons reach the state of saturation \cite{saturation} with
the characteristic transverse momentum $Q_s$ growing with energy $1/x_B$
(for reviews, see Ref. \cite{satreviews})

It is easy to see that the BK equation (\ref{bk}) is conformally invariant in the two-dimensional space. This follows from the conformal   
invariance of the light-like Wilson lines. The  Wilson line
\begin{equation}
U(x_\perp)~=~{\rm Pexp}~\Big\{-ig\!\int_{-\infty}^\infty\!dx^+~A_+(x^+,x_\perp)\Big\}
\end{equation}
is invariant under the inversion $x^\mu\rightarrow x^\mu/x^2$ (with respect to the point with zero (-) component). 
Indeed, $(x^+,x_\perp)^2=x_\perp^2$ so after the inversion $x_\perp\rightarrow x_\perp/x_\perp^2$ and $x^+\rightarrow x^+/x_\perp^2$ and
therefore
\begin{equation}
U(x_\perp)~\rightarrow~{\rm Pexp}~\Big\{-ig\!\int_{-\infty}^\infty\!d{x^+\over x_\perp^2}~A_+({x^+\over x_\perp^2},x_\perp)\Big\}
~=~U(x_\perp/x_\perp^2)
\label{inversion}
\end{equation}
Moreover, it is easy to check formally that the Wilson line operators lie in the standard representation of the conformal M\"obius group SL(2,C) 
with conformal spin 0 (see \cite{nlobksym})). It should be noted that the conformal invariance of the linear BFKL equation 
was first proved in Ref. \cite{lip86}.

The NLO evolution of color dipole in QCD is not expected to be M\"obius invariant due to the conformal anomaly leading 
to dimensional transmutation and running coupling constant. 
To understand the relation between the high-energy behavior of amplitudes and M\"obius invariance of Wilson lines, 
it is instructive to consider the conformally invariant  ${\cal N}=4$ super Yang-Mils theory. 
This theory was  intensively studied in recent years due to the fact that at large coupling constants 
it is dual to the IIB string theory in the AdS$_5$ background. In the light-cone limit,
the contribution of scalar operators to Maldacena-Wilson line \cite{mwline} vanishes so one has the usual Wilson line constructed from gauge fields and therefore the LLA evolution
equation for color dipoles in the ${\cal N}=4$ SYM has the same form as (\ref{bk}). At the NLO level, the contributions from gluino and scalar loops enter the game.

As I mentioned above, formally the light-like Wilson lines are  M\"obius invariant.
However, the light-like Wilson lines are divergent in the longitudinal direction and moreover,  it is exactly the evolution 
equation with respect to this longitudinal cutoff which governs the high-energy behavior of amplitudes. 
At present, it is not known how to find the conformally invariant cutoff in the longitudinal direction.  When we use the non-invariant cutoff 
we expect, as usual, the invariance to hold in the leading order but to be
violated in higher orders in perturbation theory. In our calculation we restrict the longitudinal momentum of the gluons composing Wilson lines, 
and with this non-invariant cutoff the NLO evolution equation in QCD has extra non-conformal parts not related to the running of coupling constant.
Similarly, there will be non-conformal parts coming from the longitudinal cutoff of Wilson lines in the ${\cal N}=4$ SYM equation.
I will demonstrate below that it is possible to construct the
``composite conformal dipole operator'' (order by order in perturbation theory) which mimics the conformal cutoff
in the longitudinal direction so the corresponding evolution equation is M\"obius invariant.
With the NLO accuracy this composite operator has the form \cite{nlobksym} ($a$ is an arbitrary constant):
\begin{eqnarray}
&&\hspace{-1mm}
[{\rm Tr}\{\hat{U}_{z_1}\hat{U}^{\dagger}_{z_2}\}\big]_{a,Y}^{\rm conf}~
\label{confodipole}\\
&&\hspace{-1mm}
=~{\rm Tr}\{\hat{U}^\sigma_{z_1}\hat{U}^{\dagger\sigma}_{z_2}\}
+{\alpha_s\over 2\pi^2}\!\int\! d^2 z_3~{z_{12}^2\over z_{13}^2z_{23}^2}
[ {\rm Tr}\{T^n\hat{U}^\sigma_{z_1}\hat{U}^{\dagger\sigma}_{z_3}T^n\hat{U}^\sigma_{z_3}\hat{U}^{\dagger\sigma}_{z_2}\}
-N_c {\rm Tr}\{\hat{U}^\sigma_{z_1}\hat{U}^{\dagger\sigma}_{z_2}\}]
\ln {4az_{12}^2\over sz_{13}^2z_{23}^2}~+~O(\alpha_s^2)
\nonumber
\end{eqnarray}
where the second term is a ``counterterm'' restoring the conformal invariance lost because of the cutoff (\ref{cutoff}). 
This is quite similar 
to the construction of the composite renormalized local operator in the case when the UV cutoff  does not respect the 
symmetries of the bare operator - in this case the symmetry of the UV-regularized operator is preserved 
order by order in perturbation theory by subtraction of the symmetry-restoring counterterms.

The NLO BK equation written in terms of these composite conformal operators is conformally invariant (in ${\cal N}=4$ SYM) and so are the impact factors \cite{nlobksym}. 
Below I  present the results for these impact factors and for the ``pomeron residue'' $F(\nu)$ in a simple case of correlator of four scalar currents in the 
Regge limit.

In QCD we do not have the conformal invariance so the corresponding composite operators are not, strictly speaking, conformal. 
Still, if we write down the NLO BK equation in terms of these operators it has a nice property of being a sum of conformal part
and the running-coupling part proportional to $b={11\over 3}N_c-{2\over 3}n_f$. I calculate the NLO coefficient function in the expansion of two electromagnetic currents
in these composite operators which determines the photon impact factor at the NLO level.

The paper is organized as follows. In  Sect. II I remind the general logic of the operator expansion using the example of light-ray OPE. 
In Sect. III I formulate the program of high-energy OPE and carry it out in subsequent Sections. In Sect IV we
calculate the high-energy amplitudes in ${\cal N}=4$ SYM in the leading order and then in the NLO.
Sect V is devoted to the NLO high-energy amplitudes in QCD and Sect. VI contains the conclusions. The details of the calculation of 
Feynman diagrams for the NLO evolution of color dipoles are not presented here (see the original publications \cite{balbel, prd75, nlobk, nlobksym}) but we 
will outline the calculation of impact factors in ${\cal N}=4$ SYM and in QCD.

 \section{The logic of OPE}
 Let me first remind the logic of usual operator expansion near the light cone. A typical example is the calculation of the
 structure functions of deep inelastic scattering (DIS) at moderate $x_B=Q^2/(2p\cdot q)$ determined by the T-product of two electromagnetic currents 
 $T\{j_\mu(x)j_\nu(y)\}$ switched between the target states. The first step is to identify the relevant operators. 
 To this end, we formally put $Q^2$ to infinity, then $(x-y)^2=0$ and we see that the relevant operators are the light-ray ones of the type of $\bar{\psi}(x)\gamma_\mu[x,y]\psi(y)$ where
 \begin{equation}
 [x,y]~\equiv~{\rm Pexp}\Big\{-ig\!\int_0^1\! du~(x-y)^\mu A_{\mu}(ux-uy+y)\Big\}
 \label{pexp}
 \end{equation}
 is a standard notation for a straight-line gauge link connecting points $x$ and $y$.
 
 Now, to get the $Q^2$ behavior of structure functions one needs to perform the following four steps:
 \begin{itemize}
 \item{} Separate relevant Feynman loop integrals over $k_\perp$ in two parts - the coefficient functions (with transverse momenta $k_\perp^2$ greater 
 than the factorization scale $\mu^2$) and parton densities - matrix elements of light-ray operators with $k_\perp^2<\mu^2$ (see Fig. 1). 
 Technically, the cutoff of the transverse momenta 
 in these matrix elements is done by adding counterterms with normalization point $\mu^2$.
 \item{} Find the evolution equations of light-ray operators with respect to $\mu^2$.
\item{} Solve of the corresponding evolution equation(s). 
\item{} Assemble the result for structure functions: take some initial conditions at low normalization
point (usually around $\mu^2$=1GeV$^2$),  evolve parton densities to high $Q^2$ and multiply the result by the coefficient function.
 \end{itemize}
 
Let me explain these steps at the NLO level in detail.
\begin{figure}[h]

\vspace{5mm}
\includegraphics[width=43mm]{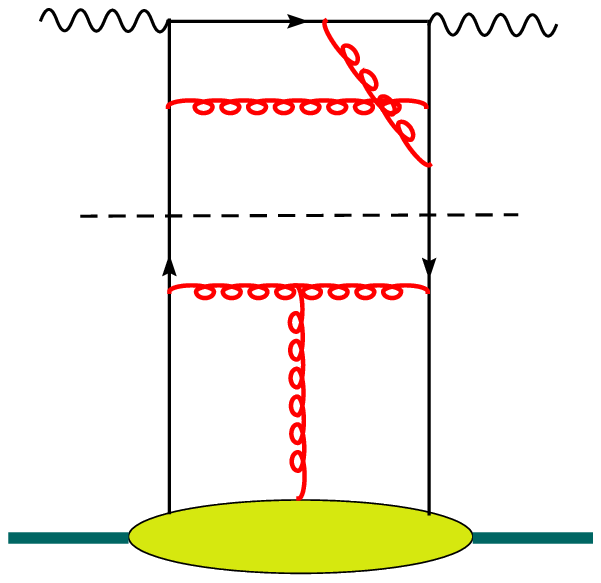}
\hspace{1cm}
\vspace{5mm}
\includegraphics[width=16mm]{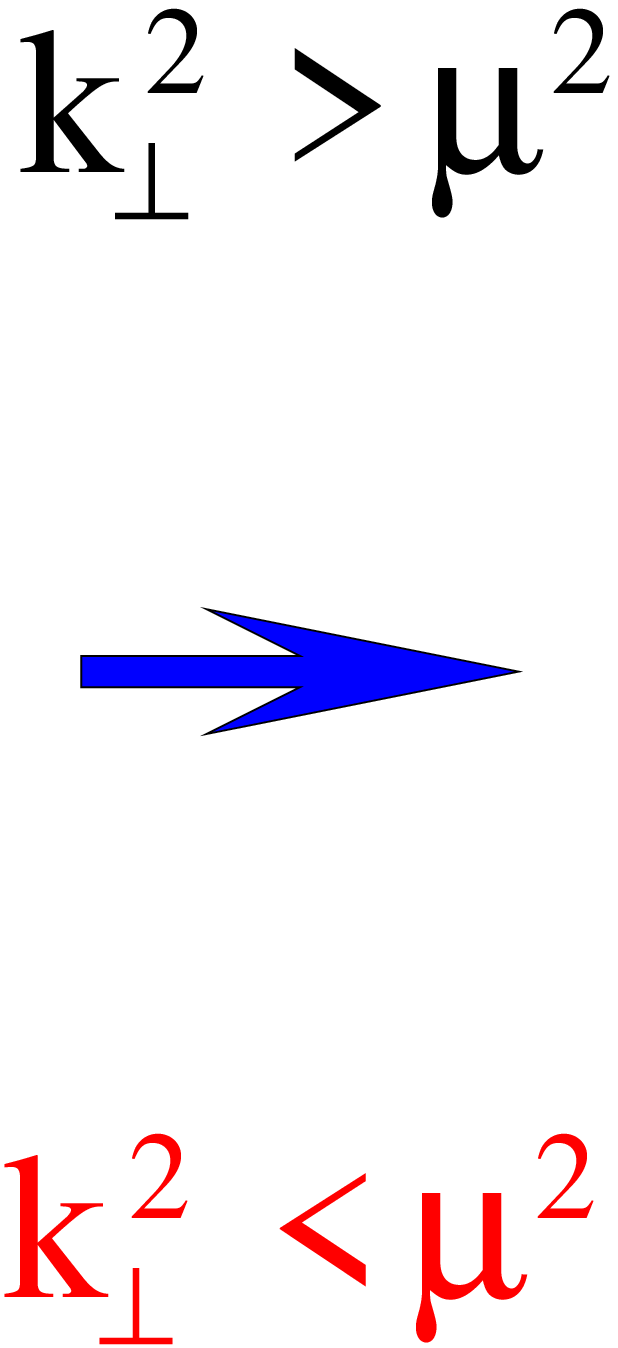}
\hspace{1cm}
\includegraphics[width=85mm]{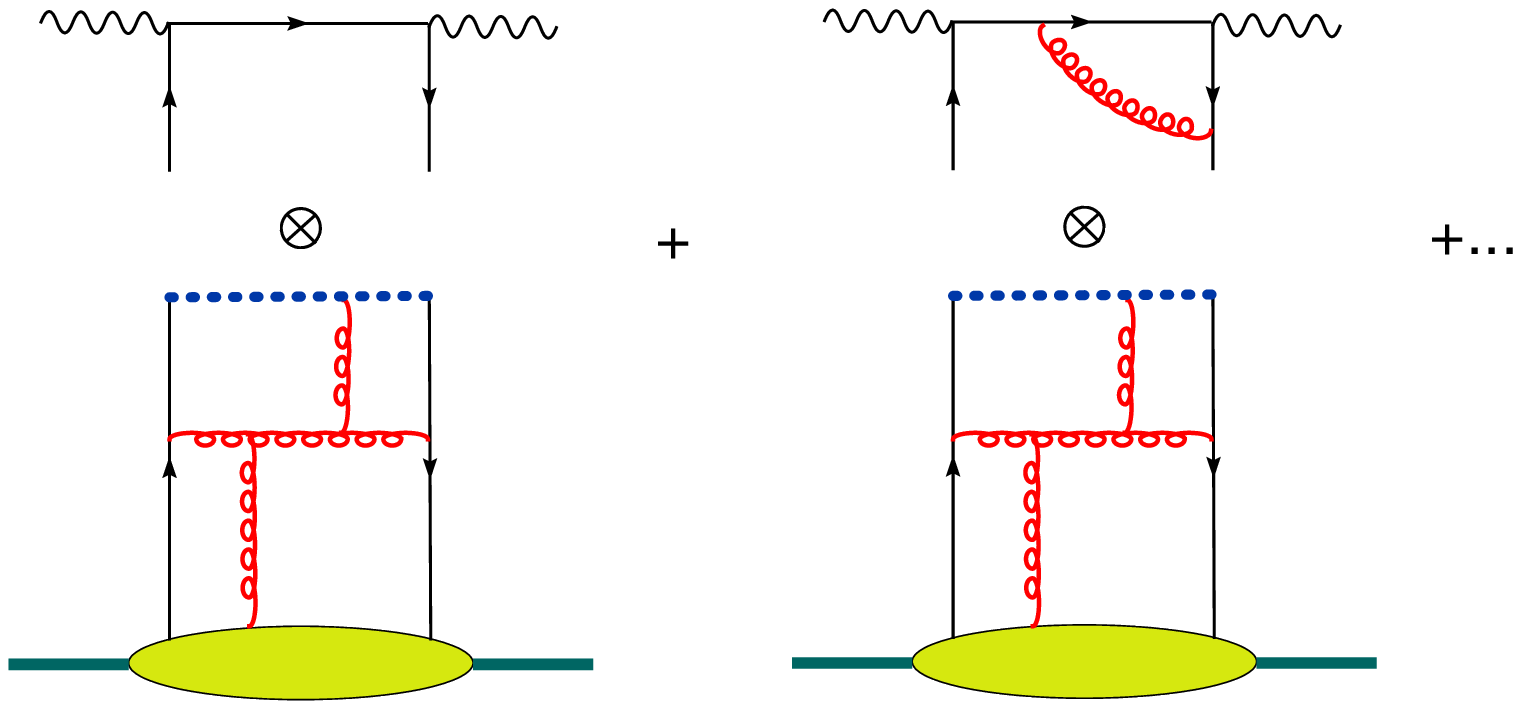}
\caption{Operator product expansion near the light cone. Gauge link is denoted by a dotted line. \label{fig:opel}}
\end{figure}

 One can calculate the coefficient functions by direct computation of Feynman diagrams with quark and gluon tails  
 or by using the Feynman diagrams in external quark and gluon fields (see e.g. \cite{strop, eveq}).  A typical NLO contribution 
 coming from the upper part of last diagram in Fig. \ref{fig:opel} has the form
 \begin{eqnarray}
&& \hspace{-2mm}
T\{j_\mu(x)j_\mu(0)\}~=~{-i\over \pi^2x^4}\!\int_0^1\! du~\Big[{1\over u}\Big]_+[\psi(x)[x,ux]\!\not\!x \psi(ux)]^{\rm l.t.}\Big(1+{\alpha_s\over 4\pi}[\ln x^2\mu^2+\ln u]\Big)+...
\label{tycoef}
 \end{eqnarray}
 where $\Big[{1\over u}\Big]_+$ is a standard ``plus'' prescription 
 \begin{equation}
 \int_0^1\! du~\Big[{1\over u}\Big]_+f(u)~\equiv~\int_0^1\! du~{f(u)-f(0)\over u}
 \label{plupres}
 \end{equation}
 and the leading-twist light-ray operator $[\psi(x)[x,ux]\!\!\!\not\!\!x \psi(ux)]^{\rm l.t.}$   is taken at $x^2=0$. Formally, one can consider the operator
 at  $\tilde{x}$ where $\tilde{x}$ is close to $x$ and light-like. Alternatively, one can define  a ``leading twist''  prescription by
 subtraction of all higher twists in the non-local form, see \cite{opeanni}. In any case, the light-ray operator $[\psi(x)[x,ux]\!\not\!x \psi(ux)]^{\rm l.t.}$ 
 has specific light-cone UV divergencies (in addition to usual UV contributions assembled to the running coupling constant) which are 
 regularized by adding light-ray counterterms.  The resulting composite operator (original operator minus counterterms) depends on the renormalization point 
 $\mu$  and the renorm-group equation for these operators leads to the DGLAP evolution
 equation for parton densities. 
 
 A typical contribution to the NLO evolution equation of the quark light-ray operator coming from the bottom part of diagram in Fig. \ref{fig:opel} has the form 
 \begin{eqnarray}
&& \hspace{-2mm}
\mu{d\over d\mu}[\psi(x)[x,0]\!\not\! x\psi(0)]_{\rm l.t.}^{\mu}
~=~{\alpha_s(\mu)\over 2\pi}c_F\!\int_0^1\! du~\Big[{1\over u}\Big]_+[\psi(x)[x,ux]\!\not\!x \psi(ux)]_{\rm l.t.}^{\mu}\Big(1+{\alpha_s\over 4\pi}[\ln x^2\mu^2+\ln u]\Big)+...
\label{tyeveq}
 \end{eqnarray}
($c_F=(N_c^2-1)/2N_c$). This completes the second step of the above four-step program.

The third step is the solution of the evolution equation of (\ref{tyeveq}) type. I will present this solution in a simple case of forward matrix elements of
quark operators for a non-singlet case determined by quark parton densities. For example, in the case of unpolarized proton we get
 \begin{eqnarray}
&& \hspace{-2mm}
\langle p|[\bar{q}(x)[x,0]\!\not\! xq(0)-\bar{q}\!\not\! x[0,x]q(x)]_{\rm l.t.}^{\mu}|p\rangle~=~
2(px)\!\int_0^1\! d\omega~{\cal D}_q(\omega)\big[e^{ip\cdot x\omega}-e^{-ip\cdot x\omega}\big]
\label{qpdf}
 \end{eqnarray}
 where $p$ is the proton momentum and $D_q(\omega)$ is a parton density of quark $q$ ($u$, $d$, or $s$). The solution of the evolution equation for parton densities
is given by the Mellin integral. In terms of light-ray operators it has the form
 \begin{equation}
\langle\psi(x)[x,0]\!\not\! x\lambda^a\psi(0)]_{\rm l.t.}^{\mu}\rangle~=~\int_{-\half-i\infty}^{-\half+i\infty}\!{d\nu\over 2\pi i}
\Big({\alpha_s(\mu)\over\alpha_s(\mu_0)}\Big)^{-\gamma_j^{(1)}\over b}e^{-{\alpha_s(\mu)-\alpha_s(\mu_0)\over 4\pi b}[\gamma_j^{(2)}+\gamma_j^{(1)} b_1]}
\!\int_0^\infty\! du~u^j\langle\psi(ux)[ux,0]\!\not\! x\lambda^a\psi(0)]_{\rm l.t.}^{\mu_0}\rangle
\label{mellinsol}
 \end{equation}
 where $\lambda^a$ is a flavor Gell-Mann matrix and $\langle \calo\rangle$ means any forward matrix element of the operator $\calo$. Also, 
 $\gamma_j(\alpha_s)=\gamma_j^{(1)}{\alpha_s\over 4\pi}+\gamma_j^{(2)}{\alpha^2_s\over 16\pi^2}+...$ is the anomalous dimension of the light-ray operator 
 $\!\int_0^\infty\! du~u^j\psi(ux)[ux,0]\!\not\! x\lambda^a\psi(0)$ and $b_1$ is the second coefficient of Gell-Mann-Low function.
 
 Now, to perform step four and get a structure function ($-3F_1+\half F_2$ in our case) one can take  parton densities
 at low $\mu_0^2\sim $ 1Gev$^2$, put it into Eq. (\ref{mellinsol}), take $\mu^2=Q^2$ and multiply the result by the coefficient function
 (\ref{tycoef}) taken at $\mu^2=Q^2$. In future, one should get the parton densities at low $\mu^2$ from non-perturbative (lattice) QCD
 but at present people use  well established models. The evolution of parton densities in QCD is calculated up to the third order in $\alpha_s$  \cite{3loops}
 and the results agree with experiment in a broad range of parameters $Q^2$ and $x_B$.

 \section{High-energy OPE in Wilson lines}
Now we want to extend the four-step program from the previous section to describe the high-energy amplitudes.
A typical process is deep inelastic scattering at small values of Bjorken $x$. Since we are interested now in $x_B$ evolution rather than 
 $Q^2$ evolution, as step one we factorize in rapidity instead of transverse momenta. As a preliminary step we identify the relevant operators.
The virtual photon splits into quark-antiquark pair, and if we set formally the energy of incoming photon to infinity we see that these
quark and antiquark travel along the light-like classical trajectories. 
 This is the well-known general result from quantum mechanics - the fast particle moves along its straight-line classical trajectory and the only quantum effect is 
the eikonal phase factor acquired along this propagation path. In QCD, for fast quark or gluon scattering 
off some target, this eikonal phase factor is a Wilson line (\ref{defU}) - an infinite gauge link ordered along the straight 
line collinear to particle's velocity $n^\mu$:
\begin{figure}[htb]
\psfig{file=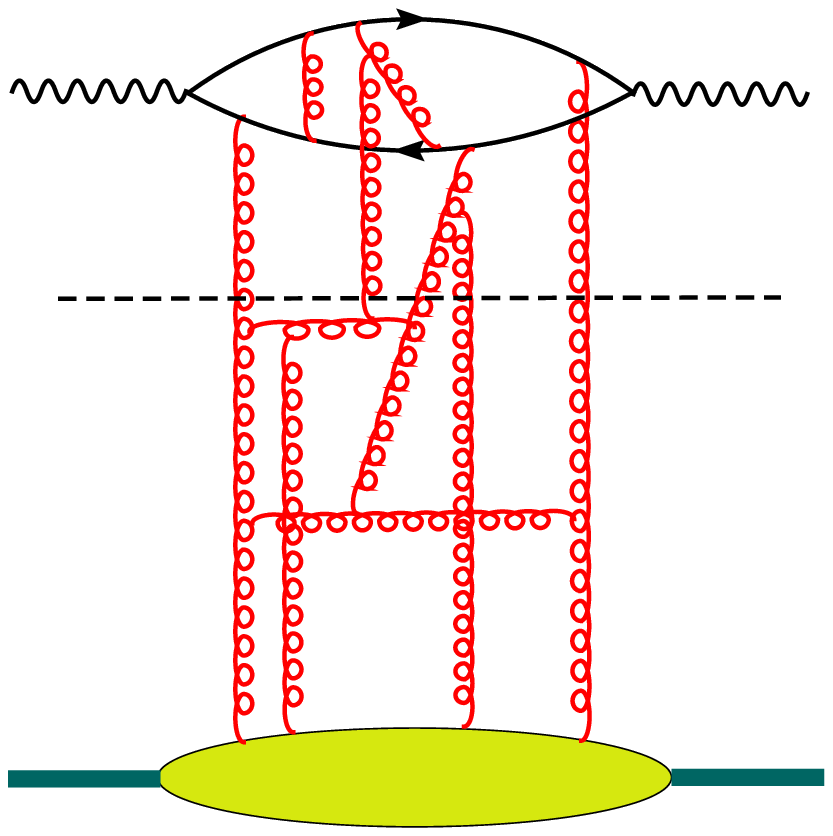,width=46mm} 
\hspace{4mm} 
\psfig{file=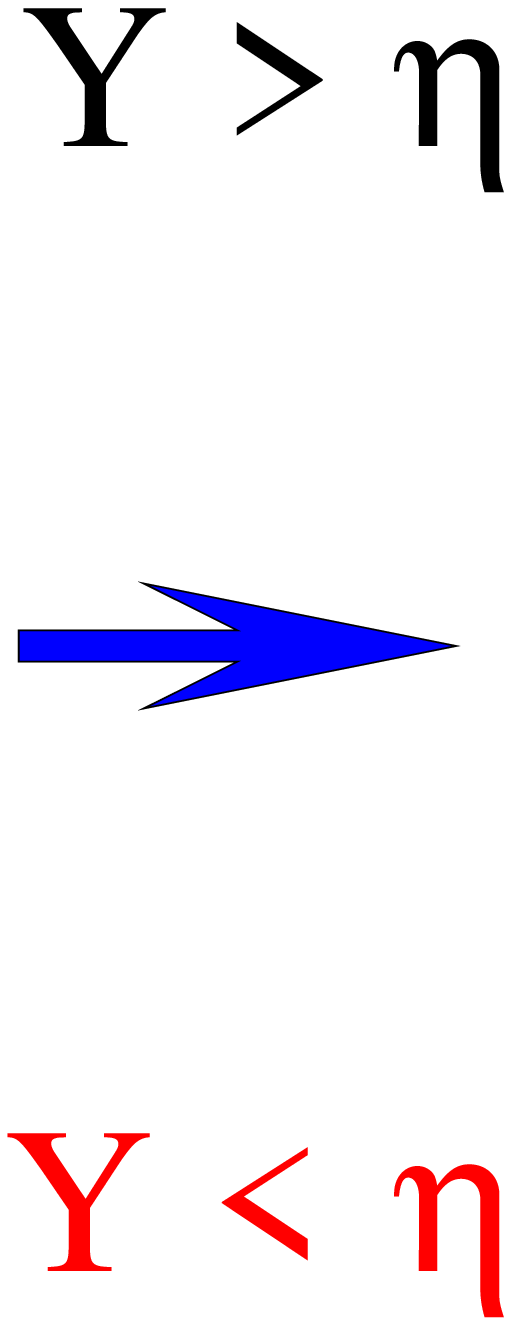,width=8mm} 
\hspace{4mm}
\psfig{file=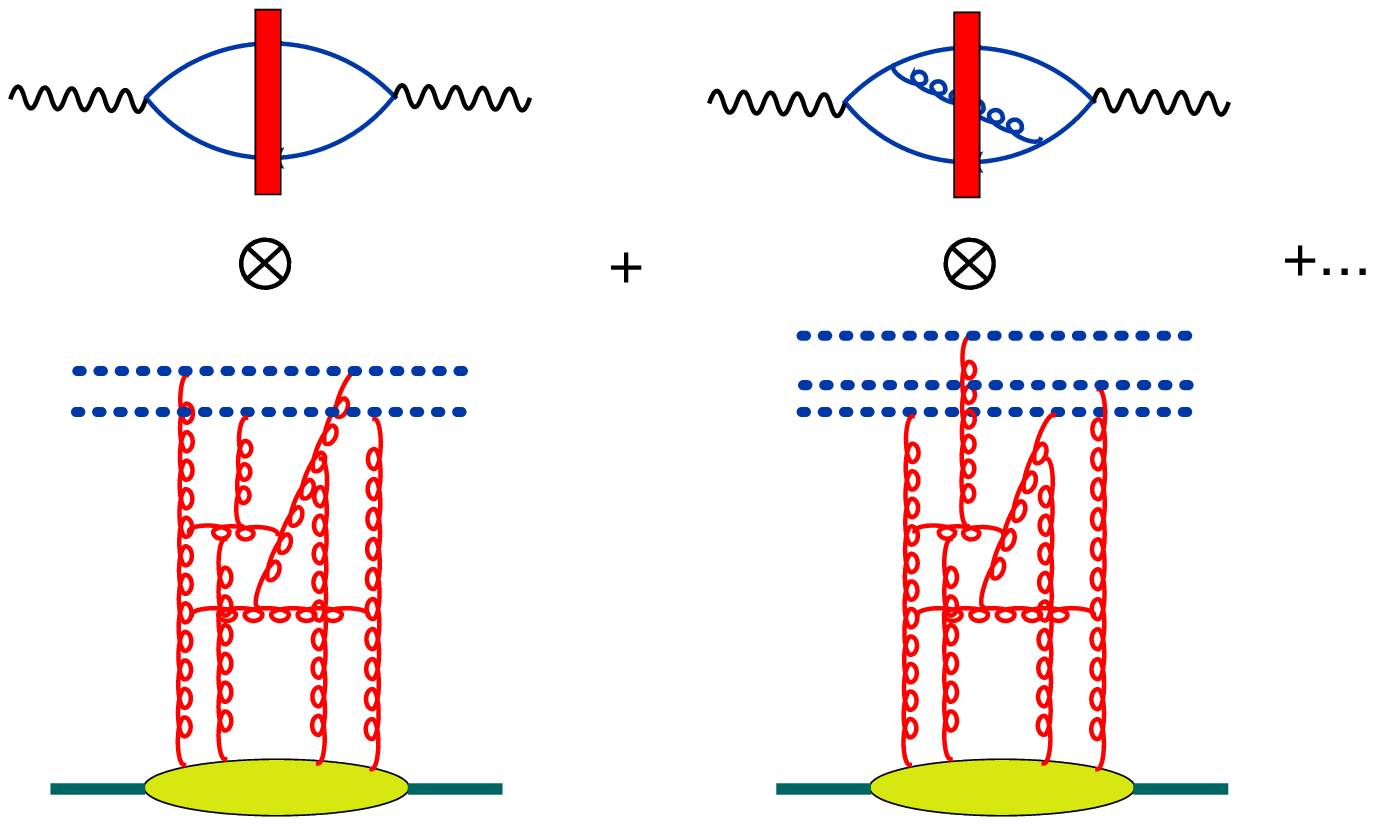,width=82mm}
\caption{High-energy operator expansion in Wilson lines}
\label{fig:opeh}
\end{figure}

Having identified the operators, we can repeat the 4 steps which give us the high-energy (small-$x$ for DIS) evolution of the amplitude.
 \begin{itemize}
 \item{} Separate the relevant Feynman loop integrals over the longitudinal momentum $\alpha$ in two parts - the coefficient functions (``impact factors'') with  $\alpha$ greater 
 than the rapidity factorization scale $\sigma=e^{\eta}$ and matrix elements of Wilson-line operators with $\alpha<\sigma$ (see Fig. \ref{fig:opeh}). We were not able to find the analog of 
 dimensional regularization for longitudinal divergence so we cut the integration over $\alpha<\sigma$ `` by hand''. 
  \item{} Find the evolution equations of color dipoles with respect to the cutoff in rapidity $Y$.
\item{} Solve of the corresponding evolution equation(s). 
\item{} Assemble the result for structure functions: take the initial conditions at low energy,  evolve color dipoles to high rapidity $Y\sim Y_A$  and multiply 
the result by the corresponding impact factor. 
\end{itemize}

Unlike the  DGLAP evolution discussed above, the BK evolution equation for color dipoles with respect to energy 
is non-linear which makes it considerably more difficult to solve. At present, for the  experimentally interesting case of DIS from proton or nucleus we have no analytical 
solution so one has to rely upon the approximate solutions and numerical simulations. One can linearize the BK equation and get the usual linear BFKL evolution 
but the validity of this linearization in the saturation region of small $x$ is questionable. Still, there are purely perturbative processes where the BFKL equation
gives the correct (pre-asymptotic) behavior at high energies: for example the scattering of virtual photons with equal (and high) virtualities, or observation
of two Mueller-Navelet jets \cite{mujets} in hadron-hadron scattering. In this case, the leading-order analysis has been performed, but the generalizations of the LO results
to the next-to-leading order has not been obtained since it is difficult to take into account the running-coupling effects in the BFKL equation.

By the same token, since in  ${\cal N}=4$ SYM the coupling does not run,  the NLO BFKL program can be performed to the very end. I will present the NLO analysis of 
high-energy ``scattering of two scalar currents'' and get the NLO result for this correlation function in an explicit form. In subsequent two chapters we will carry out steps (1-4) 
of our program for ${\cal N}=4$ SYM and then for QCD (where we'll discuss only steps 1 and 2).

\section{High-energy amplitudes in ${\cal N}=4$ SYM in the next-to-leading order}

\subsection{Regge limit and Pomeron in ${\cal N}=4$ SYM}
As we mentioned above, at first we will find the  NLO amplitudes at high energies for ${\cal N}=4$ SYM and turn to QCD later. 
We use the ${\cal N}=4$ Lagrangian in the form (see e.g. Ref. \cite{bel03}):
\begin{eqnarray}
&&\mathcal{L}~=~
-{1\over 4}F^{\mu\nu}F_{\mu\nu} + {1\over 2}\big(D^\mu \Phi^a_I\big)\big(D_\mu \Phi^a_I\big)
-{1\over 4}g^2 f^{abc}f^{lmc} \Phi^a_I \Phi^b_J \Phi^l_I \Phi^m_J 
\nonumber\\
&&~~~~~~~~+\bar\lambda_{\dot\alpha A}^a
\sigma^{\dot\alpha \beta}_\mu {\cal D}^\mu\lambda^{aA}_\beta
-i\lambda^{\alpha A}_a\bar{\Sigma}^s_{AB}\Phi^s_b\lambda^B_{\alpha k}f^{abc}
+i\bar{\lambda}^a_{\dot\alpha A}\Sigma^{s AB}\Phi^s_b\bar{\lambda}^{\dot\alpha}_{B c}f^{abc}
\label{props}
\end{eqnarray}
Here $\Phi^a_I$ are scalars, $\lambda^{\alpha A}_a$ gluinos and $\Sigma^a_{IJ}=(Y^i_{AB},i\bar{Y}^i_{AB})$,
$\bar{\Sigma}^a_{IJ}=(Y^i_{AB},-i\bar{Y}^i_{AB})$  where 
$Y^i_{AB}$ are standard `t Hooft symbols. The bare propagators are
\begin{eqnarray}
&&
\langle \Phi^a_I(x) \Phi^b_J(y)\rangle~=~i{\delta^{ab}\delta_{IJ}\!\int\!\dhd^4p {e^{-ip\cdot(x-y)}\over -p^2+i\epsilon},~~~~~~
\langle\lambda^{aI}_\beta(x)\,\bar\lambda^{bJ}_{\dot\alpha}(y)\rangle =
\int\!\dhd^4p e^{ip\cdot(x-y)}{ip_\mu\,\bar{\sigma}^\mu_{\beta\dot{\alpha}}\over -p^2+i\epsilon},                         }
\end{eqnarray}
and the vertex of gluon emission in the momentum space is proportional to $(k_1-k_2)^\mu T^a\delta_{IJ}$ for the scalars and 
$\sigma^\mu T^a$ for gluinos. Here $\sigma^\mu=(1,\vec{\sigma})$,
 $\bar{\sigma}^\mu=(-1,\vec{\sigma})$ where $\vec{\sigma}$ are usual Pauli matrices and our metric is $g^{\mu\nu}=(-1,1,1,1)$.

For simplicity, let us consider correlation function of four scalar currents
\begin{eqnarray}
A(x,y,x',y')~=~(x-y)^4(x'-y')^4\langle \calo(x) \calo^\dagger(y)\calo(x')\calo^\dagger(y') \rangle~
\label{correl1}
\end{eqnarray}
where ${\calo}\equiv{4\pi^2\sqrt{2}\over \sqrt{N_c^2-1}}{\rm Tr\{Z^2\}}$ ($Z={1\over\sqrt{2}}(\Phi_1+i\Phi_2)$) is a renorm-invariant chiral primary operator.

In a conformal theory this four-point amplitude $A(x,y;x',y')$ depends on two conformal ratios which can be chosen as
\begin{eqnarray}
&&\hspace{-1mm}
R~=~{(x-x')^2(y-y')^2\over (x-y)^2(x'-y')^2},~~~~~~~~~~~~~
r~=~R\Big[1-{(x-y')^2(y-x')^2\over (x-x')^2(y-y')^2}+{1\over R}\Big]^2
\label{cratios1}
\end{eqnarray}
We are interested in the behavior of the correlator (\ref{correl1}) in the high-energy (Regge) limit. In the coordinate space it can be achieved as follows:
\begin{eqnarray}
&&\hspace{-1mm}
x=\rho x_\ast {2\over s}p_1+x_\perp,~~~ y=\rho y_\ast {2\over s}p_1+y_\perp, ~~~~~~~~~~~~~~~~~
x'=\rho' x_\bu {2\over s}p_2 +x'_\perp,~~~  y'=\rho' y'_\bu {2\over s}p_2 +y'_\perp
\label{reggelimit}
\end{eqnarray}
 with $\rho,\rho'\rightarrow\infty$ and $x_\ast>0>y_\ast$, $x'_\bu>0>y'_\bu$. (Strictly speaking, $\rho\rightarrow\infty$ or $\rho'\rightarrow \infty$ would be sufficient to reach the Regge limit). 
 Hereafter I use the notations $x_\bu=-p_1^\mu x_\mu$, $x_\ast=-p_2^\mu x_\mu$
where $p_1$ and $p_2$  are light-like vectors normalized by $-2(p_1,p_2)=s$. These ``Sudakov variables'' are related to 
the usual light-cone coordinates
$x^\pm={1\over\sqrt{2}}(x^0\pm x^3)$ by $x_\ast=x^+\sqrt{s/2},~x_\bu=x^-\sqrt{s/2}$ so $x={2\over s}x_\ast p_1+{2\over s}x_\bu p_2+x_\perp$. The metric is  $g^{\mu\nu}$=(-1,1,1,1)   so 
$x^2=-{4\over s}x_\bu x_\ast +\vec{x}_\perp^2$.
In the Regge limit (\ref{reggelimit}) the full conformal group  reduces to M\"{o}bius subgroup SL(2,C) leaving the transverse plane $(0,0,z_\perp)$ invariant. 

 As demonstrated in Ref. \cite{cornalba}, the pomeron contribution in a conformal theory can be represented as an integral over 
 one real variable $\nu$
\begin{eqnarray}
&&\hspace{-3mm}
(x-y)^4(x'-y')^4\langle \calo(x) \calo^\dagger(y)\calo(x')\calo^\dagger(y') \rangle~
=~{i\over 2}\!\int\! d\nu~\tilde{ f}_+(\nu)
{\tanh\pi\nu\over \nu}F(\nu)
\Omega(r,\nu)R^{\half\omega(\nu)}
\label{koppinkoop}
\end{eqnarray}
Here $\omega(\nu)\equiv \omega(0,\nu)$ is the pomeron intercept, $\tilde{f}_+(\nu)\equiv \tilde{f}_+(\omega(\nu))$
where  $\tilde{f}_+(\omega)=(e^{i\pi\omega}-1)/\sin\pi\omega$ is the signature factor in the coordinate space, and
 $F(\nu)$ is the ``pomeron residue''  (strictly speaking, the product of two  pomeron residues). The conformal function $\Omega(r,\nu)$
is given by a hypergeometric function (see Ref. \cite{penecostalba}) but for our purposes it is convenient to use
 the representation in terms of the two-dimensional integral
\begin{eqnarray}
&&\hspace{-5mm}
\Omega(r,\nu)~=~{\nu^2\over\pi^3}
\!\int\! d^2z \Big[{-\kappa^2\over (-2\kappa\cdot\zeta)^2}\Big]^{\half +i\nu} \Big[{{-\kappa'}^2\over (-2\kappa'\cdot\zeta)^2}\Big]^{\half -i\nu}
\label{integral7}
\end{eqnarray}
where $\zeta\equiv{ p_1\over s}+z_{\perp}^2p_2+z_{\perp}$  and
\begin{eqnarray}
&&\hspace{-5mm}
\kappa~=~{\sqrt{s}\over 2x_\ast}({p_1\over s}+x^2p_2+x_\perp)-{\sqrt{s}\over 2y_\ast}({p_1\over s}+y^2p_2+y_\perp)
\label{kappas}\\
&&\hspace{-5mm}
\kappa'~=~{\sqrt{s}\over 2x'_\bu}({p_1\over s}+{x'}^2p_2+x'_\perp)-{\sqrt{s}\over 2y'_\bu}({p_1\over s}+{y'}^2p_2+y'_\perp)
\nonumber
\end{eqnarray}
are  two SL(2,C)-invariant vectors \cite{penecostalba} (see also \cite{nlobfklconf})
such that $\kappa^2~=~{s(x-y)^2\over 4x_\ast y_\ast}$, ${\kappa'}^2~=~{s(x'-y')^2\over 4x'_\bu y'_\bu}$ and therefore
\begin{equation}
\kappa^2{\kappa'}^2~=~{1\over R},~~~~~~~~~~~~~~4(\kappa\cdot\kappa')^2~=~{r\over R}
\label{Rr}
\end{equation}
In our limit (\ref{reggelimit}) $x^2=x_\perp^2,~{x'}^2={x'}_\perp^2$ and similarly for $y$.  
Note that all the dependence on large energy 
($\equiv$ large $\rho,\rho'$) is contained in  $R^{\half\omega(\nu)}$.

The dynamical information about the conformal theory is encoded in two functions: pomeron intercept and pomeron residue.
The pomeron intercept is known both in the small and large $\alpha_s$ limit. The NLO intercept at small $\alpha_s$
was calculated in ${\cal N}=4$ SYM by Lipatov and Kotikov \cite{lipkot}
\begin{eqnarray}
&&\hspace{-2mm} 
\omega(\nu)~=~{\alpha_s\over \pi}N_c\Big(\chi(\nu)+{\alpha_sN_c\over 4\pi}\Big[
6\zeta(3)-
{\pi^2\over 3}\chi(\nu)+\chi''(\nu)
-~2\Phi(\nu)-2\Phi(-\nu)\Big]\Big)
\label{eigen2}
\end{eqnarray}
Our main goal is the description of the amplitude in the next-to-leading order in perturbation theory, but 
it is worth noting that the pomeron intercept is known also in the limit of large 
't Hooft coupling $\lambda=4\pi\alpha_sN_c$
\begin{equation}
\hspace{-1mm}
\omega(\nu)+1~=~j(\nu)~=~2-2{\nu^2+1\over \sqrt{\lambda}}
\end{equation}
where 2 is the graviton spin and the first correction was calculated in Ref. \cite{klov,brtan}.

The pomeron residue $F(\nu)$ is known in the leading order both at small \cite{penecostalba, penecostalba2,penedones} and large \cite{cornalba} 't Hooft coupling
\begin{equation}
\hspace{-1mm}
F(\nu)~\stackrel{\lambda\rightarrow 0}{\rightarrow}~\lambda^2{\pi\sin\pi\nu \over 4\nu\cos^3\pi\nu},~~~~
F(\nu)~\stackrel{\lambda\rightarrow \infty}{\rightarrow}~{\pi^3\nu^2(1+\nu)^2\over \sinh^2\pi\nu}
\label{loif}
\end{equation}
To find the NLO amplitude, we must also calculate the ``pomeron residue'' $F(\nu)$ in the next-to-leading order. 
In the rest of this section  we will do this using the four steps of the high-energy operator product expansion in Wilson lines.

\subsection{High-energy OPE in the leading order}
\subsubsection{Leading order: impact factor}
As I discussed above, the main idea behind the high-energy operator expansion is the rapidity factorization.  At the first step, we integrate 
over gluons with rapidities $Y>\eta$ and leave the integration over $Y<\eta$ for later time, see Fig. 2. 
It is convenient to use the background field formalism: we integrate over gluons with $\alpha>\sigma$ and leave gluons with $\alpha<\sigma$ as a background field, to
be integrated over later. The result of the integration is the coefficient
function (``impact factor'') in front of the Wilson-line operators with rapidities up to $\eta=\ln\sigma$:
\begin{eqnarray}
&&\hspace{-2mm} 
 U^\sigma_x~=~{\rm Pexp}\Big[-ig\!\int_{-\infty}^\infty\!\! du ~p_1^\mu A^\sigma_\mu(up_1+x_\perp)\Big],
 ~~~~~~~~~~~~~~~
A^\sigma_\mu(x)~=~\int\!d^4 k ~\theta(\sigma-|\alpha_k|)e^{ik\cdot x} A_\mu(k)
\label{cutoff}
\end{eqnarray}
where  the  Sudakov variable $\alpha_k$ is defined as usual,  $k=\alpha_kp_1+\beta_kp_2+k_\perp$. 
The impact factor is given then by a set of Feynman diagrams in the external field of gluons with $\alpha<\sigma$. Since the rapidities of the background
gluons are very different from the rapidities of gluons in our Feynman diagrams, the background field can be taken in a form of the shock wave due to the Lorentz contraction.
It is very easy to derive the expression of a quark (or gluon) propagator in this shock-wave background.  We represent the propagator as a path integral over various trajectories,
each of them weighed with the gauge factor Pexp$(ig\int\! dx_\mu A^\mu)$ ordered along the propagation path. Now, since the shock wave is very thin, quarks (or gluons) do not
have time to deviate in transverse direction so their trajectory inside the shock wave can be approximated by a segment of the straight line. Morover, since there is no external field 
outside the shock wave 
\footnote{In principle, there may be a pure gauge field outside the shock wave. This does not change our result for the coefficient function in front of color dipole operator, see the discussion
in Ref. \cite{mobzor,npb96}}
the integral over the segment of straight line can be formally extended to $\pm\infty$ limits yielding the Wilson-line operator.

\begin{figure}[htb]
\psfig{file=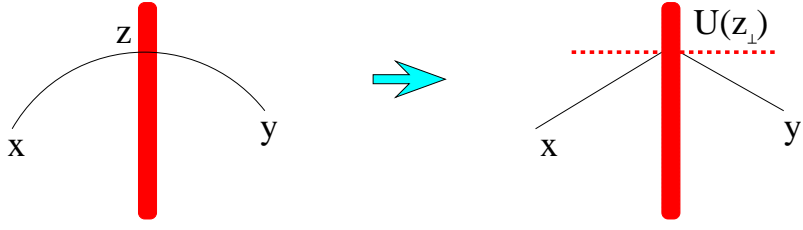,width=106mm} 
\caption{Propagator in a shock-wave background}
\label{fig:shwaveprop}
\end{figure}

Thus, the structure of the propagator in a shock-wave background is as follows: \\ 
$\big[$Free propagation from initial point $x$ to the point of intersection with the shock wave $z\big]$\\
$\times$ $\big[$Interaction
with the shock wave described by the Wilson-line operator $U_z\big]$\\
$\times$ $\big[$Free propagation from point of interaction $z$ to the final point $y\big]$. \\

The explicit form can be taken from Ref. \cite{nlobksym}
\begin{eqnarray}
&&\hspace{-4mm}
\langle \hat{\Phi}_I(x)\hat{\Phi}_J(y)\rangle_{\rm shockwave}
\stackrel{x_\ast>0>y_\ast}{=}~2i\delta^{IJ}\!\int d^4 z\delta(z_\ast) {1\over 4\pi^2[(x-z)^2+i\epsilon]}~U^{ab}_{z_\perp}~
{\partial^{(z)}_\ast}{1\over 4\pi^2[(z-y)^2+i\epsilon]}
\nonumber \\
&&\hspace{-2mm}
=~{s^2\delta^{IJ}\over 64\pi^3x_\ast y_\ast}\!\int_0^\infty\! d\alpha~\alpha e^{i\alpha{s\over 4}\calz}~=~-{\delta^{IJ}\over 4\pi^3x_\ast y_\ast\calz^2}
\label{scalar-sw1}
\end{eqnarray}
where $\calz\equiv -{4\over\sqrt{s}}(\kappa\cdot\zeta)=-{4\over s}(x-y)_\bu+{(x-z)_\perp^2\over x_\ast}-{(y-z)_\perp^2\over y_\ast}$. Note that the interaction with the shock wave does not change the 
$\alpha$-component of the scalar particle's momentum.

Let us  calculate the impact factor taking $x_\bu=y_\bu=0$ for simplicity. 
The leading-order impact factor is proportional to the product of two propagators (\ref{scalar-sw1}), see Fig. \ref{fig:loif}:
\begin{figure}[htb]
    \includegraphics[width=50mm]{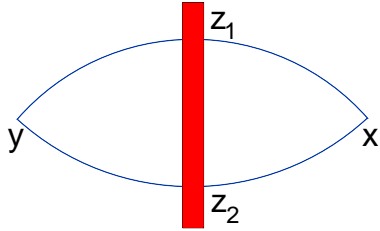}
\caption{Impact factor in the leading order. \label{fig:loif}}
\end{figure}
%
\begin{eqnarray}
&&
\langle T\{\halo(x)\halo(y)\}\rangle^{\rm LO}_{\rm shockwave}~=~{(x-y)^{-4}\over \pi^2(N_c^2-1)}\!\int\! {d^2 z_{1\perp}d^2 z_{2\perp}\over z_{12}^4}
~\calr^2~{\rm Tr}\{U_{z_1}U^\dagger_{z_2}\}
\label{loifscal}
\end{eqnarray}
where
\begin{equation}
\calr~=~-{(x-y)^2z_{12}^2\over x_\ast y_\ast \calz_1\calz_2}~=~{\kappa^2(\zeta_1\cdot\zeta_2)\over 2(\kappa\cdot\zeta_1)(\kappa\cdot\zeta_2)}
\label{calr}
\end{equation}
and  $\zeta_i\equiv {p_1\over s}+z_i^2 p_2+z_i^\perp$, $\calz_i\equiv -{4\over\sqrt{s}}(\kappa\cdot\zeta_i)={(x-z_i)^2\over x_\ast}-{(y-z_i)^2\over y_\ast}$. Note that the leading-order impact factor is 
conformally (M\"obius) invariant - it goes into itself under the inversion (\ref{inversion}). 

This formula can be promoted to the operator equation as follows
\begin{eqnarray}
&&
(x-y)^4T\{\halo(x)\halo(y)\}^{\rm LO}~=~{1\over \pi^2(N_c^2-1)}\!\int\! {d^2 z_1d^2 z_2\over z_{12}^4}
~\calr^2~{\rm Tr}\{\hat{U}^{\sigma_A}_{z_1}\hat{U}^{\dagger \sigma_A}_{z_2}\}
\label{lope}
\end{eqnarray}
where $\sigma_A\sim {\sqrt{x_\ast|y_\ast|}\over s(x-y)^2}$ is the characteristic $\alpha's$ in the scalar loop which serve as an upper bound for rapidity of Wilson-line gluons.
(Recall that $|y_\ast|$ and $x_\ast$ are of the same order of magnitude as seen from Eq. (\ref{reggelimit})).

 We'll need later the projection of the T-product in the l.h.s. of this equation onto the conformal eigenfunctions of the BFKL equation \cite{lip86}
\begin{equation}
\hspace{-0mm}
E_{\nu,n}(z_{10},z_{20})~
=~\Big[{\tilde{z}_{12}\over \tilde{z}_{10}\tilde{z}_{20}}\Big]^{\half+i\nu+{n\over 2}}
\Big[{\barz_{12}\over \barz_{10}\barz_{20}}\Big]^{\half+i\nu-{n\over 2}}
\label{eigenfunctions}
\end{equation}
(here $\tilde{z}=z_x+iz_y,\barz=z_x-iz_y$, $z_{10}\equiv z_1-z_0$ etc.). 
Since $\halo$'s are scalar operators,  the only non-vanishing contribution comes from projection on the eigenfunctions with spin $0$:
\begin{eqnarray}
&&\hspace{-2mm}
{1\over\pi^2}\!\int\! {dz_1dz_2\over z_{12}^4}~
\calr^2\Big[{z_{12}^2\over z_{10}^2z_{20}^2}\Big]^{\half+i\nu}
~=~\Big[{-\kappa^2\over (-2\kappa\cdot\zeta_0)^2}\Big]^{\half+i\nu} {\Gamma^2\big(\half-i\nu\big)\over \Gamma (1-2i\nu)}
{\big({1\over 4}+\nu^2\big)\pi\over \cosh\pi\nu}
\label{projlo}
\end{eqnarray} 
where  
$\zeta_0\equiv{ p_1\over s}+z_{0\perp}^2p_2+z_{0\perp}$.

Now, using the decomposition of the product of the transverse $\delta$-functions in conformal 3-point functions $E_{\nu,n}(z_{10},z_{20})$ \cite{lip86}
\begin{equation}
\hspace{-1mm}
\delta^{(2)}(z_1-w_1)\delta^{(2)}(z_2-w_2)~
=~\sum_{n=-\infty}^\infty\!\int\! {d\nu\over \pi^4}~{\nu^2+{n^2\over 4}\over z_{12}^2w_{12}^2}
\int\! d^2\rho~ E^\ast_{\nu,n}(w_1-\rho,w_2-\rho)E_{\nu,n}(z_1-\rho,z_2-\rho)
\label{lobzor120}
\end{equation}
we obtain (dots stand for contributions of higher spins $n$ which we do not need for our correlator (\ref{correl1}))
\begin{eqnarray}
&&\hspace{-1mm}      
(x-y)^4 T\{\halo(x)\halo^\dagger(y)\}~
~=~-\int\! d\nu\!\int\! d^2z_0~{\nu^2(1+4\nu^2)\over 4\pi\cosh\pi\nu}
{\Gamma^2\big(\half-i\nu\big)\over\Gamma(1-2i\nu)}
~\Big({-\kappa^2\over (-2\kappa\cdot\zeta_0)^2}\Big)^{\half +i\nu}
 \hat{\cal U}^{\sigma_A}(\nu,z_0)~+~...
 \label{lopeu}
 \end{eqnarray}
where 
\begin{eqnarray}
&&\hspace{-1mm}
 \hat{\cal U}^\sigma(\nu,z_0)~\equiv~{1\over \pi^2}\!\int\! {d^2z_1d^2z_2\over z_{12}^4}~
\Big({z_{12}^2\over z_{10}^2z_{20}^2}\Big)^{\half -i\nu}~ \hat{\cal U}^\sigma(z_1,z_2)
\label{opspin}
\end{eqnarray}
and $\hat{\cal U}^\sigma(z_1,z_2)$ is a ``color dipole in the adjoint representation''
 \begin{equation}
 \hat{\cal U}^\sigma(z_1,z_2)~=~1-{1\over N_c^2-1}{\rm Tr}\{ \hat{U}^\sigma_{z_1} \hat{U}^{\dagger\sigma}_{z_2}\}
 \label{colodipole}
 \end{equation}
 %

\subsubsection{Leading order: BK equation}
Next step is to obtain the evolution equation for color dipoles in the leading order in $\alpha_s$.  For the light-cone dipoles,
the contribution of scalar operators to Maldacena-Wilson line \cite{mwline} vanishes so one has the usual Wilson line constructed from gauge fields and therefore the LLA evolution
equation for color dipoles in the ${\cal N}=4$ SYM has the same form as in QCD.

\begin{figure}[htb]
\centering
\includegraphics[width=1.0\textwidth]{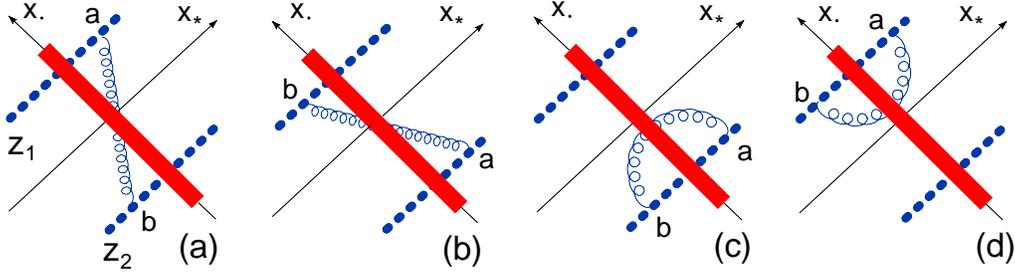}
\hspace{-4.6cm}
\vspace{-1cm}
\caption{Leading-order diagrams for the small-$x$ evolution of color dipole\label{fig:bkevol}. Gauge links are denoted by dotted lines.}
\end{figure}

To find the evolution of the color dipole (\ref{fla1}) with respect to rapidity of the 
Wilson lines in the leading log approximation
we consider the matrix element of the color dipole between (arbitrary) target states and 
integrate over the gluons with rapidities $Y_1>Y>Y_2=Y_1-\Delta Y$ leaving the gluons with $Y<Y_2$ as
a background field (to be integrated over later).
In the frame of gluons with $Y\sim Y_1$ the fields with
$Y<Y_2$ shrink to a pancake and we obtain the four diagrams shown in Fig. 
\ref{fig:bkevol}. Technically,  to find the kernel in the leading-ordrer approximation we 
write down the general form of the operator equation for the evolution of the color dipole 

\begin{eqnarray}
&&\hspace{-6mm}
{d \over dY}{\rm Tr}\{\hat{U}^Y_{z_1}\hat{U}^{\dagger Y}_{z_2}\}=
K_{\rm LO}{\rm Tr}\{\hat{U}^Y_{z_1}\hat{U}^{\dagger Y}_{z_2}\}+...
\label{eveq}
\end{eqnarray}
(where dots stand for the higher orders of the expansion) 
and calculate the l.h.s. of Eq. (\ref{eveq}) in the shock-wave background
\begin{eqnarray}
&&\hspace{-2mm}
{d \over dY}\langle{\rm Tr}\{\hat{U}^Y_{z_1}\hat{U}^{\dagger Y}_{z_2}\}\rangle_{\rm shockwave}=
\langle K_{\rm LO}{\rm Tr}\{\hat{U}^Y_{z_1}\hat{U}^{\dagger Y}_{z_2}\}\rangle_{\rm shockwave}
\label{eveqmaels}
\end{eqnarray}
In what follows we replace $\langle ...\rangle_{\rm shockwave}$ 
by $\langle ...\rangle$ for brevity.

With future NLO computation in view, we will perform the leading-order calculation in the lightcone gauge $~p_2^\mu A_\mu=0$. 
The  gluon propagator in a shock-wave external field has the form\cite{prd99,balbel} 
\begin{eqnarray}
&&\hspace{-4mm}
\langle \hat{A}^a_\mu(x)\hat{A}^b_\nu(y)\rangle
~\stackrel{x_\ast>0>y_\ast}{=}~-{i\over 2}\int d^4 z~\delta(z_\ast)~
{x_\ast g^\perp_{\mu\xi}-p_{2\mu}(x-z)^\perp_\xi \over \pi^2[(x-z)^2+i\epsilon]^2}\;U^{ab}_{z_\perp}
{1\over\partial_\ast^{(z)}}~{y_*\delta^{\perp\xi}_\nu
 - p_{2\nu}(y-z)_\perp^\xi\over \pi^2[(z-y)^2+i\epsilon]^2}
\label{gluon-sw1}
\end{eqnarray}
where ${1\over \partial_\ast}$ can be either ${1\over \partial_\ast+i\epsilon }$ or 
${1\over \partial_\ast-i\epsilon }$ which leads to the same result. (This is obvious for the leading order and 
correct in NLO after subtraction of the leading-order contribution, see Eq. (\ref{nloifscal}) below).

We obtain
\begin{equation}
\hspace{-0mm}
g^2\!\int_0^\infty \! du\! \int^0_{-\infty} \! 
dv~\langle\hat{A}^{a,Y_1}_\bu(up_1+x_\perp)
\hat{A}^{b,Y_1}_\bu(vp_1+y_\perp)\rangle_{\rm Fig. \ref{fig:bkevol}a}
~=~-4\alpha_s
\int_0^{e^{Y_1}}\!{d\alpha\over\alpha}
(x_\perp|{p_i\over p_\perp^2-i\epsilon}
U^{ab}{p_i\over p_\perp^2-i\epsilon}|y_\perp)
\label{bk1}
\end{equation}
Hereafter we use Schwinger's notations 
$(x_\perp|F(p_\perp)|y_\perp)\equiv \int\!\dhd p~e^{i(p,x-y)_\perp}F(p_\perp)$ 
(the scalar product of the four-dimensional vectors in our notations is 
$x\cdot y=-{2\over s}(x_\ast y_\bullet+x_\ast y_\bullet)+(x,y)_\perp$). 
Note that the interaction with the shock wave does not change the $\alpha$-component
of the gluon momentum, same as for the scalar propagator (\ref{scalar-sw1})

Formally, the integral over $\alpha$ diverges at the lower limit, but since we integrate over the rapidities $Y>Y_2$ in the leading log approximation, we get  ($\Delta Y\equiv Y_1-Y_2$)
\begin{eqnarray}
&&\hspace{-26mm}
g^2\!\int_0^\infty \! du \int^0_{-\infty} \! dv~\langle \hat{A}^{a,Y_1}_\bu(up_1+x_\perp)
\hat{A}^{b,Y_1}_\bu(vp_1+y_\perp)\rangle_{\rm Fig. \ref{fig:bkevol}a}
~=~-4\alpha_s\Delta Y
(x_\perp|{p_i\over p_\perp^2}U^{ab}{p_i\over p_\perp^2}|y_\perp)
\label{bk2}
\end{eqnarray}
 and therefore
\begin{eqnarray}
&&\hspace{-2mm}\langle \hat{U}^Y_{z_1}\otimes \hat{U}^{\dagger Y}_{z_2}\rangle_{\rm Fig. \ref{fig:bkevol}a}^{Y_1}
~=~-{\alpha_s\over \pi^2}\Delta Y~ (T^aU_{z_1}\otimes T^bU_{z_2}^\dagger)
\!\int\! d^2z_3 {(z_{13},z_{23})\over z_{13}^2z_{23}^2}
U_{z_3}^{ab}
\label{bk3}
\end{eqnarray}
(hereafter $T^a_{mn}\equiv-if^{amn}$).
The contribution of the diagram in Fig.  \ref{fig:bkevol}b is obtained from Eq. (\ref{bk3})
by the replacement $T^aU_{z_1}\otimes T^bU_{z_2}^\dagger \rightarrow U_{z_1} T^b\otimes U_{z_2}^\dagger T^a$, $z_2\leftrightarrow z_1$ and the two remaining diagrams are obtained from
Eq. \ref{bk2} by taking $z_2=z_1$ (Fig. \ref{fig:bkevol}c) and  $z_1=z_2$ (Fig. \ref{fig:bkevol}d).
Finally, one has
\begin{eqnarray}
&&\hspace{-2mm}
\langle{\rm Tr}\{\hat{U}^{Y_1}_{z_1} \hat{U}^{\dagger Y_1}_{z_2}\}\rangle_{\rm Fig. \ref{fig:bkevol}}
~=~{\alpha_s\Delta Y\over \pi^2} 
\!\int\! d^2z_3 {z_{12}^2\over z_{13}^2z_{23}^2}
[{\rm Tr}\{T^aU_{z_1} U^\dagger_{z_3}T^aU_{z_3} U_{z_2}^\dagger\}
-{1\over N_c}{\rm Tr}\{U_{z_1} U_{z_2}^\dagger\}]
\label{bk5}
\end{eqnarray}
There are also contributions coming from  diagrams similar to Fig. \ref{fig:bkevol} but without the gluon-shockwave intersection.
These diagrams are proportional to the original dipole ${\rm Tr}\{U_{z_1} U_{z_2}^\dagger\}$ 
and therefore the corresponding term can be derived from the contribution
of Fig. \ref{fig:bkevol} graphs using the requirement that the r.h.s. of the evolution equation
should vanish in the absence  of the shock wave (when $U\equiv 1$).
It is easy to see that this requirement leads to 
\begin{eqnarray}
&&\hspace{-2mm}
\langle{\rm Tr}\{\hat{U}^{Y_1}_{z_1} \hat{U}^{\dagger Y_1}_{z_2}\}\rangle
~=~{\alpha_s\Delta Y\over \pi^2} 
\!\int\! d^2z_3 {z_{12}^2\over z_{13}^2z_{23}^2}
[{\rm Tr}\{T^aU_{z_1} U^\dagger_{z_3}T^aU_{z_3} U_{z_2}^\dagger\}
-N_c{\rm Tr}\{U_{z_1} U_{z_2}^\dagger\}]
\nonumber
\end{eqnarray}
which gives  the BK equation for the evolution of the color dipole in the adjoint representation:.
\begin{eqnarray}
&&\hspace{-2mm}
{d\over dY}{\rm Tr}\{\hat{U}^Y_{z_1} \hat{U}^{\dagger Y}_{z_2}\}
~=~{\alpha_s\over \pi^2} 
\!\int\! d^2z_3 {z_{12}^2\over z_{13}^2z_{23}^2}
[{\rm Tr}\{T^a\hat{U}^Y_{z_1} \hat{U}^{\dagger Y}_{z_3}T^a\hat{U}^Y_{z_1} \hat{U}^{\dagger Y}_{z_2}\}
-N_c{\rm Tr}\{\hat{U}^Y_{z_1} \hat{U}^{\dagger Y}_{z_2}\}]
\label{bkadj}
\end{eqnarray}

\subsubsection{Leading order: BFKL evolution of color dipoles}
Next step is the evolution of color dipole.
To find the amplitude (\ref{correl1}) in the leading order (and NLO as well)  it is sufficient to take into account only the linear evolution of Wilson-line operators
which corresponds to taking into account only two gluons in the t-channel.
The non-linear effects in the evolution (and the production) of t-channel gluons enter the four-current amplitude (\ref{correl1}) in the form of so-called
``pomeron loops'' which start from the NNLO BFKL order.
 With this two-gluon accuracy 
\begin{eqnarray}
&&\hspace{-1mm} {1\over N_c}{\rm Tr}\{T^n\hat{U}^\sigma_{z_1}\hat{U}^{\dagger\sigma}_{z_3}T^n\hat{U}^\sigma_{z_3}\hat{U}^{\dagger\sigma}_{z_2}\}
-{\rm Tr}\{\hat{U}^\sigma_{z_1}\hat{U}^{\dagger\sigma}_{z_2}\}~=~-{1\over 2}(N_c^2-1)\big[\hat{\cal U}^\sigma(z_1,z_3)+\hat{\cal U}^\sigma(z_2,z_3)-\hat{\cal U}^\sigma(z_1,z_2)\big]
\nonumber
\end{eqnarray}
where $\hat{\cal U}^\sigma(z_1,z_2)$ is a color dipole in the adjoint representation, see Eq. (\ref{colodipole}).
 The BFKL equation 
for $\hat{\cal U}^\sigma(z_1,z_2)$  takes the form (recall that $\sigma=e^Y$)
\begin{eqnarray}
&&\hspace{-2mm}
\sigma{d\over d\sigma}\hat{\cal U}^\sigma(z_1,z_2)
~=~\!\int\!d^2z_3d^2z_4 K_{\rm LO}(z_1,z_2;z_3,z_4)~\hat{\cal U}^\sigma(z_3,z_4)
\label{evol}
\end{eqnarray}
where
\begin{eqnarray}
&&\hspace{-4mm}
K_{\rm LO}(z_1,z_2;z_3,z_4)~=~{\alpha_sN_c\over 2\pi^2}\Big[{z_{12}^2\delta^{2}(z_{13})\over z_{14}^2z_{24}^2}
+{z_{12}^2\delta^{2}(z_{24})\over z_{13}^2z_{23}^2}
-~\delta^{2}(z_{13})\delta^{2}(z_{24})\!\int\! d^2z~{z_{12}^2\over (z_1-z)^2(z_2-z)^2}\Big]
\label{klo}
\end{eqnarray}
 The solution of this equation is easily formulated in terms of $\hat{\cal U}^\sigma(\nu,z_0)$ - projection of color dipole on Lipatov's eigenfunctions (\ref{eigenfunctions}):
\begin{equation}
\hspace{-0mm}
\hat{\cal U}^\sigma(\nu,z_0)~=~(\sigma/\sigma_0)^{\omega(\nu)} \hat{\cal U}^{\sigma_0}(\nu,z_0)
\label{evolresulo}
\end{equation}
%
\subsubsection{LO: amplitude}
 The last step is a matrix element of the color dipole operator $\hat{\cal U}^{\sigma_0}(\nu,z_0)$ ``between scalar states'', i.e. the correlator
of color dipole and ``bottom pair'' of scalar operators $\calo(x')$ and $\calo(y')$ in the leading order in perturbation theory. 
The easiest way to get this matrix element is to write down the high-energy OPE for the bottom pair of operators similar to Eq. (\ref{loif})
\begin{eqnarray}
&&\hspace{-1mm}      
(x'-y')^4 T\{\halo(x')\halo^\dagger(y')\}~
=~-\!\int\! d\nu'\!\int\! d^2z'_0~{{\nu'}^2(1+4{\nu'}^2)\over 4\pi\cosh\pi\nu'}
{\Gamma^2\big(\half-i\nu'\big)\over\Gamma(1-2i\nu')}
~\Big({-{\kappa'}^2\over (-2\kappa'\cdot\zeta'_0)^2}\Big)^{\half +i\nu'}
 \hat{\cal V}^{\lambda_B}(\nu',z'_0).
 \label{lopev}
 \end{eqnarray}
Here $\zeta'_0\equiv p_1+{{z'}_{0}^2\over s}p_2+z'_{0\perp}$,  
$b_0={\kappa'}^{-2}+i\epsilon={4 x'_\bu y'_\bu\over s(x'-y')^2}+i\epsilon$, and
\begin{equation}
 \hat{\cal V}^{\lambda}(\nu',z'_0)~=~{1\over\pi^2}\!\int\! {d^2z_1d^2z_2\over z_{12}^4}~
\Big({z_{12}^2\over z_{10}^2z_{20}^2}\Big)^{\half -i\nu'}~ \hat{\cal V}^\lambda(z_1,z_2),
\label{voper}
\end{equation}
where the operator  is made from the dipoles  
$
 \hat{\cal V}^\lambda(z'_1,z'_2)~=~1-{1\over N_c^2-1}{\rm Tr}\{ \hat{V}^\lambda_{z'_1} \hat{V}^{\dagger\lambda}_{z'_2}\}
 $
(cf. Eq. (\ref{colodipole}))  ordered along the straight line $\parallel~p_2$ with the rapidity restriction
\begin{eqnarray}
&&\hspace{-2mm} 
 V^\lambda_x~=~{\rm Pexp}\Big[-ig\!\int_{-\infty}^\infty\!\! du~ p_1^\mu A^\lambda_\mu(up_2+x_\perp)\Big],
~~~~~~~~~~~~~~~
A^\lambda_\mu(x)~=~\int\! d^4 k ~\theta(\lambda-|\beta_k|)e^{ik\cdot x} A_\mu(k)
\label{cutoffv}
\end{eqnarray}
Similarly to the case of  the upper impact factor discussed above, the cutoff $\lambda$  for $\beta$ integration in Eq. (\ref{cutoffv}) should be chosen
of order of characteristic $\beta$'s in the lower impact factor so $\lambda_B\sim {\sqrt{x'_\bu |y'_\bu|}\over s(x'-y')^2}$.

In the leading order in perturbation theory
\begin{eqnarray}
&&\hspace{-2mm}
\langle\calu(z_1,z_2)\calv(w_1,w_2)\rangle~=~-{\alpha_s^2\pi^2N_c^2\over 2(N_c^2-1)}\ln^2{(z_1-w_1)^2(z_2-w_2)^2\over (z_1-w_2)^2(z_2-w_1)^2}~
\label{nadop1}
\end{eqnarray}
which will be true in the LLA as long as the $\alpha$ and $\beta$ cutoffs do not allow large logarithms $\ln {\alpha\beta s\over k_\perp^2}$ where 
$k_\perp^2$ is characteristic transverse momentum in gluon ladder describing the BFKL evolution. 
 (This is similar to taking $\mu^2$ around 1 GeV for the initial point of the DGLAP evolution so the logarithms $\ln{\mu^2\over m_p^2}$ can be neglected). Thus, if we choose the final point of evolution 
(\ref{evolresulo}) to be $\sigma_0\sim {k_\perp^2\over\lambda_Bs}\sim (|x-y|_\perp |x'-y'|_\perp\lambda_Bs)^{-1}$, the correlator of color dipoles
 $\langle\calu^{\sigma_0}(z_1,z_2)\calv^{\lambda_B}(w_1,w_2)\rangle$ will be given by Eq. (\ref{nadop1}) which translates to 
\begin{equation}
\hspace{-0mm}
\langle\hat{\cal U}^{\sigma_0}(\nu,z_0)\hat{\cal V}^{\lambda_B}(\nu',z'_0)\rangle~
=~-{\alpha_s^2N_c^2\over N_c^2-1}
{16\pi^2\over \nu^2(1+4\nu^2)^2}
\Big[\delta(z_0-z'_0)\delta(\nu+\nu')
+~{2^{1-4i\nu}\delta(\nu-\nu')\over \pi|z_0-z'_0|^{2-4i\nu}}
{\Gamma\big({1\over 2}+i\nu\big)\Gamma(1-i\nu)\over\Gamma(i\nu)\Gamma\big(\half-i\nu\big)}\Big]
\label{fla59}
 \end{equation}
where we used the following orthogonality condition for the eigenfunctions (\ref{eigenfunctions}), see Ref. \cite{lip86}:
\begin{eqnarray}
&&\hspace{-1mm}
\int\! {d^2z_1 d^2z_2\over z_{12}^4}~ E^\ast_{\nu',m}(z_1-z'_0,z_2-z'_0)E_{\nu,n}(z_1-z_0,z_2-z_0)~=~{\pi^4\over 2\big(\nu^2+{n^2\over 4}\big)}\Bigg[\delta(\nu-\nu')\delta_{m,n}\delta^{(2)}(z_0-z'_0)
\label{ortho}\\
&&\hspace{-1mm}
+~\delta(\nu+\nu')\delta_{m,-n}(\tilde{z}_0-\tilde{z}'_0)^{-1+n-2i\nu}(\bar{z}_0-\bar{z}'_0)^{-1-n-2i\nu}{2^{4i\nu+1}\over \pi}
\Big({|n|\over 2}+i\nu\Big){\Gamma\big({1+|n|\over 2}-i\nu\big)\Gamma\big({|n|\over 2}+i\nu\big)\over \Gamma\big({1+|n|\over 2}+i\nu\big)\Gamma\big({|n|\over 2}-i\nu\big)}\Bigg]
\nonumber
 \end{eqnarray}
With our choice of $\sigma_0$ for the endpoint of the evolution of the color dipole (\ref{evolresulo}) the correlator of two color dipoles $\hat{\cal U}^{\sigma_A}(\nu,z_0)$ and 
$\hat{\cal V}^{\lambda_B}(\nu',z'_0)$ takes the form
\begin{eqnarray}
&&\hspace{-1mm}
\langle\hat{\cal U}^{\sigma_A}(\nu,z_0)\hat{\cal V}^{\lambda_B}(\nu',z'_0)\rangle~
\label{confdipsca}\\
&&\hspace{-1mm}
=~-{\alpha_s^2N_c^2\over N_c^2-1}{32\pi^2\over \nu^2(1+4\nu^2)^2}\Big({x_\ast y_\ast x_\bu y_\bu\over s(x-y)_\perp^2(x'-y')_\perp^2}\Big)^{\omega(\nu)\over 2}
\Big[\delta(z_0-z'_0)\delta(\nu+\nu')+{2^{1-4i\nu}\delta(\nu-\nu')\over \pi|z_0-z'_0|^{2-4i\nu}}
{\Gamma\big({1\over 2}+i\nu\big)\Gamma(1-i\nu)\over\Gamma(i\nu)\Gamma\big(\half-i\nu\big)}\Big].
\nonumber
 \end{eqnarray}
Combining now Eqs. (\ref{lopeu}),  (\ref{lopev}), and the above equation we see that the  leading-order amplitude is given by Eq. (\ref{koppinkoop}) 
with $\omega_0(\nu)={\alpha_2N_c\over\pi}\chi(\nu)$ and 
\begin{eqnarray}
&&\hspace{-0mm}
F_0(\nu)~=~{N_c^2\over N_c^2-1}{4\pi^4\alpha_s^2\over\cosh^2\pi\nu}~
\end{eqnarray}
which agrees with the leading-order impact factor calculated in Refs. \cite{penecostalba,penecostalba2}.
Here we used the integral
\begin{eqnarray}
&&\hspace{-2mm}
\int\! {d^2z'_0\over [(z_0-z'_0)^2]^{1-2i\nu}}
\Big[{{-\kappa'}^2\over (-2\kappa'\cdot\zeta'_0)^2}\Big]^{\half+i\nu}
~=~{\pi\over 2i\nu}\Big[{{-\kappa'}^2\over (-2\kappa'\cdot\zeta_0)^2}\Big]^{\half-i\nu}
\end{eqnarray}
%

\subsection{NLO: Rapidity factorization}
Now we repeat the same four steps of operator expansion at the NLO accuracy. 
A general form of the expansion of  T-product of the currents $\calo(x)$ and $\calo(y)$ 
 in color dipoles looks as follows:
\begin{eqnarray}
&&\hspace{-1mm}
 T\{\halo(x)\halo(y)\}~=~\int\! d^2z_1d^2z_2~I^{\rm LO}(z_1,z_2)
 {\rm Tr}\{\hat{U}^Y_{z_1}\hat{U}^{\dagger Y}_{z_2}\}
 \nonumber\\
&&\hspace{-1mm}
+\int\! d^2z_1d^2z_2d^2z_3~I^{\rm NLO}(z_1,z_2,z_3)
[ {\rm Tr}\{T^n\hat{U}^Y_{z_1}\hat{U}^{\dagger Y}_{z_3}T^n\hat{U}^Y_{z_3}\hat{U}^{\dagger Y}_{z_2}\}
 -N_c {\rm Tr}\{\hat{U}^Y_{z_1}\hat{U}^{\dagger Y}_{z_2}\}]
 \label{opeq}
 \end{eqnarray}
(structure of the NLO contribution is clear from the topology of diagrams in the shock-wave background, see Fig. \ref{fig:nloif} below). 

The NLO impact factor for two $Z^2$ currents is given by the diagrams shown in Fig. \ref{fig:nloif}.
\begin{figure}[htb]
\includegraphics[width=0.9\textwidth]{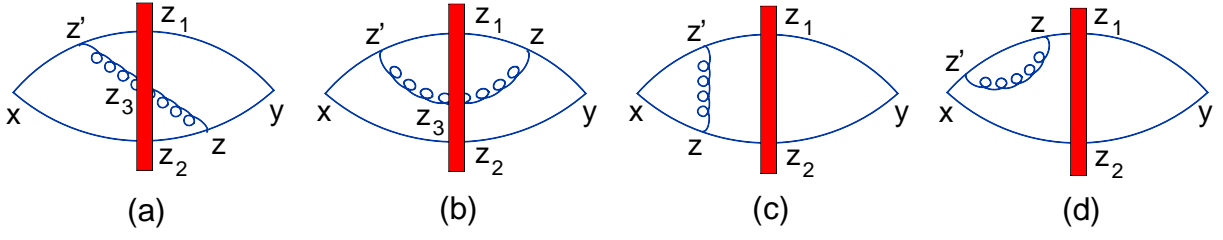}
\caption{Diagrams for the NLO impact factor. \label{fig:nloif}}
\end{figure}
The gluon propagator in the shock-wave background at $x_\ast>0>y_\ast$ in the light-like gauge $p_2^\mu A_\mu=0$ is given by Eq. (\ref{gluon-sw1})
To calculate the next-to-leading impact factor we also need the three-point scalar-scalar-gluon vertex Green function (vertex with tails) which is proportional to
\begin{equation}
\hspace{-2mm}
\int\! d^4z
~\Big[{1\over (x-z)^2}\stackrel{\leftrightarrow}{\partial_\mu}
{1\over(z-y)^2}\Big]{z_\nu\over z^4}-\mu\leftrightarrow\nu~=~4i\pi^2{x_\mu y_\nu-x_\nu y_\mu\over x^2y^2(x-y)^2}
\label{conformalinteg}
\end{equation}
Using this formula, one obtains the contribution of Fig. \ref{fig:nloif}a,b diagrams in the form (the details of the calculation are presented in Ref. \cite{nlobksym})
\begin{equation}
\hspace{1mm}
\langle T\{\halo(x)\halo^\dagger(y)\}\rangle^{\rm Fig. \ref{fig:nloif}a,b}
~=~                     
{\alpha_s\over (N^2_c-1)\pi^4x_\ast^2 y_\ast^2}\!\int\! {d^2z_1d^2z_2 \over\calz_1^2\calz_2^2}\!\int\! d^2z_3
{z_{12}^2\over z_{13}^2z_{23}^2}
{\rm Tr}\{T^n U_{z_1}U^\dagger_{z_3}T^nU_{z_3}U^\dagger_{z_2}\} 
\!\int_0^\infty\!{d\alpha\over\alpha}~e^{i\alpha {s\over 4}\calz_3}         
\label{ot5}           
\end{equation}

Let us discuss now the contribution of Fig. \ref{fig:nloif}c,d diagrams. Since this contribution is proportional to 
${\rm Tr}\{U_{z_1}U^\dagger_{z_2}\} $ it can be restored from the comparison of Eq. (\ref{ot5}) 
with the pure perturbative series for the correlator $\langle T\{\halo(x)\halo(y)\}\rangle$. If we 
switch off the shock wave the contribution of the Fig. \ref{fig:nloif} diagrams is given by the second term in 
Eq. (\ref{ot5}) (with $U,U^\dagger$ repaced by 1). On the other hand, perturbative series for the 
correlator $\langle T\{\halo(x)\halo(y)\}\rangle$ vanishes \cite{vanish} and therefore the contribution of the 
Fig. \ref{fig:nloif}c,d diagrams should be equal to the second term in 
the r.h.s. Eq. (\ref{ot5}) with opposite sign. 
Thus, we get
\begin{eqnarray}
&&\hspace{-1mm}
\langle T\{\halo(x)\halo^\dagger(y)\}\rangle_A^{\rm Fig. \ref{fig:nloif}}
\nonumber\\
&&\hspace{-1mm}=~                     
{\alpha_s\over (N^2_c-1)\pi^4x_\ast^2 y_\ast^2}\!\int\! {d^2z_1d^2z_2 \over\calz_1^2\calz_2^2}\!\int\! d^2z_3
{z_{12}^2\over z_{13}^2z_{23}^2}\Big[
{\rm Tr}\{T^n U_{z_1}U^\dagger_{z_3}T^nU_{z_3}U^\dagger_{z_2}\} 
-N_c{\rm Tr}\{U_{z_1}U^\dagger_{z_2}\} \Big]
\!\int_0^\infty\!{d\alpha\over\alpha}~e^{i\alpha {s\over 4}\calz_3}   
\label{ot5and6}           
\end{eqnarray}

The integral over $\alpha$ in the r.h.s. of Eq. (\ref{ot5and6}) diverges. This divergence reflects the fact that the r.h.s.  of Eq. (\ref{ot5and6}) is not exactly the NLO impact factor since we must subtract the 
matrix element of the leading-order contribution. Indeed,  the NLO impact factor is a coefficient function defined according to Eq. (\ref{opeq}).
To find the NLO impact factor, we consider the operator equation (\ref{opeq}) in 
the shock-wave background (in the leading order $\langle\hat{U}_{z_3}\rangle_A=U_{z_3}$):
\begin{eqnarray}
&&\hspace{-1mm}
 \langle T\{\halo(x)\halo(y)\}\rangle_A~-\int\! d^2z_1d^2z_2~I^{\rm LO}(x,y;z_1,z_2)
  \langle{\rm Tr}\{\hat{U}^Y_{z_1}\hat{U}^{\dagger Y}_{z_2}\}\rangle_A
 \nonumber\\
&&\hspace{-1mm}
=~\int\! d^2z_1d^2z_2d^2z_3~I^{\rm NLO}(x,y;z_1,z_2,z_3;Y)
[ {\rm Tr}\{T^nU_{z_1}U^\dagger_{z_3}T^nU_{z_3}U^\dagger_{z_2}\}
 -N_c {\rm Tr}\{U_{z_1}U^\dagger_{z_2}\}]
 \label{opeq1}
 \end{eqnarray}
The NLO matrix element $ \langle T\{\halo(x)\halo(y)\}\rangle_A$ is given by Eq. (\ref{ot5and6}) 
while 
\begin{eqnarray}
&&\hspace{-1mm}
\int\! d^2z_1d^2z_2~I^{\rm LO}(x,y;z_1,z_2)
  \langle{\rm Tr}\{\hat{U}^Y_{z_1}\hat{U}^{\dagger Y}_{z_2}\}\rangle_A
 \nonumber\\
&&\hspace{-1mm}
=~{(x_\ast y_\ast)^{-2}\over \pi^2(N_c^2-1)}\!\int\! d^2z_1d^2z_2~
{1\over\calz_1^2\calz_2^2}{\alpha_s\over\pi^2}\!\int_0^\sigma\!{d\alpha\over\alpha}\!\int\! d^2z_3
{z_{12}^2\over z_{13}^2z_{23}^2}
[ {\rm Tr}\{T^nU_{z_1}U^\dagger_{z_3}T^nU_{z_3}U^\dagger_{z_2}\}
 -N_c {\rm Tr}\{U_{z_1}U^\dagger_{z_2}\}]
 \label{mael1}
 \end{eqnarray}
as follows from Eqs. (\ref{loifscal}) and (\ref{bkadj}).
The $\alpha$ integration is cut from above by $\sigma=e^Y$ in accordance with the definition of operators $\hat{U}^Y$  (\ref{cutoff}). Subtracting  (\ref{mael1}) from  Eq. (\ref{ot5and6}) we get
\begin{eqnarray}
&&\hspace{-1mm}
I^{\rm NLO}(x,y;z_1,z_2,z_3;Y)~=~{\alpha_s(x_\ast y_\ast)^{-2}\over\pi^4(N_c^2-1)}
{z_{13}^2\over z_{12}^2z_{23}^2\calz_1^2\calz_2^2}
\Big[\!\int_0^\infty\!{d\alpha\over\alpha}~e^{i\alpha {s\over 4}\calz_3}   
-\!\int_0^\sigma\!{d\alpha\over\alpha}\Big]
 \nonumber\\
&&\hspace{-1mm}
=~-{\alpha_s(x_\ast y_\ast)^{-2}\over\pi^4(N_c^2-1)}
{z_{13}^2\over z_{12}^2z_{23}^2\calz_1^2\calz_2^2}
\Big[\ln{\sigma s\over 4}\calz_3-{i\pi\over 2}+C\Big]
 \label{nloifscal}
 \end{eqnarray}

Let us rewrite the operator expansion (\ref{opeq}) in the explicit form \cite{nlobksym}:
\begin{eqnarray}
&&\hspace{-5mm}
(x-y)^4 T\{\halo(x)\halo^\dagger(y)\}~
=~{1\over \pi^2(N_c^2-1)}\!\int\! {d^2 z_1d^2 z_2\over z_{12}^4}~
\calr^2~\Big[{\rm Tr}\{\hat{U}^\sigma_{z_1}\hat{U}^{\dagger\sigma}_{z_2}\}
 \label{ope1}\\
&&\hspace{-5mm}
-~{\alpha_s\over\pi^2}\!\int\! d^2 z_3~{z_{12}^2\over z_{13}^2z_{23}^2}
\Big[\ln{ s\over 4}\sigma\calz_3-{i\pi\over 2}+C\Big]
[ {\rm Tr}\{T^n\hat{U}^\sigma_{z_1}\hat{U}^{\dagger\sigma}_{z_3}T^n\hat{U}^\sigma_{z_3}\hat{U}^{\dagger\sigma}_{z_2}\}
 -N_c {\rm Tr}\{\hat{U}^\sigma_{z_1}\hat{U}^{\dagger\sigma}_{z_2}\}]
\nonumber
 \end{eqnarray}
Note that the l.h.s. of the Eq. (\ref{ope1}) is conformally invariant while the coefficient function in the r.h.s. is not (due to $[\ln{ s\over 4}\sigma\calz_3$ factor).
The reason for that is the cutoff in the longitudinal direction (\ref{cutoff}). Indeed, 
we consider the light-like dipoles (in the $p_1$ direction) and impose the cutoff
 on the maximal $\alpha$ emitted by any gluon from the Wilson lines.
Formally, the light-like Wilson lines are  M\"obius invariant.
However, the light-like Wilson lines are divergent in the longitudinal direction and moreover,  it is exactly the evolution 
equation with respect to this longitudinal cutoff which governs the high-energy behavior of amplitudes. 
At present, it is not known how to find the conformally invariant cutoff in the longitudinal direction.  When we use the non-invariant cutoff 
we expect, as usual, the invariance to hold in the leading order but be
violated in higher orders in perturbation theory. In our calculation we restrict the longitudinal momentum of the gluons composing Wilson lines, 
and with this non-invariant cutoff the NLO evolution equation in QCD has extra non-conformal parts not related to the running of coupling constant.
Similarly, there will be non-conformal parts coming from the longitudinal cutoff of Wilson lines in the ${\cal N}=4$ SYM equation.
We will demonstrate below that it is possible to construct the
``composite conformal dipole operator'' (order by order in perturbation theory) which mimics the conformal cutoff
in the longitudinal direction so the corresponding evolution equation has no extra non-conformal parts. This is similar 
to the construction of the composite renormalized local operator in the case when the UV cutoff  does not respect the 
symmetries of the bare operator - in this case the symmetry of the UV-regularized operator is preserved 
order by order in perturbation theory by subtraction of the symmetry-restoring counterterms.
Following Ref. \cite{nlobksym} we choose the conformal composite operator in the form (\ref{confodipole})
\begin{eqnarray}
&&\hspace{-1mm}
[{\rm Tr}\{\hat{U}_{z_1}\hat{U}^{\dagger}_{z_2}\}\big]_{a,Y}^{\rm conf}~
\nonumber\\
&&\hspace{-1mm}
=~{\rm Tr}\{\hat{U}^\sigma_{z_1}\hat{U}^{\dagger\sigma}_{z_2}\}
+{\alpha_s\over 2\pi^2}\!\int\! d^2 z_3~{z_{12}^2\over z_{13}^2z_{23}^2}
[ {\rm Tr}\{T^n\hat{U}^\sigma_{z_1}\hat{U}^{\dagger\sigma}_{z_3}T^n\hat{U}^\sigma_{z_3}\hat{U}^{\dagger\sigma}_{z_2}\}
-N_c {\rm Tr}\{\hat{U}^\sigma_{z_1}\hat{U}^{\dagger\sigma}_{z_2}\}]
\ln {4az_{12}^2\over sz_{13}^2z_{23}^2}~+~O(\alpha_s^2)
\nonumber
\end{eqnarray}
where $a$ is an arbitrary constant. It is convenient to choose the rapidity-dependent constant 
$a\rightarrow ae^{-2Y}$ so that the 
$[{\rm Tr}\{\hat{U}^\sigma_{z_1}\hat{U}^{\dagger\sigma}_{z_2}\}\big]_a^{\rm conf}$ 
does not depend on $Y=\ln\sigma$ and all the rapidity dependence is encoded into $a$-dependence:
\begin{eqnarray}
&&\hspace{-1mm}
[{\rm Tr}\{\hat{U}_{z_1}\hat{U}^{\dagger}_{z_2}\}\big]_a^{\rm conf}~
\nonumber\\
&&\hspace{-1mm}
=~{\rm Tr}\{\hat{U}^\sigma_{z_1}\hat{U}^{\dagger\sigma}_{z_2}\}
+~{\alpha_s\over 2\pi^2}\!\int\! d^2 z_3~{z_{12}^2\over z_{13}^2z_{23}^2}
[ {\rm Tr}\{T^n\hat{U}^\sigma_{z_1}\hat{U}^{\dagger\sigma}_{z_3}T^n
\hat{U}^\sigma_{z_3}\hat{U}^{\dagger\sigma}_{z_2}\}
-N_c {\rm Tr}\{\hat{U}^\sigma_{z_1}\hat{U}^{\dagger\sigma}_{z_2}\}]
\ln {4az_{12}^2\over \sigma^2 sz_{13}^2z_{23}^2}~+~O(\alpha_s^2)
\label{confodipola}
\end{eqnarray}
Using the leading-order evolution equation (\ref{bkadj})
it is easy to see that ${d\over dY}[{\rm Tr}\{\hat{U}_{z_1}\hat{U}^{\dagger}_{z_2}\}\big]_a^{\rm conf}~=~0$ (with our $O(\alpha_s^2)$ accuracy). 

Rewritten in terms of conformal dipoles (\ref{confodipola}), the operator expansion (\ref{ope1})  takes the form:
\begin{eqnarray}
&&\hspace{-1mm}
(x-y)^4 T\{\halo(x)\halo^\dagger(y)\}~
=~{1\over \pi^2(N_c^2-1)}\!\int\! {d^2 z_1 d^2 z_2\over z_{12}^4}~\calr^2~\Big\{[{\rm Tr}\{\hat{U}_{z_1}\hat{U}^{\dagger}_{z_2}\}]_a^{\rm conf}
 \nonumber\\
&&\hspace{-1mm}
-~{\alpha_s\over 2\pi^2}\!\int\!  d^2z_3
{z_{12}^2\over z_{13}^2z_{23}^2}
\Big(\ln{ asz_{12}^2\over 4 z_{13}^2z_{23}^2}
\calz_3^2-i\pi+2C\Big)
[ {\rm Tr}\{T^n\hat{U}_{z_1}\hat{U}^{\dagger}_{z_3}T^n\hat{U}_{z_3}\hat{U}^{\dagger}_{z_2}\}
 -N_c {\rm Tr}\{\hat{U}_{z_1}\hat{U}^{\dagger}_{z_2}\}]_a\Big\}
 \label{opeconfa}
 \end{eqnarray}
We need to choose the new ``rapidity cutoff'' $a$ in such a way that all the energy dependence is included in the matrix element(s) of 
Wilson-line operators so the impact factor should not depend on energy ( $\equiv$ should not scale with $\rho$ as $\rho\rightarrow\infty$).  A suitable
choice of $a$ is given by $a_0=\kappa^{-2}+i\epsilon={ 4x_\ast y_\ast\over s(x-y)^2}+i\epsilon$ so we obtain
\begin{eqnarray}
&&\hspace{-1mm}
(x-y)^4 T\{\halo(x)\halo^\dagger(y)\}~
=~{1\over \pi^2(N_c^2-1)}\!\int\! {d^2 z_1 d^2 z_2\over z_{12}^4}~\calr^2~\Big\{[{\rm Tr}\{\hat{U}^\sigma_{z_1}\hat{U}^{\dagger\sigma}_{z_2}\}]_{a_0}^{\rm conf}
 \nonumber\\
&&\hspace{-1mm}
-~{\alpha_s\over 2\pi^2}\!\int\!  d^2z_3
{z_{12}^2\over z_{13}^2z_{23}^2}
\Big[\ln{-x_\ast y_\ast z_{12}^2\over (x-y)^2 z_{13}^2z_{23}^2}
\calz_3^2+2C\Big]
[ {\rm Tr}\{T^n\hat{U}^\sigma_{z_1}\hat{U}^{\dagger\sigma}_{z_3}
T^n\hat{U}^\sigma_{z_3}\hat{U}^{\dagger\sigma}_{z_2}\}
 -N_c {\rm Tr}\{\hat{U}^\sigma_{z_1}\hat{U}^{\dagger\sigma}_{z_2}\}]\Big\}
 \label{opeconf}
 \end{eqnarray}
where the conformal dipole $[{\rm Tr}\{\hat{U}^\sigma_{z_1}\hat{U}^{\dagger\sigma}_{z_2}\}]^{\rm conf}$ is given by Eq. (\ref{confodipola}) with
$a_0={4 x_\ast y_\ast\over s(x-y)^2}$.   

Now it is evident that the impact factor in the r.h.s. of this equation is M\"obius invariant and does not scale with $\rho$ so 
Eq. (\ref{confodipola}) gives conformally invariant operator up to $\alpha_s^2$ order. In higher orders, one should expect the correction 
terms with more Wilson lines.

 With the two-gluon accuracy integration over one $z$ in the r.h.s. of Eq. (\ref{opeconf}) can be performed explicitly 
so that the resulting operator expansion takes the form
\begin{eqnarray}
&&\hspace{-1mm}      
(x-y)^4 T\{\halo(x)\halo^\dagger(y)\}~
\label{resope}\\
&&\hspace{-1mm} 
=~-{1\over\pi^2}\!\int\! {dz_1dz_2\over z_{12}^4}~
\hat{\cal U}_{\rm conf}^{a_0}(z_1,z_2)\calr^2\Big\{1-{\alpha_sN_c\over 2\pi}
\Big[\ln^2\calr-{\ln\calr\over \calr} -2C\big(\ln \calr-{1\over\calr}+2\big)+~2{\rm Li}_2(1-\calr)-{\pi^2\over 3}\Big]
\Big\}
\nonumber
 \end{eqnarray}

 We need the projection of the T-product in the l.h.s. of this equation onto the conformal eigenfunctions of the BFKL equation with spin 0 (cf. Eq. (\ref{projlo})
\begin{eqnarray}
&&\hspace{-2mm}
{1\over\pi^2}\!\int\! {dz_1dz_2\over z_{12}^4}~
\calr^2\Big\{1-{\alpha_sN_c\over 2\pi}
\Big[\ln^2\calr-{\ln\calr\over \calr} -2C\big(\ln \calr-{1\over\calr}+2\big)+~2{\rm Li}_2(1-\calr)-{\pi^2\over 3}\Big]
\Big\}\Big[{z_{12}^2\over z_{10}^2z_{20}^2}\Big]^{\half+i\nu}
\nonumber\\
&&\hspace{-2mm}
=~\Big[{-\kappa^2\over (-2\kappa\cdot\zeta_0)^2}\Big]^{\half+i\nu} {\Gamma^2\big(\half-i\nu\big)\over \Gamma (1-2i\nu)}
{\big({1\over 4}+\nu^2\big)\pi\over \cosh\pi\nu}
\Big\{1+{\alpha_sN_c\over 2\pi}\Phi_1(\nu)\Big\}
\label{projnlo}
\end{eqnarray} 
where  
\begin{equation}
\hspace{-0mm}
\Phi_1(\nu)~=~-2\psi'\big(\half+i\nu\big)-2\psi'\big(\half-i\nu\big)+{2\pi^2\over 3}
+{\chi(\nu)-2\over \nu^2+{1\over 4}}+2C\chi(\nu)
\label{Fi}
\end{equation}
and
$\zeta_0\equiv{ p_1\over s}+z_{0\perp}^2p_2+z_{0\perp}$.

Now, using the decomposition (\ref{lobzor120}) of the product of the transverse $\delta$-functions in conformal 3-point functions $E_{\nu,n}(z_{10},z_{20})$  
 we obtain 
\begin{eqnarray}
&&\hspace{-1mm}      
(x-y)^4 T\{\halo(x)\halo^\dagger(y)\}~
\nonumber\\
&&\hspace{-1mm}=~-\int\! d\nu\!\int\! d^2z_0~{\nu^2(1+4\nu^2)\over 4\pi\cosh\pi\nu}
{\Gamma^2\big(\half-i\nu\big)\over\Gamma(1-2i\nu)}
~\Big({-\kappa^2\over (-2\kappa\cdot\zeta_0)^2}\Big)^{\half +i\nu}\big[1+{\alpha_sN_c\over 2\pi}\Phi_1(\nu)\big]
 \hat{\cal U}_{\rm conf}^a(\nu,z_0)
 \label{opeu}
 \end{eqnarray}
where 
\begin{eqnarray}
&&\hspace{-1mm}
 \hat{\cal U}_{\rm conf}^a(\nu,z_0)~\equiv~{1\over \pi^2}\!\int\! {d^2z_1d^2z_2\over z_{12}^4}~
\Big({z_{12}^2\over z_{10}^2z_{20}^2}\Big)^{\half -i\nu}~ \hat{\cal U}_{\rm conf}^a(z_1,z_2)
\label{opconfspin}
\end{eqnarray}
is a conformal dipole in the $z_0,\nu$ representation.
\subsection{NLO BK in ${\cal N}=4$ SYM}
\begin{figure}[htb]
\includegraphics[width=1.2\textwidth]{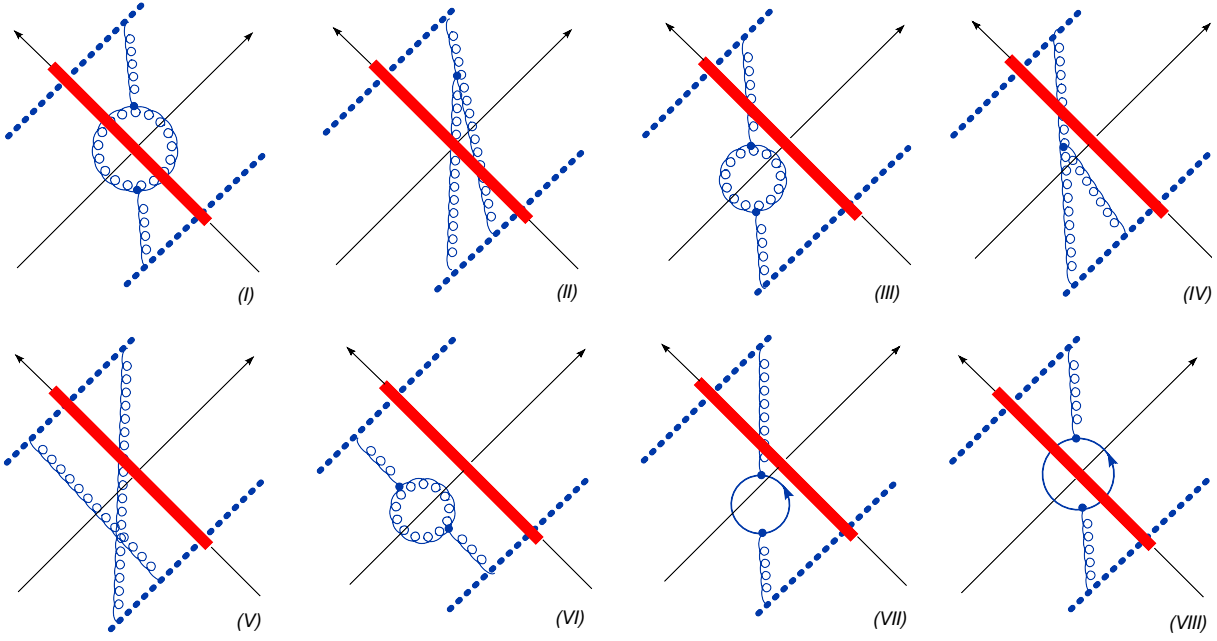}

\vspace{-32mm}
\caption{Different types of diagrams for the NLO evolution of color dipole. \label{nlofigs}}
\end{figure}

The typical diagrams for the NLO evolution of color dipole are shown Fig. \ref{nlofigs}. Here
solid line depicts either scalar particle or gluino.
\hspace{11mm}
\vspace{-0mm}
The contribution of these diagrams is calculated using the gluon propagator in a shock-wave background (\ref{gluon-sw1}) and usual Yang-Mills vertices.
The typical contribution has a form of an integral over Feynman parameter which is a fraction of momentum $\alpha$ carried by one of the gluon 
after splitting described by the three- gluon vertex. For example,
\begin{eqnarray}
&&\hspace{-4mm}
\langle {\rm Tr}\{\hat{U}_x\hat{U}^{\dagger}_y\}
\rangle_{\rm Fig. \ref{nlofigs}I}~=~
{\alpha_s^2\over 8\pi^4}\ln\Delta Y
\int_0^1 du~u\bar{u}\!\int \!d^2 zd^2 z'~U_z^{bb'}U_{z'}^{cc'}
\nonumber\\ 
&&\hspace{-4mm}
\times~{\rm Tr}\Big\{T^af^{abc}\Big[
{Z_{ij}\over (z-z')^2[uz_{13}^2+\bar{u}z_{14}^2]}
-(z_1\leftrightarrow z_2)\Big]
\Big\}U_x
\Big\{T^{a'}f^{a'b'c'}\Big[{Z'_{ij}\over (z-z')^2[uz_{23}^2+\bar{u}z_{24}^2]}
-(z_1\leftrightarrow z_2)\Big]
\Big\}U^{\dagger}_y
\label{2cutsum}
\end{eqnarray}
where 
\begin{equation}
Z^{ij}~=~(z_{13}^2-z_{14}^2)g^{ij}+{2\over u}z_{34}^iz_{14}^j+{2\over\bar{u}}z_{13}^iz_{34}^j,~~~~~~
{Z'}^{ij}~=~(z_{23}^2-z_{24}^2)g^{ij}+{2\over u}z_{34}^iz_{24}^j+
{2\over \bar{u}}z_{23}^iz_{34}^j
\label{XY}
\end{equation}
and $\baru\equiv 1-u$. It is easy to see that integral (\ref{2cutsum}) diverges as $u\rightarrow 0$ and $u\rightarrow 1$. 
This divergence comes in the momentum space from the integrals of the type 
$$
\int_0^\infty\! d\alpha_1 d\alpha_2~{1\over\alpha_1(k_{1\perp}^2\alpha_2+k_{1\perp}^2\alpha_2)}
$$
where  $\alpha_ip_1$ and $k_{i\perp}$ are longitudinal and transverse momenta carried by gluons in the loop in Fig. \ref{nlofigs}I.
If we put a lower cutoff $\alpha_i>\sigma'$ on the 
$\alpha_i$ integrals we would get a contribution $\sim \ln^2{\sigma\over\sigma'}$ coming from the region $\alpha_2\gg\alpha_1>\sigma'$ (or $\alpha_1\gg\alpha_2>\sigma'$ )
which corresponds to the the square of the leading-order BK kernel rather than to the NLO
kernel. To get the NLO kernel we need to subtract this $(LO)^2$ contribution. 
Indeed, the operator form of the evolution equation for the color dipole up to the 
next-to-leading order looks like
\begin{eqnarray}
&&\hspace{-6mm}
{d \over dY}{\rm Tr}\{\hat{U}_x\hat{U}^\dagger_y\}=
K_{\rm LO}{\rm Tr}\{\hat{U}_x\hat{U}^\dagger_y\}+K_{\rm NLO}{\rm Tr}\{\hat{U}_x\hat{U}^\dagger_y\}
\label{nloeveq}
\end{eqnarray}
where $Y=\ln\sigma$. 
Our goal is to find $K_{\rm NLO}$ by considering the l.h.s. of this equation in the 
external shock-wave background so 
\begin{eqnarray}
&&\hspace{-6mm}
 \langle K_{\rm NLO}{\rm Tr}\{\hat{U}_x\hat{U}^\dagger_y\}\rangle_{\rm shockwave}=
{d \over dY}\langle{\rm Tr}\{\hat{U}_x\hat{U}^\dagger_y\}\rangle_{\rm shockwave}-
\langle K_{\rm LO}{\rm Tr}\{\hat{U}_x\hat{U}^\dagger_y\}\rangle_{\rm shockwave}
\label{subtraction}
\end{eqnarray}
The subtraction (\ref{subtraction}) leads to the $\Big[{1\over u}\Big]_+$ prescription (\ref{plupres})
for the terms divergent as ${1\over u}$ 
(similarly, ${1\over \bar{u}}\rightarrow \Big[{1\over \bar{u}}\Big]_+$ for the contribution divergent as $u\rightarrow 1$). 
With this prescription, the integrals over the Feynman parameter converge. The calculation of diagrams is carried out in Refs. \cite{nlobk,nlobksym} and the result has the form
\begin{eqnarray}
&&\hspace{-4mm}
{d\over dY}{\rm Tr}\{\hat{U}^Y_{z_1} \hat{U}^{\dagger Y}_{z_2}\}~ =~{\alpha_s\over \pi^2}
\!\int\!d^2z_3~
{ z_{12}^2\over z_{13}^2 z_{23}^2}\Big\{1-{\alpha_sN_c\over 4\pi}\Big[{\pi^2\over 3}
+2\ln{z_{13}^2\over z_{12}^2}\ln{z_{23}^2\over z_{12}^2}\Big]\Big\}
[{\rm Tr}\{T^a\hat{U}^Y_{z_1} \hat{U}^{\dagger Y}_{z_3}T^a\hat{U}^Y_{z_3} \hat{U}^{\dagger Y}_{z_2}\}-N_c{\rm Tr}\{\hat{U}^Y_{z_1} \hat{U}^{\dagger Y}_{z_2}\}]
\nonumber\\
&&\hspace{-4mm} 
-~{\alpha_s^2\over 4\pi^4}
~\!\int \!{d^2 z_3d^2 z_4\over z_{34}^4}~
{ z_{12}^2z_{34}^2\over z_{13}^2 z_{24}^2}
\Big[1+
{ z_{12}^2z_{34}^2\over z_{13}^2z_{24}^2-z_{23}^2z_{14}^2}\Big]
\ln{z_{13}^2z_{24}^2\over z_{14}^2z_{23}^2}
\nonumber\\
&&\hspace{32mm} 
\times~
{\rm Tr}\{[T^a,T^b]\hat{U}^Y_{z_1}T^{a'}T^{b'}\hat{U}_{z_2}^{\dagger Y}+T^bT^a\hat{U}^Y_{z_1}[T^{b'},T^{a'}]
\hat{U}_{z_2}^{\dagger Y}\}(\hat{U}^Y_{z_3})^{aa'}(\hat{U}^Y_{z_4}-\hat{U}^Y_{z_3})^{ bb'}.
\label{nlobksym}
\end{eqnarray}
All terms in the r.h.s. of this equation are M\"obius invariant except the double-log term 
proportional to $\ln{z_{13}^2\over z_{12}^2}\ln{z_{23}^2\over z_{12}^2}$. As was discussed in the Introduction, the reason
for this non-invariance is the cutoff in the longitudinal direction which violates the formal
invariance of the non-cut Wilson lines.

It is worth noting that conformal and non-conformal terms come from graphs with different topology: the conformal terms come from 1$\rightarrow$3 dipoles diagrams (see Fig. 6 in Ref. \cite{nlobk}) which describe the dipole creation while the non-conformal double-log term comes from the 
1$\rightarrow$2 dipole transitions (Fig.9 in Ref. \cite{nlobk})) which can be regarded as a combination of dipole creation and dipole recombination.

Our aim is to rewrite the evolution equation in terms of  composite conformal dipoles  (\ref{confodipola}). 
In the next-to-leading order the  conformal dipole has the form
\begin{eqnarray}
&&\hspace{-5mm}
\big[{\rm Tr}\{\hat{U}_{z_1}\hat{U}^{\dagger}_{z_2}\}\big]_a^{\rm conf}~=~{\rm Tr}\{\hat{U}^Y_{z_1}\hat{U}^{\dagger Y}_{z_2}\}
\label{confodipole1}\\
&&\hspace{-5mm}
+~{\alpha_s\over 2\pi^2}
\!\int\!d^2z_3~
{z_{12}^2\over z_{13}^2 z_{23}^2}\Big[1-
{\alpha_sN_c\over 4\pi}{\pi^2\over 3}\Big]\big[{\rm Tr}\{T^a\hat{U}^Y_{z_1}\hat{U}^{\dagger Y}_{z_3}T^a\hat{U}^Y_{z_3}\hat{U}^{\dagger Y}_{z_2}\} 
-N_c {\rm Tr}\{\hat{U}^Y_{z_1}\hat{U}^{\dagger Y}_{z_2}\}\big]\ln {4az_{12}^2\over \sigma^2sz_{13}^2z_{23}^2}
\nonumber\\
&&\hspace{-5mm} 
-~{\alpha_s^2\over 8\pi^4}
\int \!d^2 z_3 d^2 z_4 {z_{12}^2\over z_{13}^2z_{24}^2z_{34}^2}
\Big\{2\ln{z_{12}^2z_{34}^2\over z_{14}^2z_{23}^2}
+\Big[1+{z_{12}^2z_{34}^2\over z_{13}^2z_{24}^2-z_{14}^2z_{23}^2}\Big]\ln{z_{13}^2z_{24}^2\over z_{14}^2z_{23}^2}\Big\}\ln {4a\over\sigma^2s}f(z_i)
\nonumber\\ 
&&\hspace{-5mm}
\times~ {\rm Tr}\{[T^a,T^b]\hat{U}^Y_{z_1}T^{a'}T^{b'}\hat{U}^{\dagger Y}_{z_2}
 + T^bT^a\hat{U}^Y_{z_1} [T^{b'},T^{a'}]\hat{U}^{\dagger Y}_{z_2}\}
 [(\hat{U}^Y_{z_3})^{aa'}(\hat{U}^Y_{z_4})^{bb'}-(z_4\rightarrow z_3)]
\nonumber\\
&&\hspace{-5mm}
+~{\alpha_s\over 32\pi^4}
\!\int\!d^2z_3~
{z_{12}^2\over z_{13}^2 z_{23}^2}\!\int\! d^2z_4(\hat{U}^Y_{z_3})^{aa'}
\nonumber\\
&&\hspace{-5mm}
\times~
 \Big\{
 {\rm Tr}\{T^aT^b\hat{U}^Y_{z_1}T^{a'}T^{b'}\hat{U}^{\dagger Y}_{z_2}
 + T^bT^a\hat{U}^Y_{z_1}T^{b'}T^{a'}\hat{U}^{\dagger Y}_{z_2}\}
 (2\hat{U}^Y_{z_4}-\hat{U}^Y_{z_1}-\hat{U}^Y_{z_2})^{bb'} 
{z_{12}^2\over z_{14}^2z_{24}^2}
 \ln^2\big({az_{12}^2\over \sigma^2s z_{14}^2z_{24}^2}\big)
 \nonumber\\
&&\hspace{-5mm}
 -~
 {\rm Tr}\{T^aT^b\hat{U}^Y_{z_1}[T^{a'},T^{b'}]\hat{U}^{\dagger Y}_{z_2}
 +[T^b,T^a]\hat{U}^Y_{z_1}T^{b'}T^{a'}\hat{U}^{\dagger Y}_{z_2}\}
 (2\hat{U}^Y_{z_4}-\hat{U}^Y_{z_1}-\hat{U}^Y_{z_3})^{bb'}{z_{13}^2\over z_{14}^2z_{34}^2}
 \ln^2\big({4az_{13}^2\over \sigma^2sz_{14}^2z_{34}^2}\big)
 \nonumber\\
&&\hspace{-5mm}
-~
 {\rm Tr}\{[T^a,T^b]\hat{U}^Y_{z_1}T^{a'}T^{b'}\hat{U}^{\dagger Y}_{z_2}
 + T^bT^a\hat{U}^Y_{z_1}[T^{b'},T^{a'}]\hat{U}^{\dagger Y}_{z_2}\}
 (2\hat{U}^Y_{z_4}-\hat{U}^Y_{z_2}-\hat{U}^Y_{z_3})^{bb'} 
 {z_{23}^2\over z_{24}^2z_{34}^2}\ln^2\big({4az_{23}^2\over \sigma^2sz_{24}^2z_{34}^2}\big)
 \Big\}
  \nonumber\\
&&\hspace{-5mm}
-~{\alpha_s^2N_c\over 8\pi^4}\!\int\!d^2z_3~
{z_{12}^2\over z_{13}^2 z_{23}^2}\!\int\! d^2z_4{z_{12}^2\over z_{14}^2z_{24}^2}
({\rm Tr}\{T^aU_{z_1}U_{z_4}^\dagger T^a U_{z_4}U_{z_2}^\dagger\}
-N_c{\rm Tr}\{U_{z_1}U_{z_2}^\dagger\})\ln^2\big({4a z_{12}^2\over \sigma^2sz_{14}^2z_{24}^2}\big)
\nonumber
 \end{eqnarray}
where function $f$ is undetermined in the NLO (to fix it one needs the NNLO accuracy).
Using the equations for ${d\over dY}{\rm Tr}\{\hat{U}^Y_{z_1}\hat{U}^{\dagger Y}_{z_2}\}$
and  ${d\over dY}\big[{\rm Tr}\{T^a\hat{U}^Y_{z_1}\hat{U}^{\dagger Y}_{z_3}T^a\hat{U}_{z_3}\hat{U}^{\dagger Y}_{z_2}\} 
-N_c {\rm Tr}\{\hat{U}^Y_{z_1}\hat{U}^{\dagger Y}_{z_2}\}\big]$  from Ref. \cite{nlobksym}
one can demonstrate that  ${d\over dY}\big[{\rm Tr}\{\hat{U}_{z_1}\hat{U}^{\dagger}_{z_2}\}\big]_a^{\rm conf}~=~0$.

Differentiating now with respect to $a$ we get
\begin{eqnarray}
&&\hspace{-5mm}
2a{d\over da}\big[{\rm Tr}\{\hat{U}_{z_1}\hat{U}^{\dagger}_{z_2}\}\big]_a^{\rm conf}~
=~{\alpha_s\over \pi^2}
\!\int\!d^2z_3~
{z_{12}^2\over z_{13}^2 z_{23}^2}\Big[1-
{\alpha_sN_c\over 4\pi}{\pi^2\over 3}\Big]\big[{\rm Tr}\{T^a\hat{U}_{z_1}\hat{U}^{\dagger }_{z_3}T^a\hat{U}_{z_3}\hat{U}^{\dagger }_{z_2}\} 
-N_c {\rm Tr}\{\hat{U}_{z_1}\hat{U}^{\dagger }_{z_2}\}\big]_a^{\rm conf}
\nonumber\\
&&\hspace{-2mm} 
-~{\alpha_s^2\over 4\pi^4}
\int \!d^2 z_3 d^2 z_4 {z_{12}^2\over z_{13}^2z_{24}^2z_{34}^2}
\Big\{2\ln{z_{12}^2z_{34}^2\over z_{14}^2z_{23}^2}
+\Big[1+{z_{12}^2z_{34}^2\over z_{13}^2z_{24}^2-z_{14}^2z_{23}^2}\Big]\ln{z_{13}^2z_{24}^2\over z_{14}^2z_{23}^2}\Big\}
\nonumber\\ 
&&\hspace{-2mm}
\times~ {\rm Tr}\{[T^a,T^b]\hat{U}_{z_1}T^{a'}T^{b'}\hat{U}^{\dagger }_{z_2}
 + T^bT^a\hat{U}_{z_1} [T^{b'},T^{a'}]\hat{U}^{\dagger }_{z_2}\}
 [(\hat{U}_{z_3})^{aa'}(\hat{U}_{z_4})^{bb'}-(z_4\rightarrow z_3)]
 \label{nlobksymconf1}
\end{eqnarray}
where 
\begin{eqnarray}
&&\hspace{-1mm}
[{\rm Tr}\{T^a\hat{U}_{z_1}\hat{U}^{\dagger}_{z_3}T^a\hat{U}_{z_3}\hat{U}^{\dagger}_{z_2}\}
-N_c{\rm Tr}\{\hat{U}_{z_1}\hat{U}^{\dagger}_{z_2}\}]^{\rm conf}_a
~ =~{\rm Tr}\{T^a\hat{U}^Y_{z_1}\hat{U}^{\dagger Y}_{z_3}T^a\hat{U}^Y_{z_3}\hat{U}^{\dagger Y}_{z_2}\}
-N_c{\rm Tr}\{\hat{U}^Y_{z_1}\hat{U}^{\dagger Y}_{z_2}\}
 ~\nonumber\\
&&\hspace{-1mm}
+~{\alpha_s\over 8\pi^2}\!\int\! d^2z_4(\hat{U}^Y_{z_3})^{aa'}
 \Big[
 {\rm Tr}\{T^aT^b\hat{U}^Y_{z_1}T^{a'}T^{b'}\hat{U}^{\dagger Y}_{z_2}
 + T^bT^a\hat{U}^Y_{z_1}T^{b'}T^{a'}\hat{U}^{\dagger Y}_{z_2}\}
 (2\hat{U}^Y_{z_4}-\hat{U}^Y_{z_1}-\hat{U}^Y_{z_2})^{bb'}{z_{12}^2\over z_{14}^2z_{24}^2}\ln{4az_{12}^2\over s\sigma^2z_{14}^2z_{24}^2}
  \nonumber\\
&&\hspace{-1mm}
 -~
 {\rm Tr}\{T^aT^b\hat{U}^Y_{z_1}[T^{a'},T^{b'}]\hat{U}^{\dagger Y}_{z_2}
 +[T^b,T^a]\hat{U}^Y_{z_1}T^{b'}T^{a'}\hat{U}^{\dagger Y}_{z_2}\}
 (2\hat{U}^Y_{z_4}-\hat{U}^Y_{z_1}-\hat{U}^Y_{z_3})^{bb'}{z_{13}^2\over z_{14}^2z_{34}^2}\ln{4az_{13}^2\over s\sigma^2z_{14}^2z_{34}^2}
 \nonumber\\
&&\hspace{-1mm}
-~
 {\rm Tr}\{[T^a,T^b]\hat{U}^Y_{z_1}T^{a'}T^{b'}\hat{U}^{\dagger Y}_{z_2}
 + T^bT^a\hat{U}^Y_{z_1}[T^{b'},T^{a'}]\hat{U}^{\dagger Y}_{z_2}\}
 (2\hat{U}^Y_{z_4}-\hat{U}^Y_{z_2}-\hat{U}^Y_{z_3})^{bb'} 
 {z_{23}^2\over z_{24}^2z_{34}^2}\ln{4az_{23}^2\over s\sigma^2z_{24}^2z_{34}^2}
 \Big]
  \nonumber\\
&&\hspace{-1mm}
-~{\alpha_sN_c\over 2\pi^2}\!\int\! d^2z_4{z_{12}^2\over z_{14}^2z_{24}^2}
({\rm Tr}\{T^aU_{z_1}U_{z_4}^\dagger T^a U_{z_4}U_{z_2}^\dagger\}
-N_c{\rm Tr}\{U_{z_1}U_{z_2}^\dagger\})\ln{4a z_{12}^2\over s\sigma^2z_{14}^2z_{24}^2}~+~O(\alpha_s^2)
\label{confopera1}
 \end{eqnarray}
is a conformal $Y$-independent operator found in Ref. \cite{nlobksym}.
One sees now that the  evolution equation  with respect to parameter $a$ (\ref{nlobksymconf1}) is 
obviously M\"obius invariant.

\subsection{NLO evolution of the conformal dipole}
With out two-gluon accuracy the evolution equation (\ref{nlobksymconf1}) reduces to
\begin{eqnarray}
&&\hspace{-2mm}
2a{d\over da}\hat{\cal U}^a_{\rm conf}(z_1,z_2)~
~=~
\!\int\!d^2z_3d^2z_4 [K_{\rm LO}(z_1,z_2;z_3,z_4)+K_{\rm NLO}(z_1,z_2;z_3,z_4)]~
\hat{\cal U}^a_{\rm conf}(z_3,z_4)
\label{nlobfklconf}
\end{eqnarray}
where the kernel $K_{\rm NLO}(z_1,z_2;z_3,z_4)$ in the first two orders has the form \cite{nlobk, nlobfklconf} (see also Ref. \cite{fadin10})
\begin{eqnarray}
&&\hspace{-3mm}
K_{\rm NLO}(z_1,z_2;z_3,z_4)~=~-{\alpha_sN_c\over 4\pi}{\pi^2\over 3}K_{\rm LO}(z_1,z_2;z_3,z_4)
\nonumber\\
&&\hspace{-3mm}
+~{\alpha_s^2N_c^2\over 8\pi^4 z_{34}^4}~\Bigg[{z_{12}^2z_{34}^2\over z_{13}^2z_{24}^2}
\Big\{\Big(
1+{z_{12}^2z_{34}^2\over z_{13}^2z_{24}^2- z_{14}^2z_{23}^2}\Big)
\ln{z_{13}^2z_{24}^2\over z_{14}^2z_{23}^2}
+2\ln{z_{12}^2z_{34}^2\over z_{14}^2z_{23}^2}\Big\}
+12\pi^2\zeta(3)z_{34}^4\delta(z_{13})\delta(z_{24})
\Bigg]
\label{nlokonf}
\end{eqnarray}
Once we know that the kernel $K_{\rm NLO}$ is conformal, its eigenfunctions are fixed by conformal invariance (see Eq. (\ref{eigenfunctions}))
and calculation of the eigenvalues reproduces the pomeron intercept (\ref{eigen2}).
\begin{equation}
\int\!d^2z_3d^2z_4 ~K(z_1,z_2;z_3,z_4) E_{\nu,n}(z_{30},z_{40})~=~\omega(n,\nu)E_{\nu,n}(z_{10},z_{20})
\end{equation}
For the composite operators with definite conformal spin (\ref{opconfspin}) the 
evolution equation (\ref{nlobfklconf})  takes the simple form
\begin{equation}
\hspace{-0mm}
2a{d\over da} \hat{\cal U}_{\rm conf}^a(\nu,z_0)~=~\omega(\nu) \hat{\cal U}_{\rm conf}^a(\nu,z_0)
\label{evolv}
\end{equation}
The result of the evolution (\ref{evolv}) is 
\begin{equation}
\hspace{-0mm}
\hat{\cal U}_{\rm conf}^a(\nu,z_0)~=~(a/\tilde{a})^{\half\omega(\nu)} \hat{\cal U}_{\rm conf}^{a_0}(\nu,z_0)
\label{evolresult}
\end{equation}
where the endpoint of the evolution $\tilde{a}$ should be taken from the requirement that the amplitude of scattering of conformal dipoles
with ``normalization points'' $\tilde{a}$ and $b$ should not contain large logarithms of energy so it will serve as the initial point
of the evolution (cf. Eq. (\ref{fla59}) for the leading order). 

\subsection{NLO scattering of conformal dipoles and the NLO amplitude}
The last step of our program is the ``matrix element'' of the conformal dipole 
\begin{equation}
\hspace{5mm}
\langle T\{\hat{\cal U}_{\rm conf}^{\tilde a}(\nu,z_0)\calo(x')\calo^\dagger(y')\}\rangle
\label{mael}
\end{equation}
As we have done for the leading order, we start with the high-energy OPE for $\calo(x')\calo^\dagger(y')$
\begin{eqnarray}
&&\hspace{-1mm}      
(x'-y')^4 T\{\halo(x')\halo^\dagger(y')\}~
\nonumber\\
&&\hspace{-1mm}
=~-\!\int\! d\nu'\!\int\! d^2z'_0~{{\nu'}^2(1+4{\nu'}^2)\over 4\pi\cosh\pi\nu'}
{\Gamma^2\big(\half-i\nu'\big)\over\Gamma(1-2i\nu')}
~\Big({-{\kappa'}^2\over (-2\kappa'\cdot\zeta'_0)^2}\Big)^{\half +i\nu'}\big[1+{\alpha_sN_c\over 2\pi}\Phi_1(\nu')\big]
 \hat{\cal V}_{\rm conf}^{b_0}(\nu',z'_0).
 \label{opev}
 \end{eqnarray}
where $b_0=1/{\kappa'}^2$and  the conformal operator  $\hat{\cal V}_{\rm conf}^b(z_1,z_2)$ is given by Eq. (\ref{voper}).

Now we multiply Eq. (\ref{opev}) by the NLO  amplitude of scattering of two conformal dipoles \cite{confamp}:
\begin{eqnarray}
&&\hspace{-5mm}
\langle\hat{\cal U}_{\rm conf}^{\tilde a}(\nu,z_0)\hat{\cal V}_{\rm conf}^b(\nu',z'_0)\rangle~
=~-{\alpha_s^2N_c^2\over N_c^2-1}
{16\pi^2\over \nu^2(1+4\nu^2)^2}
\Big[\delta(z_0-z'_0)\delta(\nu+\nu')
\label{confdipscatlow}\\
&&\hspace{-5mm}
+{2^{1-4i\nu}\delta(\nu-\nu')\over \pi|z_0-z'_0|^{2-4i\nu}}
{\Gamma\big({1\over 2}+i\nu\big)\Gamma(1-i\nu)\over\Gamma(i\nu)\Gamma\big(\half-i\nu\big)}\Big]\Big[1+{\alpha_sN_c\over 2\pi}\Big(\chi(\nu)\big[\ln \tilde{a}b -i\pi-4C
-{2\over\nu^2+{1\over 4}}\big]-{\pi^2\over 3}\Big)\Big].
\nonumber
 \end{eqnarray}

From this equation we see that we need to stop the evolution  (\ref{evolresult}) at $\tilde{a}=1/b$. With this choice of $\tilde{a}$ the correlation function 
takes the form:
\begin{eqnarray}
&&\hspace{-5mm}
\langle T\{\hat{\cal U}_{\rm conf}^{\tilde a}(\nu,z_0)\calo(x')\calo^\dagger(y')\}\rangle~=~-{\alpha_s^2N_c^2\over N_c^2-1}
\label{maele}\\
&&\hspace{-1mm}
\times~{8\pi\Gamma^2\big(\half+i\nu\big)\over  (1+4\nu^2)\cosh\pi\nu \Gamma(1+2i\nu)}
~\Big({-{\kappa'}^2\over (-2\kappa'\cdot\zeta_0)^2}\Big)^{\half +i\nu}\big[1+{\alpha_sN_c\over 2\pi}\Phi_1(\nu)\big]\Big[1-{\alpha_sN_c\over 2\pi}\Big(\chi(\nu)\big[ i\pi+4C
+{2\over\nu^2+{1\over 4}}\big]+{\pi^2\over 3}\Big)\Big].
\nonumber
 \end{eqnarray}
Finally, substituting this amplitude in Eq. (\ref{opeu}) we obtain Eq. (\ref{koppinkoop}) with 
\begin{eqnarray}
&&\hspace{-0mm}
F(\nu)~=~{N_c^2\over N_c^2-1}{4\pi^4\alpha_s^2\over\cosh^2\pi\nu}~
\big[1+{\alpha_sN_c\over 2\pi}\Phi_1(\nu)\big]^2~
\Big\{1-{\alpha_sN_c\over 2\pi}\Big[\chi(\nu)\Big(4C+{8\over 1+4\nu^2}\Big)
+{\pi^2\over 3}\Big]+O(\alpha_s^2)\Big\}
\nonumber\\
&&\hspace{-0mm}
=~{N_c^2\over N_c^2-1}{4\pi^4\alpha_s^2\over\cosh^2\pi\nu}~
\Big\{1+{\alpha_sN_c\over \pi}\Big[-2\psi'\big(\half+i\nu\big)-2\psi'\big(\half-i\nu\big)
+{\pi^2\over 2}-{8\over 1+4\nu^2}\Big]+O(\alpha_s^2)\Big\}
\label{pomrez}
\end{eqnarray}
which gives the pomeron residue in the next-to-leading order.

\begin{figure}[htb]
\psfig{file=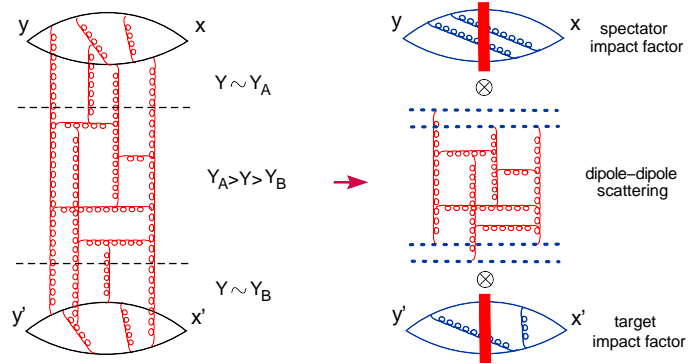,width=90mm}
\caption{High-energy factorization into the product of two impact factors and dipole-dipole scattering}
\label{aba:fig1}
\end{figure}

It is instructive to represent this result in a way symmetric between projectile and target as a product of spectator impact factor, target impact factor, and 
scattering of two (conformal) dipoles as shown in Fig. \ref{aba:fig1}.

Combining the operator expansions (\ref{opeu}) and (\ref{opev}) we get
\begin{eqnarray}
&&\hspace{-1mm}      
(x-y)^4 (x'-y')^4\langle T\{\halo(x)\halo^\dagger(y)\halo(x')\halo^\dagger(y')\}\rangle~
\label{opeuv}\\
&&\hspace{-1mm}
=~\!\int\! d\nu d\nu'\!\int\! d^2z_0d^2z'_0~{\nu^2(1+4\nu^2)\over 4\pi\cosh\pi\nu}
{\Gamma^2\big(\half-i\nu\big)\over\Gamma(1-2i\nu)}\Big({-\kappa^2\over (-2\kappa\cdot\zeta_0)^2}\Big)^{\half +i\nu}\big[1+{\alpha_sN_c\over 2\pi}\Phi_1(\nu)\big]
\nonumber\\
&&\hspace{-1mm}
\times~
{{\nu'}^2(1+4{\nu'}^2)\over 4\pi\cosh\pi\nu'}
{\Gamma^2\big(\half-i\nu'\big)\over\Gamma(1-2i\nu')}
\Big({-{\kappa'}^2\over (-2\kappa'\cdot\zeta'_0)^2}\Big)^{\half +i\nu'}
\big[1+{\alpha_sN_c\over 2\pi}\Phi_1(\nu')\big]
\langle \hat{\cal U}_{\rm conf}^{a_0}(\nu,z_0) \hat{\cal V}_{\rm conf}^{b_0}(\nu',z'_0)\rangle
\nonumber
 \end{eqnarray}
The  scattering of two conformal dipoles  is obtained from Eq. (\ref{confdipscatlow}) by evolution (\ref{evolresult}):
\begin{eqnarray}
&&\hspace{-1mm}
\langle\hat{\cal U}_{\rm conf}^a(\nu,z_0)\hat{\cal V}_{\rm conf}^b(\nu',z'_0)\rangle~
\label{confdipscat}\\
&&\hspace{-1mm}
=~i{\alpha_s^2N_c^2\over N_c^2-1}
{(a+i\epsilon)^{\omega(\nu)\over 2}(b+i\epsilon)^{\omega(\nu)\over 2}-(-a-i\epsilon)^{\omega(\nu)\over 2}(-b-i\epsilon)^{\omega(\nu)\over 2}\over\sin\pi\omega(\nu)}
\Big[1-{\alpha_sN_c\over 2\pi}\Big(\chi(\gamma)\big[4C+{8\over 1+4\nu^2}\big]+{\pi^2\over 3}\Big)\Big]
\nonumber\\
&&\hspace{-3mm}
\times~{16\pi^2\over \nu^2(1+4\nu^2)^2}
\Big[\delta(z_0-z'_0)\delta(\nu+\nu')+{2^{1-4i\nu}\delta(\nu-\nu')\over \pi|z_0-z'_0|^{2-4i\nu}}
{\Gamma\big({1\over 2}+i\nu\big)\Gamma(1-i\nu)\over\Gamma(i\nu)\Gamma\big(\half-i\nu\big)}\Big].
\nonumber
 \end{eqnarray}
Substituting this amplitude with  $a_0=\kappa^{-2}+i\epsilon$ and $b_0={\kappa'}^{-2}+i\epsilon$ in Eq. (\ref{opeuv})  we reobtain  Eq. (\ref{koppinkoop}) with the pomeron residue 
(\ref{pomrez}).

\section{Next-to-leading order in QCD}
In ${\cal N}=4$ SYM we were able to perform all four steps of our program and arrive at explicit result (\ref{koppinkoop}) with $\omega(\nu)$ and $F(\nu)$ given by 
Eqs. (\ref{eigen1}) and (\ref{pomrez}), respectively. In QCD, we do not have conformal invariance so there is no general formula for the pomeron contribution
similar to Eq. (\ref{koppinkoop}). Moreover, by the same reason (absence of Mobius invariance) we cannot solve the equation for evolution of color dipoles analytically. 
Here I will describe the first two steps of OPE program - (1) operator expansion in color dipoles and (2) evolution equation for color dipoles.

\subsection{High-energy OPE and photon impact factor}
I will consider the process of deep inelastic scattering (DIS) at small Bjorken $x$. As I discussed in Sect. III, the virtual photon splits in to $q\bar{q}$ pair which 
moves in an ``external'' field of target's gluons. To calculate the leading-order impact factor, we need the quark propagator in the shock-wave background. 
It has the form \cite{npb96}
\begin{eqnarray}
&&\hspace{-11mm}
\langle\hat{\psi}(x)\hat{\bar\psi}(y)\rangle~=~\theta(x_\ast y_\ast){(\not\!x -\not\!y)\over 2\pi^2(x-y)^4}
+i\theta(x_\ast)\theta(-y_\ast)\!\int\! d^4z~\delta(z_\ast){(\not\!x-\not\!z)\over 2\pi^2(x-z)^4}\not\!p_2U_z{(\not\!z-\not\!y)\over 2\pi^2(y-z)^4}
\nonumber\\
&&\hspace{11mm}
-~i\theta(-x_\ast)\theta(y_\ast)\!\int\! d^4z~\delta(z_\ast){(\not\!x-\not\!z)\over 2\pi^2(x-z)^4}\not\!p_2U^\dagger_z{(\not\!z-\not\!y)\over 2\pi^2(x-z)^4}
\label{kvprop}
\end{eqnarray}
where $\bar\psi=-i\psi^\dagger\gamma_0$. We use Dirac matrices $\gamma^\mu=-i\Big(\begin{array}{cc} 0&\sigma^\mu\\ -\bar{\sigma}_\mu&0\end{array}\Big)$ where $\sigma^\mu$ are defined in Eq. (\ref{props}).
The LO impact factor is a product of two propagators (\ref{kvprop}), see Fig. \ref{fig:loif} with solid lines representing quarks
\begin{eqnarray}
&&\hspace{-1mm}
-(x-y)^4\langle\hat{\bar\psi}(x)\gamma_\mu\hat{\psi}(x)\hat{\bar\psi}(y)\gamma_\nu\hat{\psi}(y)\rangle~\stackrel{x_\ast>0>y_\ast}{=}
\nonumber\\
&&\hspace{-1mm}
~{1\over 16\pi^8}\!\int\! d^4z_1d^4z_2~\delta(z_{1\ast})\delta(z_{2\ast}){(x-y)^4\over X_1^4Y_1^4X_2^4Y_2^4}~
{\rm tr}^{\rm spin}\{\gamma_\mu\! \!\not\!X_1\!\!\not\!p_2\!\!\not\!Y_1\gamma_\nu\!\!\not\!Y_2\!\not\!p_2\!\not\!\!X_2\}~{\rm tr}\{U(z_{1\perp})U^\dagger(z_{2\perp})\}
\nonumber\\
&&\hspace{-1mm}
=~-{1\over 2\pi^6}\int {d^2z_{1\perp}d^2z_{2\perp}\over z_{12}^4}
{\calr^2\over(\kappa\cdot\zeta_1)(\kappa\cdot\zeta_2)}{\partial^2\over\partial x^\mu\partial y^\nu}
\big[-2(\kappa\cdot\zeta_1)(\kappa\cdot\zeta_2)+\kappa^2(\zeta_1\cdot\zeta_2) \big]{\rm tr}\{U_{z_{1\perp}}U^\dagger_{z_{2\perp}}\}
\label{fimpactorlo}
\end{eqnarray}
Hereafter I use ${\rm tr}\{...\}$ to denote the color trace in the fundamental representation (and reserve the notation ${\rm Tr}\{...\}$ for the trace in the adjoint representation).
 The above equation is explicitly M\"obius invariant. In addition, it is easy to check that ${\partial\over\partial x_\mu}$(r.h.s)=0.

Feynman diagrams for the photon impact factor are the same as in Fig. \ref{fig:nloif} but with the solid lines representing quarks.
The calculation of the NLO impact factor is similar to the scalar case. The master formula is (cf. Eq. (\ref{conformalinteg}))
\begin{eqnarray}
&&\hspace{-12mm}
\int\! d^4z                                                             
~{\not\!{x}-\not\!{z}\over (x-z)^4}\gamma_\mu
{\not\!{z}-\not\!{y}\over(z-y)^4}{z_\nu\over z^4}-\mu\leftrightarrow\nu~
\nonumber\\
&&\hspace{-2mm} 
=~{\pi^2\over x^2y^2(x-y)^2}\Big[-\!\not\! x\gamma_\mu\!\not\! y\Big({x_\nu\over x^2}+{y_\nu\over y^2}\Big)-\half(\!\not\! x\gamma_\mu\gamma_\nu-\gamma_\mu\gamma_\nu\!\not\! y)
-2x_\mu y_\nu{\!\not\! x-\!\not\! y\over (x-y)^2}\Big]~-~\mu\leftrightarrow\nu
\label{confintegral}
\end{eqnarray}
which gives the 3-point $\psi\Bar{\psi}F_{\mu\nu}$ Green function in the leading order in $g$. Using this formula one easily obtains
\begin{eqnarray}
&&\hspace{-2mm}
-\langle\bar\psi(x)\gamma_\mu\psi(x)\bar\psi(y)\gamma_\nu\psi(y)\rangle~
\stackrel{x_\ast>0>y_\ast}{=}{s\over 64\pi^8x_\ast^2 y_\ast^2}\alpha_s
\!\int\! d^2z_1{d^2z_2 d^2z_3\over z_{13}^2z_{23}^2}
[{\rm tr}\{U_{z_1}U^\dagger_{z_3}\}{\rm tr}\{U_{z_3}U^\dagger_{z_2}\}
-N_c{\rm tr}\{U_{z_1}U^\dagger_{z_2}\}]
\nonumber\\
&&\hspace{-2mm}
\times~\!\int\! dz_{1\bu}dz_{2\bu}dz_{3\bu}dz'_{3\bu}\theta(z_3-z'_3)_\bu
~\Big[{x_\ast \over (x-z_1)^2}{(\!\not\! x-\!\not\! z_3)\over (x-z_3)^4}\gamma_i\!\not\! z_{13}
-2{(\!\not\! x-\!\not\! z_1)x_\ast\over (x-z_1)^4(x-z_3)^2}z^{13}_i\Big] \!\not\! p_2{\!\not\! z_1-\!\not\! y\over (z_1-y)^4}
\nonumber\\
&&\hspace{-2mm}
\times~\gamma_\nu\Big[{y_\ast \over (y-z_2)^2}{(\!\not\! y-\!\not\! z_3)\over (y-z'_3)^4}\gamma^i\!\not\! z_{23}
-2{(\!\not\! y-\!\not\! z_2)y_\ast\over (y-z_2)^4(y-z'_3)^2}z_{23}^i\Big] \!\not\! p_2{\!\not\! z_2-\!\not\! x\over (z_2-x)^4}\gamma_\mu
\label{fla520}
\end{eqnarray}
Performing integrals over $z_\bu$'s and taking traces one gets after some algebra the NLO contribution in the form (recall that $z_{ij}^2=-2(\zeta_i\cdot\zeta_j)$ and $\calz_i=-{4\over\sqrt{s}}(\kappa\cdot\zeta_i)$)
\begin{eqnarray}
&&\hspace{-1mm}
-\hat{\bar\psi}(x)\gamma_\mu\hat{\psi}(x)\hat{\bar\psi}(y)\gamma_\nu\hat{\psi}(y)~
\nonumber\\
&&\hspace{-1mm}
\stackrel{x_\ast>0>y_\ast}{=}~
-{1\over \pi^6}\!\int\! {d^2z_1d^2z_2\over z_{12}^4}~
{ \calr^2\over (\kappa\cdot\zeta_1)(\kappa\cdot\zeta_2)}\Bigg[
[{\rm tr}\{\hat{U}^Y_{z_1} \hat{U}^{\dagger Y}_{z_2}\}]
 {\partial^2\over\partial x^\mu\partial y^\nu}
\big[-(\kappa\cdot\zeta_1)(\kappa\cdot\zeta_2)+\half\kappa^2(\zeta_1\cdot\zeta_2) \big]
\nonumber\\
&&\hspace{-1mm}
+~{\alpha_s\over 16\pi^2}
\!\int\! d^2z_3
[{\rm tr}\{\hat{U}^Y_{z_1}\hat{U}^{\dagger Y}_{z_3}\}{\rm tr}\{\hat{U}^Y_{z_3}\hat{U}^{\dagger Y}_{z_2}\}
-N_c{\rm tr}\{\hat{U}^Y_{z_1}\hat{U}^{\dagger Y}_{z_2}\}]
\nonumber\\
&&\hspace{-1mm}
\times~
\Big\{
{(\zeta_1\cdot\zeta_2)\over (\zeta_1\cdot\zeta_3) (\zeta_1\cdot\zeta_3)}
\big[\ln\sigma^2 s(\kappa\cdot\zeta_3)^2-i\pi+2C\big] {\partial^2\over\partial x^\mu\partial y^\nu}
\big[-2(\kappa\cdot\zeta_1)(\kappa\cdot\zeta_2)+\kappa^2(\zeta_1\cdot\zeta_2) \big]
\nonumber\\
&&\hspace{-1mm}
+~{(\kappa\cdot\zeta_2)\over(\kappa\cdot\zeta_3)}
{\partial^2\over\partial x^\mu\partial y^\nu}
\Big[-{(\kappa\cdot\zeta_1)^2\over (\zeta_1\cdot\zeta_3)}
+{(\kappa\cdot\zeta_1)(\kappa\cdot\zeta_3)(\zeta_1\cdot\zeta_2)\over  (\zeta_1\cdot\zeta_3) (\zeta_2\cdot\zeta_3)}
+{(\kappa\cdot\zeta_1)(\kappa\cdot\zeta_2)\over (\zeta_2\cdot\zeta_3)}-{\kappa^2(\zeta_1\cdot\zeta_2)\over (\zeta_2\cdot\zeta_3)}\Big]
\nonumber\\
&&\hspace{-1mm}
+~{(\kappa\cdot\zeta_1)\over (\kappa\cdot\zeta_3)}
{\partial^2\over\partial x^\mu\partial y^\nu}
\Big[-{(\kappa\cdot\zeta_2)^2\over (\zeta_2\cdot\zeta_3)}
+{(\kappa\cdot\zeta_2)(\kappa\cdot\zeta_3)(\zeta_1\cdot\zeta_2)\over  (\zeta_1\cdot\zeta_3) (\zeta_2\cdot\zeta_3)}
+{(\kappa\cdot\zeta_1)(\kappa\cdot\zeta_2)\over (\zeta_1\cdot\zeta_3)}-{\kappa^2(\zeta_1\cdot\zeta_2)\over (\zeta_1\cdot\zeta_3)}\Big]
\nonumber\\
&&\hspace{-1mm}
+~          
{(\kappa\cdot\zeta_1)^2\over (\kappa\cdot\zeta_3)^2}
{\partial^2\over\partial x^\mu\partial y^\nu}
\Big[{(\kappa\cdot\zeta_2)(\kappa\cdot\zeta_3)\over (\zeta_1\cdot\zeta_3)}-{\kappa^2(\zeta_2\cdot\zeta_3)\over 2(\zeta_1\cdot\zeta_3)}\Big]                          
+{(\kappa\cdot\zeta_2)^2\over (\kappa\cdot\zeta_3)^2}
{\partial^2\over\partial x^\mu\partial y^\nu}
\Big[{(\kappa\cdot\zeta_1)(\kappa\cdot\zeta_3)\over (\zeta_2\cdot\zeta_3)}-{\kappa^2(\zeta_1\cdot\zeta_3)\over 2(\zeta_2\cdot\zeta_3)}\Big]\Big\}\Bigg]
\label{nloif1}
\end{eqnarray}
where we have promoted the equation to the operator form. It can be demonstrated that ${\partial\over\partial x^\mu}$[r.h.s. of Eq. (\ref{nloif1})] = 0 which reflects the electromagnetic gauge invariance.

As for the N=4 case, it is convenient to re-expand $T\{j_\mu(x) j_\nu(y)\}$  in composite operators obtained by replacement $T^a\rightarrow t^a={\lambda^a\over 2}$ (and Tr $\rightarrow$ tr) in Eq. (\ref{confodipole1})
\begin{equation}
\big[{\rm tr}\{\hat{U}_{z_1}\hat{U}^{\dagger}_{z_2}\}\big]_a~
=~{\rm tr}\{\hat{U}^\sigma_{z_1}\hat{U}^{\dagger\sigma}_{z_2}\}
+{\alpha_s\over 4\pi^2}\!\int\! d^2 z_3~{z_{12}^2\over z_{13}^2z_{23}^2}
[ {\rm tr}\{\hat{U}^\sigma_{z_1}\hat{U}^{\dagger\sigma}_{z_3}\}{\rm tr}\{\hat{U}^\sigma_{z_3}\hat{U}^{\dagger\sigma}_{z_2}\}
-N_c {\rm tr}\{\hat{U}^\sigma_{z_1}\hat{U}^{\dagger\sigma}_{z_2}\}]
\ln {4az_{12}^2\over s\sigma^2z_{13}^2z_{23}^2}~+~...
\label{compodipole}
\end{equation}
where dots stand for $\alpha_s^2$ terms not displayed here for brevity.
In QCD, this composite operator is no longer conformal but still more convenient than the original dipole since the evolution equation
for this operator splits into the sum of the conformal part and the running-coupling part proportional to $b={11\over 3}N_c-{2\over 3}n_f$ (see the next Section). 
The final expansion of $T\{j_\mu(x) j_\nu(y)\}$ in composite dipoles (\ref{compodipole}) has the form (cf. Eq. (\ref{opeconf}))
\begin{eqnarray}
&&\hspace{-1mm}
-\hat{\bar\psi}(x)\gamma_\mu\hat{\psi}(x)\hat{\bar\psi}(y)\gamma_\nu\hat{\psi}(y)~
\nonumber\\
&&\hspace{-1mm}
\stackrel{x_\ast>0>y_\ast}{=}~
-{1\over \pi^6}\!\int\! {d^2z_1d^2z_2\over z_{12}^4}~
{ \calr^2\over (\kappa\cdot\zeta_1)(\kappa\cdot\zeta_2)}\Bigg[
[{\rm tr}\{\hat{U}_{z_1} \hat{U}^\dagger_{z_2}\}]_{a_0}
 {\partial^2\over\partial x^\mu\partial y^\nu}
\big[-(\kappa\cdot\zeta_1)(\kappa\cdot\zeta_2)+\half\kappa^2(\zeta_1\cdot\zeta_2) \big]
\nonumber\\
&&\hspace{-1mm}
+~{\alpha_s\over 16\pi^2}
\!\int\! d^2z_3
[{\rm tr}\{\hat{U}_{z_1}\hat{U}^\dagger_{z_3}\}{\rm tr}\{\hat{U}_{z_3}\hat{U}^\dagger_{z_2}\}
-N_c{\rm tr}\{\hat{U}_{z_1}\hat{U}^\dagger_{z_2}\}]_{a_0}
\nonumber\\
&&\hspace{-1mm}
\times~
\Big\{
{(\zeta_1\cdot\zeta_2)\over (\zeta_1\cdot\zeta_3) (\zeta_1\cdot\zeta_3)}
\big[\ln{-x_\ast y_\ast z_{12}^2\calz_3^2\over (x-y)^2 z_{13}^2z_{23}^2}
+2C\big] {\partial^2\over\partial x^\mu\partial y^\nu}
\big[-2(\kappa\cdot\zeta_1)(\kappa\cdot\zeta_2)+\kappa^2(\zeta_1\cdot\zeta_2) \big]
\nonumber\\
&&\hspace{-1mm}
+~{(\kappa\cdot\zeta_2)\over(\kappa\cdot\zeta_3)}
{\partial^2\over\partial x^\mu\partial y^\nu}
\Big[-{(\kappa\cdot\zeta_1)^2\over (\zeta_1\cdot\zeta_3)}
+{(\kappa\cdot\zeta_1)(\kappa\cdot\zeta_3)(\zeta_1\cdot\zeta_2)\over  (\zeta_1\cdot\zeta_3) (\zeta_2\cdot\zeta_3)}
+{(\kappa\cdot\zeta_1)(\kappa\cdot\zeta_2)\over (\zeta_2\cdot\zeta_3)}-{\kappa^2(\zeta_1\cdot\zeta_2)\over (\zeta_2\cdot\zeta_3)}\Big]
\nonumber\\
&&\hspace{-1mm}
+~{(\kappa\cdot\zeta_1)\over (\kappa\cdot\zeta_3)}
{\partial^2\over\partial x^\mu\partial y^\nu}
\Big[-{(\kappa\cdot\zeta_2)^2\over (\zeta_2\cdot\zeta_3)}
+{(\kappa\cdot\zeta_2)(\kappa\cdot\zeta_3)(\zeta_1\cdot\zeta_2)\over  (\zeta_1\cdot\zeta_3) (\zeta_2\cdot\zeta_3)}
+{(\kappa\cdot\zeta_1)(\kappa\cdot\zeta_2)\over (\zeta_1\cdot\zeta_3)}-{\kappa^2(\zeta_1\cdot\zeta_2)\over (\zeta_1\cdot\zeta_3)}\Big]
\nonumber\\
&&\hspace{-1mm}
+~          
{(\kappa\cdot\zeta_1)^2\over (\kappa\cdot\zeta_3)^2}
{\partial^2\over\partial x^\mu\partial y^\nu}
\Big[{(\kappa\cdot\zeta_2)(\kappa\cdot\zeta_3)\over (\zeta_1\cdot\zeta_3)}-{\kappa^2(\zeta_2\cdot\zeta_3)\over 2(\zeta_1\cdot\zeta_3)}\Big]                          
+{(\kappa\cdot\zeta_2)^2\over (\kappa\cdot\zeta_3)^2}
{\partial^2\over\partial x^\mu\partial y^\nu}
\Big[{(\kappa\cdot\zeta_1)(\kappa\cdot\zeta_3)\over (\zeta_2\cdot\zeta_3)}-{\kappa^2(\zeta_1\cdot\zeta_3)\over 2(\zeta_2\cdot\zeta_3)}\Big]\Big\}\Bigg]
\label{nloifoton}
\end{eqnarray}
where we set $a_0=\kappa^{-2}+i\epsilon={ 4x_\ast y_\ast\over s(x-y)^2}+i\epsilon$ same as for the ${\cal N}=4$ case.  
The explicit expression for four-Wilson-line composite operator 
$[{\rm tr}\{\hat{U}_{z_1}\hat{U}^\dagger_{z_3}\}{\rm tr}\{\hat{U}_{z_3}\hat{U}^\dagger_{z_2}\}]_a$ can be obtained from Eq.
(\ref{confopera1}) by usual substitution $T^a\rightarrow t^a$ and Tr $\rightarrow$ tr.
(It is worth noting that 
at large $N_c$ $[{\rm tr}\{\hat{U}_{z_1}\hat{U}^\dagger_{z_3}\}{\rm tr}\{\hat{U}_{z_3}\hat{U}^\dagger_{z_2}\}]_a
=[{\rm tr}\{\hat{U}_{z_1}\hat{U}^\dagger_{z_3}\}]_a[{\rm tr}\{\hat{U}_{z_3}\hat{U}^\dagger_{z_2}\}]_a$.  )
 It is easy to see now
that the NLO impact factor (\ref{nloifoton}) is  M\"obius invariant.

\subsection{Evolution of color dipoles in QCD}

Second step is the evolution equation for composite operator (\ref{compodipole}). The types of Feynman diagrams (in the shock-wave background)
in QCD are the same as in ${\cal N}=4$ SYM (see Fig. \ref{nlofigs}) but now the solid lines in Figs. \ref{nlofigs} VII and VIII  denote quarks rather than 
scalar or gluinos. The result of the evolution of the color dipole with rapidity cutoff (\ref{cutoff}) is \cite{nlobk}:  
\begin{eqnarray}
&&\hspace{-2mm}
{d\over dY}{\rm tr}\{\hat{U}_{z_1} \hat{U}^{\dagger}_{z_2}\}~
\label{nlobk}\\
&&\hspace{-2mm}
=~{\alpha_s\over 2\pi^2}
\!\int\!d^2z~
{z_{12}^2\over z_{13}^2z_{23}^2}\Big\{1+{\alpha_s\over 4\pi}\Big[b\ln z_{12}^2\mu^2
-b{z_{13}^2-z_{23}^2\over z_{12}^2}\ln{z_{13}^2\over z_{23}^2}+
({67\over 9}-{\pi^2\over 3})N_c-{10\over 9}n_f
\nonumber\\
&&\hspace{62mm} 
-~
2N_c\ln{z_{13}^2\over z_{12}^2}\ln{z_{23}^2\over z_{12}^2}\Big]\Big\}
~[{\rm tr}\{\hat{U}_{z_1} \hat{U}^{\dagger}_{z_3}\}{\rm tr}\{\hat{U}_{z_3} \hat{U}^{\dagger}_{z_2}\}
-N_c{\rm tr}\{\hat{U}_{z_1} \hat{U}^{\dagger}_{z_2}\}]   
\nonumber\\
&&\hspace{-2mm} 
+~{\alpha_s^2\over 16\pi^4}
\int \!d^2 z_3d^2 z_4
\Bigg[
\Big(-{4\over z_{34}^4}+\Big\{2{z_{13}^2z_{24}^2+z_{14}^2z_{23}^2-4z_{12}^2z_{34}^2\over  z_{34}^4[z_{13}^2z_{24}^2-z_{14}^2z_{23}^2]}
+~{z_{12}^4\over z_{13}^2z_{24}^2-z_{14}^2z_{23}^2}\Big[
{1\over z_{13}^2z_{24}^2}+{1\over z_{23}^2z_{14}^2}\Big]
\nonumber\\ 
&&\hspace{-2mm}
+~{z_{12}^2\over z_{34}^2}\Big[{1\over z_{13}^2z_{24}^2}-{1\over z_{14}^2z_{23}^2}\Big]\Big\}
\ln{z_{13}^2z_{24}^2\over z_{14}^2z_{23}^2}\Big)
[{\rm tr}\{\hat{U}_{z_1} \hat{U}^\dagger_{z_3}\}{\rm tr}\{\hat{U}_{z_3}\hat{U}^\dagger_{z_4}\}{\rm tr}\{\hat{U}_{z_4}\hat{U}^\dagger_{z_2}\}
-{\rm tr}\{\hat{U}_{z_1} \hat{U}^\dagger_{z_3} \hat{U}_{z_4}U^\dagger_{z_2}\hat{U}_{z_3}\hat{U}^\dagger_{z_4}\}-(z_4\rightarrow {z_3})]
\nonumber\\ 
&&\hspace{-2mm}
+~\Big\{{z_{12}^2\over z_{34}^2 }\Big[{1\over z_{13}^2z_{24}^2}+{1\over z_{23}^2z_{14}^2}\Big]
-{z_{12}^4\over  z_{13}^2z_{24}^2z_{14}^2z_{23}^2}\Big\}\ln{z_{13}^2z_{24}^2\over z_{14}^2z_{23}^2}
~{\rm tr}\{\hat{U}_{z_1} \hat{U}^\dagger_{z_3}\}{\rm tr}\{\hat{U}_{z_3}\hat{U}^\dagger_{z_4}\}{\rm tr}\{\hat{U}_{z_4}\hat{U}^\dagger_{z_2}\}
\nonumber\\
&&\hspace{-2mm}
+~4n_f
\Big\{{4\over z_{34}^4}
-2{z_{14}^2z_{23}^2+z_{24}^2z_{13}^2-z_{12}^2z_{34}^2\over z_{34}^4(z_{13}^2z_{24}^2-z_{14}^2z_{23}^2)}
\ln{z_{13}^2z_{24}^2\over z_{14}^2z_{23}^2}\Big\}{\rm tr}\{t^a\hat{U}_{z_1}t^b\hat{U}^{\dagger}_{z_2}\}
[{\rm tr}\{t^a
\hat{U}_{z_3}t^b \hat{U}^\dagger_{z_4}\}-(z_4\rightarrow z_3)]\Bigg]
\nonumber
\end{eqnarray}
where we use the  $\overline{MS}$ scheme. The NLO kernel is a sum of the running-coupling part (proportional to $b$), the non-conformal  
double-log 
term $\sim \ln{z_{12}^2\over z_{13}^2} \ln{z_{12}^2\over z_{13}^2}$ (same as in ${\cal N}=4$ case) and the three conformal terms which 
depend on the two four-point conformal ratios ${z_{13}^2z_{24}^2\over z_{14}^2z_{23}^2}$ 
and ${z_{12}^2z_{34}^2\over z_{13}^2 z_{24}^2}$.  

A natural guess for QCD result as opposed to ${\cal N}=4$ answer would be some sort of tree-level conformal structure which ``get dressed'' by  the 
running coupling constant. At the NLO level, this would correspond to the sum of the conformal part and the running-coupling part proportional to $b$.
As we see, the Eq. (\ref{nlobk}) does not quite look like such sum due to the presence of the double-log term. It turns out, however, that if one rewrites the evolution equation
(\ref{nlobk}) in terms of composite operators (\ref{compodipole}), the NLO kernel separates into the sum of conformal and running-coupling parts \cite{nlobksym}:
\begin{eqnarray}
&&\hspace{-2mm}
2a{d\over da}\big[{\rm tr}\{ \hat{U}_{z_1} \hat{U}^{\dagger}_{z_2}\}\big]_a~
=~{\alpha_s\over 2\pi^2}
\!\int\!d^2z_3~
{z_{12}^2\over z_{13}^2z_{23}^2}\big[{\rm tr}\{\hat{U}_{z_1}\hat{U}^{\dagger}_{z_3}\}{\rm tr}\{\hat{U}_{z_3}\hat{U}^{\dagger}_{z_2}\}
-N_c {\rm tr}\{\hat{U}_{z_1}\hat{U}^{\dagger}_{z_2}\}\big]_a  
\nonumber\\ 
&&\hspace{-2mm}
\times~\Big\{1+{\alpha_s\over 4\pi}\Big[b(\ln {z_{12}^2\mu^2\over 4}+2C)
-b{z_{13}^2-z_{23}^2\over  z_{12}^2}\ln{z_{13}^2\over z_{23}^2}+
\big({67 \over 9}-{\pi^2\over 3}\big)N_c-{10\over 9}n_f\Big]\Big\}
\nonumber\\
&&\hspace{-2mm} 
+~{\alpha_s^2\over 16\pi^4}
\int \!{d^2 z_3d^2 z_4\over z_{34}^4}\Bigg[ \Big\{\Big(
-2+2{z_{12}^2z_{34}^2\over z_{13}^2z_{24}^2}\ln{z_{12}^2z_{34}^2\over z_{14}^2z_{23}^2}
+
\Big[{z_{12}^2z_{34}^2\over z_{13}^2z_{24}^2}
\big(1+{z_{12}^2z_{34}^2\over z_{13}^2z_{24}^2-z_{14}^2z_{23}^2}\big)
+{2z_{13}^2z_{24}^2-4z_{12}^2z_{34}^2\over z_{13}^2z_{24}^2-z_{14}^2z_{23}^2}
\Big]\ln{z_{13}^2z_{24}^2\over z_{14}^2z_{23}^2}\Big)
\nonumber\\ 
&&\hspace{-2mm}
+~\big(z_3\leftrightarrow z_4\big)\Big\}~\big[\big({\rm tr}\{\hat{U}_{z_1}\hat{U}^{\dagger}_{z_3}\}{\rm tr}\{\hat{U}_{z_3}\hat{U}^{\dagger}_{z_4}\}
{\rm tr}\{\hat{U}_{z_4}\hat{U}^{\dagger}_{z_2}\}
-{\rm tr}\{\hat{U}_{z_1}\hat{U}^{\dagger}_{z_3}\hat{U}_{z_4}
\hat{U}^{\dagger}_{z_2}\hat{U}_{z_3}\hat{U}^{\dagger}_{z_4}\}
\big)
-(z_4\rightarrow z_3)\big]_a
\nonumber\\
&&\hspace{-2mm} 
+~{z_{12}^2z_{34}^2\over z_{13}^2z_{24}^2}
\Big\{2\ln{z_{12}^2z_{34}^2\over z_{14}^2z_{23}^2}
+\Big[1+{z_{12}^2z_{34}^2\over z_{13}^2z_{24}^2-z_{14}^2z_{23}^2}\Big]\ln{z_{13}^2z_{24}^2\over z_{14}^2z_{23}^2}\Big\}
\big[{\rm tr}\{\hat{U}_{z_1}\hat{U}^{\dagger}_{z_3}\}{\rm tr}\{\hat{U}_{z_3}\hat{U}^{\dagger}_{z_4}\}
{\rm tr}\{\hat{U}_{z_4}\hat{U}^{\dagger}_{z_2}\}
-z_3\leftrightarrow z_4\big]_a\Bigg]
\nonumber\\
&&\hspace{-2mm} 
+~{\alpha^2_sn_f \over 2\pi^4}\!\int\!{d^2z_3 d^2z_4\over z_{34}^4}
\Big\{2-{z_{13}^2z_{24}^2+z_{23}^2z_{14}^2- z_{12}^2z_{34}^2\over z_{13}^2z_{24}^2-z_{14}^2z_{23}^2}
\ln{z_{13}^2 z_{24}^2\over z_{14}^2z_{23}^2}\Big\}
[{\rm tr}\{t^a\hat{U}_{z_1}t^b\hat{U}^{\dagger}_{z_2}\}
{\rm tr}\{t^a\hat{U}_{z_3}t^b(\hat{U}^{\dagger}_{z_4}-\hat{U}_{z_3})\}]_a.
\nonumber\\
&&\hspace{26mm}
\label{nlobkqcd}
\end{eqnarray}
Following the analysis of Ref. \cite{nlobk} I will outline how the above kernel reproduces the NLO BFKL eigenvalues \cite{nlobfkl1} (for details see the Appendix).

In the two-gluon approximation we get
\begin{equation}
\hspace{0mm}  
{\rm tr}\{\hat{U}^Y_{z_1}\hat{U}^{\dagger Y}_{z_3}\}{\rm tr}\{\hat{U}^Y_{z_3}\hat{U}^{\dagger Y}_{z_2}\}-N_c{\rm tr}\{\hat{U}^Y_{z_1}\hat{U}^{\dagger Y}_{z_2}\}~
=~-N_c\big[\hat{\cal U}^Y(z_1,z_3)+\hat{\cal U}^Y(z_2,z_3)-\hat{\cal U}^Y(z_1,z_2)\big]
\label{confoper4lin}
\end{equation}
where 
\begin{equation}
\hat{\cal U}^Y(x_\perp,y_\perp)=1-{1\over N_c}
{\rm tr}\{\hat{U}^Y(x_\perp)\hat{U}^{\dagger Y}(y_\perp)\}
\label{fundipole}
\end{equation}
is the color dipole in the fundamental representation. 
In this approximation the composite dipole (\ref{compodipole}) reduces to (cf. Eq. (\ref{confodipola}))
\begin{eqnarray}
&&\hspace{-2mm}
\hat{\cal U}_a(z_1,z_2)~
~=~\calu^Y(z_1,z_2)+~{\alpha_sN_c\over 4\pi^2}
\!\int\!d^2z_3~
{z_{12}^2\over z_{13}^2 z_{23}^2}
[\hat\calu^Y(z_1,z_3)+\hat\calu^Y(z_2,z_3)-\hat\calu^Y(z_1,z_2)]\ln\big({4a z_{12}^2\over \sigma^2sz_{13}^2z_{23}^2}\big)
+~{\alpha_s^2N_c^2\over 16\pi^4}
\nonumber\\
&&\hspace{-2mm}
\times~
\!\int\!d^2z_3~
{z_{12}^2\over z_{13}^2 z_{23}^2}\Big[{b\over N_c}\big(\ln z_{12}^2\mu^2
-{z_{13}^2-z_{23}^2\over z_{12}^2}\ln{z_{13}^2\over z_{23}^2}\big)+
{67\over 9}-{\pi^2\over 3}-{10n_f\over 9N_c}\Big]
[\hat{\cal U}^Y(z_1,z_3)+\hat{\cal U}^Y(z_2,z_3)-\hat{\cal U}^Y(z_1,z_2)]  \ln{4a\over\sigma^2s} 
\nonumber\\
&&\hspace{-2mm} 
+~{\alpha_s^2N_c^2\over 16\pi^4}\!
\int \!{d^2 z_3d^2 z_4\over z_{34}^4}
\Big\{
-2+{z_{13}^2z_{24}^2+z_{14}^2z_{23}^2-4z_{12}^2z_{34}^2\over  z_{13}^2z_{24}^2-z_{14}^2z_{23}^2}\ln{z_{13}^2z_{24}^2\over z_{14}^2z_{23}^2}
-~{n_f\over N_c^3}
\Big[2
-{z_{14}^2z_{23}^2+z_{24}^2z_{13}^2-z_{12}^2z_{34}^2\over z_{14}^2z_{23}^2-z_{24}^2z_{13}^2}
\ln{z_{14}^2z_{23}^2\over z_{24}^2z_{13}^2}\Big]
\nonumber\\ 
&&\hspace{-2mm}
+~{z_{12}^2z_{34}^2\over z_{13}^2z_{24}^2}\Big[2\ln{z_{12}^2z_{34}^2\over z_{14}^2z_{23}^2}+
\Big(1+{ z_{12}^2z_{34}^2\over z_{13}^2z_{24}^2- z_{14}^2z_{23}^2}\Big)
\ln{z_{13}^2z_{24}^2\over z_{14}^2z_{23}^2}\Big]
\Big\}~\hat{\cal U}^Y(z_3,z_4)\ln{4a\over\sigma^2s}
+~{\alpha_s^2N_c^2\over 4\pi^2} \zeta(3)~\hat{\cal U}^Y(z_1,z_2)\ln{4a\over\sigma^2s}
\nonumber\\
&&\hspace{-2mm} 
+~{\alpha_s^2N_c^2\over 32\pi^4}
\!\int\!d^2z_3d^2 z_4~{z_{12}^2\over  z_{13}^2 z_{23}^2}\Big\{
{z_{13}^2\over z_{14}^2 z_{34}^2}
[\hat\calu^Y({z_1,z_4})+\hat\calu^Y({z_3,z_4})-\hat\calu^Y({z_1,z_3})]
\ln^2\big({4a z_{13}^2\over \sigma^2sz_{14}^2z_{34}^2}\big)~
+~{z_{23}^2\over z_{34}^2 z_{24}^2}~
[\hat\calu^Y({z_3,z_4})
\nonumber\\
&&\hspace{-2mm}
+~\hat\calu^Y({z_2,z_4})-\hat\calu^Y({z_2,z_3})]
\ln^2\big({4a z_{23}^2\over \sigma^2sz_{24}^2z_{34}^2}\big)
-{z_{12}^2\over z_{14}^2 z_{24}^2}
[\hat\calu^Y({z_1,z_4})+\hat\calu^Y({z_2,z_4})-\hat\calu^Y({z_1,z_2})]
\ln^2\big({4a z_{12}^2\over \sigma^2sz_{14}^2z_{24}^2}\big)\Big\}
\label{compopelin}
\end{eqnarray}
It is easy to see that ${d\over dY}$ ($\equiv \sigma{d\over d\sigma}$) of the r.h.s. vanishes.

The evolution equation (\ref{nlobkqcd}) turns into
\begin{eqnarray}
&&\hspace{-5mm}
2a{d\over da}\hat{\cal U}_a(z_1,z_2)~
=~{\alpha_sN_c\over 2\pi^2}
\!\int\!d^2z_3~
{z_{12}^2\over z_{13}^2z_{23}^2}\Big[1
\nonumber\\ 
&&\hspace{-2mm}
+~{\alpha_s\over 4\pi}\Big[b\big({\ln z_{12}^2\mu^2\over 4}+2C\big)
-b{z_{13}^2-z_{23}^2\over  z_{12}^2}\ln{z_{13}^2\over z_{23}^2}+
\big({67 \over 9}-{\pi^2\over 3}\big)N_c-{10\over 9}n_f\Big]
\big[\hat{\cal U}_a(z_1,z_3)+\hat{\cal U}_a(z_2,z_3)-\hat{\cal U}_a(z_1,z_2)\big]
\nonumber\\
&&\hspace{-2mm} 
+~{\alpha_s^2N_c^2\over 8\pi^4}
\int \!{d^2 z_3d^2 z_4\over z_{34}^2} \Big\{
2{z_{12}^2z_{34}^2\over z_{13}^2z_{24}^2}\ln{z_{12}^2z_{34}^2\over z_{14}^2z_{23}^2}
+
{z_{12}^2z_{34}^2\over z_{13}^2z_{24}^2}
\Big(1+{z_{12}^2z_{34}^2\over z_{13}^2z_{24}^2-z_{14}^2z_{23}^2}\Big)\ln{z_{13}^2z_{24}^2\over z_{14}^2z_{23}^2}
-{3z_{12}^2z_{34}^2\over z_{13}^2z_{24}^2-z_{14}^2z_{23}^2}
\ln{z_{13}^2z_{24}^2\over z_{14}^2z_{23}^2}
\nonumber\\ 
&&\hspace{-2mm}
+~\big(1+{n_f\over N_c^3}\big)\Big({z_{13}^2z_{24}^2+z_{14}^2z_{23}^2-z_{12}^2z_{34}^2\over z_{13}^2z_{24}^2-z_{14}^2z_{23}^2}
\ln{z_{13}^2z_{24}^2\over z_{14}^2z_{23}^2}-2\Big)
\Big\}
\hat{\cal U}_a(z_3,z_4)+{3\alpha_s^2N_c^2\over 2\pi^2}\zeta(3)\hat{\cal U}_a(z_1,z_2)
\label{nlolin}
\end{eqnarray}
where we used formula \cite{nlobksym}
\begin{eqnarray}
&&\hspace{-5mm}
\int \!d^2 z_4~\Big\{{z_{12}^2\over z_{13}^2z_{24}^2z_{34}^2}
\Big(2\ln{z_{12}^2z_{34}^2\over z_{14}^2z_{23}^2}
+\Big[1+{z_{12}^2z_{34}^2\over z_{13}^2z_{24}^2-z_{14}^2z_{23}^2}\Big]\ln{z_{13}^2z_{24}^2\over z_{14}^2z_{23}^2}\Big)
-z_3\leftrightarrow z_4\Big\}
~=~12\pi\zeta(3)[\delta(z_{23})-\delta(z_{13})]
\label{formula85}
\end{eqnarray}

For the case of forward scattering $\langle\hat{\cal U}(x,y)\rangle={\cal U}(x-y)$ and the linearized 
equation (\ref{nlolin}) can be reduced to an integral equation with respect to one variable $z\equiv z_{12}$. Using
integrals (104)-(106) from Ref. \cite{nlobk} and the integral
\begin{eqnarray}
&&\hspace{-5mm}
\int\! d\tilde{z}{1\over \tilde{z}^2(z-z'-\tilde{z})^2}\ln{z^2{z'}^2\over (z-\tilde{z}^2)(z'-\tilde{z}^2)}~=~-{\pi\over (z-z')^2}\ln^2{z^2\over {z'}^2}
\nonumber
\end{eqnarray}
we obtain
\begin{eqnarray}
&&\hspace{-2mm}
2a{d\over da}{\cal U}_a(z)~
~
=~{\alpha_sN_c\over 2\pi^2}
\!\int\!d^2z~
{z^2\over (z-z')^2 {z'}^2}\Big\{1
\nonumber\\
&&\hspace{-2mm} 
+~{\alpha_s\over 4\pi}\Big[b\big(\ln {z^2\mu^2\over 4} +2C\big)
-b{(z-z')^2-{z'}^2\over z^2}\ln{(z-z')^2\over {z'}^2}+
({67\over 9}-{\pi^2\over 3})N_c-{10\over 9}n_f\Big]
[{\cal U}_a(z-z')+{\cal U}_a(z')-{\cal U}_a(z)]
\nonumber\\
&&\hspace{-2mm} 
+~{\alpha_s^2N_c^2\over 4\pi^3}\!\int\! d^2z'~ {z^2\over {z'}^2}
\Big[-{1\over (z-z')^2}\ln^2{z^2\over {z'}^2}+F(z,z')+\Phi(z,z')\Big]~{\cal U}_a(z') +3{\alpha_s^2N_c^2\over 2\pi^2}\zeta(3){\cal U}_a(z)
\label{nlobkernel}
\end{eqnarray}
where 
\begin{eqnarray}
&&\hspace{-2mm}
F(z,z')~=~\Big(1+{n_f\over N_c^3}\Big){3(z,z')^2-2z^2{z'}^2\over16z^2{z'}^2}
\Big({2\over z^2}+{2\over {z'}^2}+{z^2-{z'}^2\over z^2{z'}^2}\ln{z^2\over {z'}^2}\Big)
\nonumber\\
&&\hspace{-1mm}
-~\Big[3+\Big(1+{n_f\over N_c^3}\Big)\Big(1-{(z^2+{z'}^2)^2\over 8z^2{z'}^2}
+{3z^4+3{z'}^4-2z^2{z'}^2\over 16z^4{z'}^4}(z,z')^2\Big)\Big]\!\int_0^\infty\!dt{1\over z^2+t^2{z'}^2}\ln{1+t\over |1-t|} 
\label{F}
\end{eqnarray}
and 
\begin{eqnarray}
&&\hspace{-2mm}
\Phi(z,z')~=~{(z^2-{z'}^2)\over (z-z')^2(z+z')^2}
\Big[\ln{z^2\over {z'}^2}\ln{z^2{z'}^2(z-z')^4\over (z^2+{z'}^2)^4}
+2{\rm Li_2}\Big(-{{z'}^2\over z^2}\Big)-2{\rm Li_2}\Big(-{z^2\over {z'}^2}\Big)\Big]
\nonumber\\
&&\hspace{-1mm}
-~\Big(1-{(z^2-{z'}^2)^2\over (z-z')^2(z+z')^2}\Big)\Big[\!\int_0^1-\int_1^\infty\Big]
{du\over (z-z'u)^2}\ln{u^2{z'}^2\over z^2}
\label{Phi}
\end{eqnarray}
The function $-{1\over (q-q')^2}\ln^2{q^2\over {q'}^2}+F(q,q')+\Phi(q,q')$ enters the NLO BFKL equation in the momentum space \cite{nlobfkl1}
and since the eigenfunctions of the forward BFKL equation are powers both in the coordinate and momentum space, 
it is clear that the corresponding eigenvalues coincide. Moreover, it can be demonstrated explicitly that
\begin{equation}
{1\over 4\pi^2}\!\int\! d^2qd^2q'~e^{i(q,z)-i(q',z')}\Big[-{1\over (q-q')^2}\ln^2{q^2\over {q'}^2}+F(q,q')+\Phi(q,q')\big]~=~-{1\over (z-z')^2}\ln^2{z^2\over {z'}^2}+F(z,z')+\Phi(z,z')
\label{furye}
\end{equation}
so the conformal (${\cal N}=4$) part of the forward kernel looks the same in the coordinate and in the momentum representation. 
As to the running-coupling part (the first term in the r.h.s. of Eq. (\ref{nlobfkernel})),  in the Appendix we demonstrate that it also agrees with the eigenvalues (\ref{eigen1}). 

\subsection{Argument of the coupling constant in the BK equation}
I will not discuss here the steps 3 and 4 of our OPE program. The reason is that in DIS from nucleon or nucleus the 
evolution of color dipoles is non-linear and the analytic solution is not known at the present time.  Even in the simpler case of 
forward $\gamma^\ast\gamma^\ast$ or onium-onium scattering where the dipole evolution is described by linear NLO BFKL equation,
it is impossible to solve this equation since we do not know the argument of the coupling constant. Moreover, it is not known how
to solve analytically this equation even if we take some simple model for the argument of coupling constant like the size of the parent dipole. 

\begin{figure}
\includegraphics[width=0.7\textwidth]{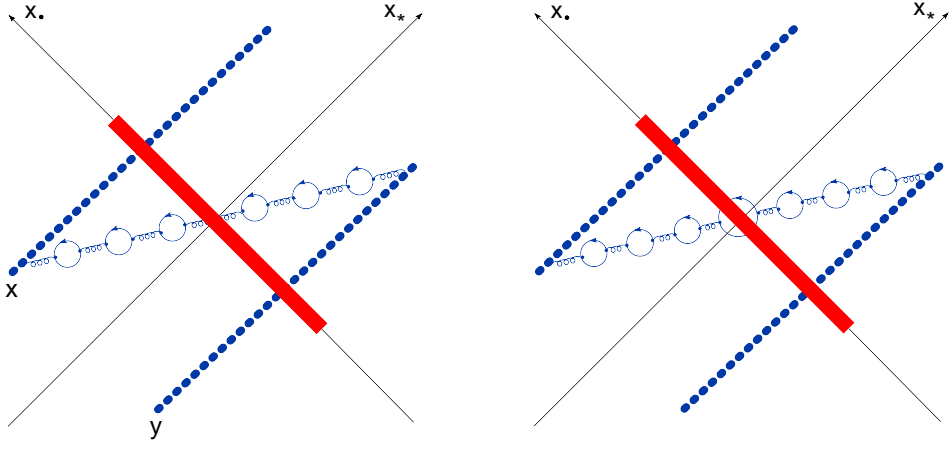}
\caption{Renormalon bubble chain of quark loops\label{quarkloops}.}
\end{figure}

Still, the first step towards the solution would be to figure out the argument of coupling constant in the NLO BFKL equation.
To get an argument of coupling constant we can use the renormalon-based approach (for a review, see Ref. \cite{reno}) and  trace the quark part of the 
$\beta$-function proportional to $n_f$.  In the leading log approximation $\alpha_s\ln {p^2\over\mu^2}\sim 1, ~\alpha_s\ll 1$ the quark part 
 of the $\beta$-function comes from the bubble chain of quark loops in the shock-wave background. 
 We can either have no intersection of quark loop with the shock wave (see  Fig. \ref{quarkloops}a) or we may have one of the loops in the shock-wave background
 (see Fig. \ref{quarkloops}b).

The sum of these diagrams yields
\begin{eqnarray}
&&\hspace{-16mm}
{d\over dY}\langle{\rm Tr}\{\hat{U}_{z_1}\hat{U}^\dagger_{z_2}\}\rangle~=~-2\alpha_s{\rm Tr}\{t^aU_{z_1}t^bU^\dagger_{z_2}\}
\int\! \dhd^2p\dhd^2l~[e^{i(p,{z_1})_\perp}-e^{i(p,{z_2})_\perp}]
[e^{-i(p-l,{z_1})_\perp}-e^{-i(p-l,{z_2})_\perp}]
\nonumber\\
&&\hspace{-16mm}
\times~
{1\over p^2(1+{\alpha_s\over 6\pi}\ln{\mu^2\over p^2})}
\Big(1-{\alpha_sn_f\over 6\pi}\ln{l^2\over \mu^2}\Big)\partial_\perp^2U^{ab}(l)
{1\over (p-l)^2(1+{\alpha_s\over 6\pi}\ln{\mu^2\over (p-l)^2})}
\label{bubblechain1}
\end{eqnarray}
where we have left only the $\beta$-function part of the quark loop.
Replacing the quark part of the $\beta$-function 
$-{\alpha_s\over 6\pi}n_f\ln{p^2\over\mu^2}$ by
the total contribution ${\alpha_s\over 4\pi}b\ln{p^2\over\mu^2}$  we get
\begin{eqnarray}
&&\hspace{-16mm}
{d\over dY}\langle{\rm Tr}\{\hat{U}_{z_1}\hat{U}^\dagger_{z_2}\}\rangle~=~-2{\rm Tr}\{t^aU_{z_1}t^bU^\dagger_{z_2}\}
\nonumber\\
&&\hspace{-16mm}
\times\int\! \dhd^2p\dhd^2q~[e^{i(p,{z_1})_\perp}-e^{i(p,{z_2})_\perp}]
[e^{-i(p-l,{z_1})_\perp}-e^{-i(p-l,{z_2})_\perp}]
{\alpha_s(p^2)\over p^2}
\alpha_s^{-1}(l^2)\partial_\perp^2U^{ab}(q)
{\alpha_s((p-l)^2)\over (p-l)^2}
\label{bubblechain2}
\end{eqnarray}
In principle, one should also include the ``renormalon dressing''  of  $\alpha_s^2$ 
in Eq. (\ref{nlobkqcd}).   We think, however, that they form a separate contribution 
which has nothing to do with the argument of the BK equation.

To go to the coordinate space, we expand the coupling constants in Eq. (\ref{bubblechain2}) 
in powers of $\alpha_s=\alpha_s(\mu^2)$, i.e. return back to Eq. (\ref{bubblechain1}) 
with ${\alpha_s\over 6\pi}n_f\rightarrow -b{\alpha_s\over 4\pi}$.
Unfortunately, the Fourier transformation to the coordinate space can be performed explicitly only for a couple of  first terms of the expansion 
  $\alpha_s(p^2)\simeq \alpha_s-{b\alpha_s\over 4\pi}\ln p^2/\mu^2+({b\alpha_s\over 4\pi}\ln p^2/\mu^2)^2$. With this accuracy \cite{prd75, kw1}
\begin{eqnarray}
&&\hspace{-6mm}
{d\over dY}{\rm Tr}\{\hat{U}_{z_1} \hat{U}^{\dagger}_{z_2}\}~
=~{\alpha_s(z_{12}^2)\over 2\pi^2}
\!\int\!d^2z~[{\rm Tr}\{\hat{U}_{z_1} \hat{U}^{\dagger}_{z_3}\}{\rm Tr}\{\hat{U}_{z_3} \hat{U}^{\dagger}_{z_2}\}
-N_c{\rm Tr}\{\hat{U}_{z_1} \hat{U}^{\dagger}_{z_2}\}]
\label{runningcoupling}\\
&&\hspace{-6mm}
\times~\Big[{z_{12}^2\over z_{13}^2 z_{23}^2}
+{1\over z_{13}^2}\Big({\alpha_s(z_{13}^2)\over\alpha_s(z_{23}^2)}-1\Big)
+{1\over z_{23}^2}\Big({\alpha_s(z_{23}^2)\over\alpha_s(z_{13}^2)}-1\Big)\Big]+...
\end{eqnarray}
where dots stand for the remaining $\alpha_s^2$  terms. (Here we promoted Wilson lines in the r.h.s. to operators). 

When the sizes of the dipoles are very different the kernel of the above equation reduces to 
\begin{eqnarray}
&\hspace{-6mm}
{\alpha_s(z_{12}^2)\over 2\pi^2}{z_{12}^2\over z_{13}^2 z_{23}^2}
  &|z_{12}|\ll |z_{13}|,|z_{23}|
\nonumber\\
&\hspace{-6mm}
{\alpha_s(z_{13})^2)\over 2\pi^2 z_{13}^2 }
  &|z_{13}\ll |z_{12}|,|z_{23}|
\nonumber\\
&\hspace{-6mm}
{\alpha_s(z_{23})^2)\over 2\pi^2 z_{23}^2 }
  &|z_{23}|\ll |z_{12}|,|z_{13}|
\label{limits}
\end{eqnarray}
so the argument of the coupling constant is the size of smallest dipole. The numerical approach to solution of the  the NLO BK equation
with this running coupling constant is presented in Ref. \cite{kovalb}.

\section{Conclusions}
The main conclusion is that the rapidity factorization and high-energy operator expansion in color dipoles works at the NLO level. 
There are many examples of the factorization which are fine at the leading order but fail at the NLO level. I believe that the high-energy
OPE has the same status as usual light-cone expansion in light-ray operators so one can calculate the high-energy amplitudes level by level in 
perturbation theory. 
As an outlook my collaborator G. Chirilli and I intend to apply the NLO high-energy operator expansion for the description of QCD amplitudes.  
(The intermediate result for the impact factor (\ref{nloifoton}) is the first step of that program). 
There are many papers devoted to analysis of  the high-energy amplitudes in QCD at the NLO level 
(see e.g. Refs. \cite{bart1,bart2,ivanov}) but all of them use traditional calculation of Feynman diagrams in momentum space. 
In our opinion, the high-energy OPE in color dipoles is technically more simple and gives us an opportunity to use an approximate 
tree-level conformal invariance in QCD. Moreover, the exact prescription for separating the coefficient functions 
(impact factors) and matrix elements  is somewhat tricky in the traditional approach while it comes naturally in the
framework of OPE logic. 
When finished,  the calculation of  the photon impact factor for the 
structure function $F_2(x)$ of deep inelastic scattering  will compete the analysis of the small-$x$ behavior of DIS structure 
functions at the NLO level. The study is in progress.

The author is grateful to  L.N. Lipatov for more than thirty years of valuable discussions and guidance. 
This work was supported by contract
 DE-AC05-06OR23177 under which the Jefferson Science Associates, LLC operate the Thomas Jefferson National Accelerator Facility.

\section{Appendix: Comparison to NLO BFKL in QCD}

We will compare our evolution equation for $\calu^Y$ to similar equation obtained from the NLO BFKL momentum-space  analysis \cite{nlobfkl1,nlobfkl2}.
 The linearized version of the Eq. (\ref{nlobk}) 
has the form (cf Eq. (\ref{nlobkernel}):
\begin{eqnarray}
&&\hspace{-2mm}
{d\over dY}{\cal U}^Y(z)~
=~{\alpha_sN_c\over 2\pi^2}
\!\int\!d^2z~
{z^2\over (z-z')^2 {z'}^2}\Big\{1
+~{\alpha_s\over 4\pi}\Big[b\big(\ln {z^2\mu^2\over 4} +2C\big)
\nonumber\\
&&\hspace{-2mm} 
-~
b{(z-z')^2-{z'}^2\over z^2}\ln{(z-z')^2\over {z'}^2}+
({67\over 9}-{\pi^2\over 3})N_c-{10\over 9}n_f-2N_c\ln{z^2\over {z'}^2}\ln{z^2\over (z-z')^2}\Big]
[{\cal U}^Y(z-z')+{\cal U}^Y(z')-{\cal U}^Y(z)]
\nonumber\\
&&\hspace{-2mm} 
+~{\alpha_s^2N_c^2\over 4\pi^3}\!\int\! d^2z'~ {z^2\over {z'}^2}
[F(z,z')+\Phi(z,z')]~{\cal U}^Y(z') +{\alpha_s^2N_c^2\over 2\pi^2}\zeta(3){\cal U}^Y(z)
\label{nlobfkernel}
\end{eqnarray}
where $F(z,z')$ and $\Phi(z,z')$ are given by Eqs. (\ref{F}) and (\ref{Phi}), respectively.

To compare the eigenvalues of the Eq. (\ref{nlobfkernel}) with NLO BFKL we expand
${\cal U}^Y(x,0)$ in eigenfunctions (\ref{eigenfunctions})
\begin{equation}
\langle\hat{\cal U}^Y(x_\perp,0)\rangle=\sum_{n=-\infty}^\infty \!\int_{-{1\over 2}-i\infty}^{-{1\over 2}+i\infty}
\! {d\gamma\over 2\pi i} ~e^{in\phi}(x_\perp^2\mu^2)^\gamma ~\langle\hat{\cal U}^Y(n,\gamma)\rangle~,
\label{expansion}
\end{equation}
compute the evolution of $\langle \hat{\cal U}(n,\gamma)\rangle$ from Eq. (\ref{nlobfkernel}) and compare it to the calculation based on the NLO BFKL results from \cite{nlobfkl1, lipkot}. 
(Here we will not consider the quark part of the NLO BK  kernel -  the agreement of that part with the $n_f$ term in the NLO BFKL kernel was proved in Ref. \cite{kw2}).

The relevant integrals have the form
\begin{eqnarray}
&&\hspace{-1mm}
{1\over 2\pi }\!\int\! d^2z~[2 (z^2/x^2)^\gamma e^{in\phi}-1] {x^2\over(x-z)^2z^2}
~=~\chi(n,\gamma)
\nonumber\\
&&\hspace{-1mm}
{1\over \pi }\!\int\! d^2z~
[2 (z^2/x^2)^\gamma e^{in\phi}-1] \Big({1\over(x-z)^2}- {1\over z^2}\Big)\ln{(x-z)^2\over z^2}
~=~\chi^2(n,\gamma)-\chi'(n,\gamma)-
{4\gamma\chi(\gamma)\over\gamma^2-{n^2\over 4}}
\nonumber\\
&&\hspace{-1mm}
{1\over \pi }\!\int\! d^2z ~ (z^2/x^2)^\gamma
{x^2\over(x-z)^2z^2}
e^{in\phi}\ln{(x-z)^2\over x^2}\ln{z^2\over x^2}
~=~{1\over 2}\chi''(n,\gamma)+\chi'(n,\gamma)\chi(n,\gamma)
\label{relintegral1}
\end{eqnarray}
where 
$\chi(n,\gamma)=-2C-\psi(\gamma+{n\over 2}) -\psi(1-\gamma+{n\over 2})$, 
and 
\begin{eqnarray}
&&\hspace{-1mm}
{1\over \pi}\int\! d^2z'~({z'}^2/z^2)^{\gamma-1}e^{in\phi'}
F(z,z')
\nonumber\\
&&\hspace{-1mm}
=~\Big\{-\Big[3+\Big(1+{n_f\over N_c^3}\Big)
{2+3\gamma\bar{\gamma}\over (3-2\gamma)(1+2\gamma)}\Big]\delta_{0n}
+\Big(1+{n_f\over N_c^3}\Big)
{\gamma\bar{\gamma}\over 2(3-2\gamma)(1+2\gamma)}\delta_{2n}\Big\}
{\pi^2\cos\pi\gamma\over(1-2\gamma)\sin^2\pi\gamma}
~\equiv~F(n,\gamma)
\nonumber
\end{eqnarray}
\begin{eqnarray}
&&\hspace{-1mm}
{1\over 2\pi}\int\! d^2z~({z'}^2/z^2)^{\gamma-1}e^{in\phi'}
\Phi(z,z')~=~-\Phi(n,\gamma)-\Phi(n,1-\gamma)
\label{relintegral3}
\end{eqnarray}
where \cite{lipkot} 
\begin{eqnarray}
&&\hspace{-1mm}
\Phi(n,\gamma)~=~\int_0^1\!{dt\over 1+t}~t^{\gamma-1+{n\over 2}}
\Big\{{\pi^2\over 12}-{1\over 2}\psi'\Big({n+1\over 2}\Big)-{\rm Li}_2(t)-{\rm Li}_2(-t)
\nonumber\\
&&\hspace{-1mm}
-~\Big(\psi(n+1)-\psi(1)+\ln(1+t)+\sum_{k=1}^\infty{(-t)^k\over k+n}\Big)\ln t
-\sum_{k=1}^\infty{t^k\over (k+n)^2}[1-(-1)^k]\Big\}
\label{fin}
\end{eqnarray}
(note that $\chi(0,\half+i\nu)\equiv \chi(\nu)$ and $\Phi(0,\half+i\nu)\equiv \Phi(\nu)$, see Eq. (\ref{fi})).
The convenient way to  calculate the integrals over angle $\phi$  is to represent 
$\cos n\phi$ as $T_n(\cos\phi)$ and use formulas for the integration of Chebyshev polynomials from Ref. \cite{lipkot}.

Using integrals (\ref{relintegral1}) - (\ref{relintegral3}) one easily obtains the  evolution equation for ${\cal U}(n,\gamma)$ in the form
\begin{eqnarray}
&&\hspace{-5mm}
{d\over dY}\langle\hat{\cal U}^Y(n,\gamma)\rangle~
\nonumber\\
&&\hspace{-5mm}=~{\alpha_s N_c\over \pi}
\Big\{\Big[1-{b\alpha_s\over 4\pi}\big({d\over d\gamma}+\ln 4-2C\big)
+{\alpha_sN_c\over 4\pi}\Big({67\over 9}-{\pi^2\over 3}-{10\over 9}{n_f\over N_c^3}\Big)\Big]\chi(n,\gamma)
+{\alpha_s b\over 4\pi}
\Big[{1\over 2}\chi^2(n,\gamma)-{1\over 2}\chi'(n,\gamma)
-{2\gamma\chi(n,\gamma)\over\gamma^2-{n^2\over 4} }\Big]
\nonumber\\
&&\hspace{-5mm}
+~{\alpha_s N_c\over 4\pi}\Big[-\chi"(n,\gamma)-2\chi(n,\gamma)\chi'(n,\gamma)+6\zeta(3)+F(n,\gamma)
-2\Phi(n,\gamma)-2\Phi(n,1-\gamma)\Big]\Big\}\langle\hat{\cal U}^Y(n,\gamma)\rangle
\label{eigenvalue}
\end{eqnarray}
where $\chi'(n,\gamma)\equiv{d\over d\gamma}\chi(n,\gamma)$ etc.

Next we calculate the same thing using NLO BFKL results \cite{nlobfkl1,lipkot}.
The  impact factor $\Phi_A(q)$ for the color dipole ${\cal U}(x,y)$ is proportional to
$\alpha_s(q)(e^{iqx}-e^{iqy})(e^{-iqx}-e^{-iqy})$ so one obtains the cross section of the scattering of color dipole in the form
\begin{eqnarray}
&&\hspace{-1mm}
\langle\hat{\cal U}(x,0)\rangle~=~{1\over 4\pi^2}\!\int\!{d^2 q\over q^2}
{d^2 q'\over {q'}^2}\alpha_s(q)
(e^{iqx}-1)(e^{-iqx}-1)\Phi_B(q')
\!\int_{a-i\infty}^{a+i\infty}\!{d\omega\over 2\pi i}\Big({s\over qq'}\Big)^\omega
G_\omega(q,q')
\label{lip1}
\end{eqnarray}
where $G_\omega(q,q')$ is the partial wave of the forward reggeized gluon scattering amplitude satisfying the equation
\begin{equation}
\omega G_\omega(q,q')=\delta^{(2)}(q-q')+\int\! d^2p K(q,p)G_\omega(p,q') 
\label{wgw}
\end{equation}
and $\Phi_B(q')$ is the target impact factor. The kernel $K(q,p)$ is symmetric with respect to $q\leftrightarrow p$  and the eigenvalues are
\begin{eqnarray}
&&\hspace{-1mm}
\int\! d^2p \Big({p^2\over q^2}\Big)^{\gamma-1}e^{in\phi}K(q,p)~=~
{\alpha_s(q)\over \pi}N_c\Big[\chi(n,\gamma)+{\alpha_sN_c\over 4\pi}\delta(n,\gamma)
\Big],
\label{lip3}\\
&&\hspace{-1mm}
\delta(n,\gamma)~=~-{b\over 2N_c}[\chi'(n,\gamma)+\chi^2(n,\gamma)]
+\Big({67\over 9}-{\pi^2\over 3}-{10\over 9}{n_f\over N_c^3}\Big)\chi(n,\gamma)
+6\zeta(3)
\nonumber\\
&&\hspace{35mm}-\chi''(n,\gamma)+F(n,\gamma)
-2\Phi(n,\gamma)-2\Phi(n,1-\gamma)\Big\}
\nonumber
\end{eqnarray}

The corresponding expression for 
$\langle\hat{\cal U}(n,\gamma)\rangle$ takes the form
\begin{equation}
\hspace{-0mm}
\langle \hat{\cal U}(n,\gamma)\rangle~=~-{1\over 2\pi^2}
\cos{\pi n\over 2}{\Gamma(-\gamma+{n\over 2})\over \Gamma(1+\gamma+{n\over 2})}
\!\int\! {d^2 q\over q^2}
{d^2 q'\over {q'}^2}e^{-in\theta}\alpha_s(q)\Big({q^2\over 4\mu^2}\Big)^\gamma
\Phi_B(q')\!\int_{a-i\infty}^{a+i\infty}\!{d\omega\over 2\pi i}\Big({s\over qq'}\Big)^\omega
G_\omega(q,q')
\label{lip4}
\end{equation}
where $\theta$ is the angle between $\vec{q}$ and $x$ axis.
Using Eq. (\ref{wgw}) we obtain
\begin{eqnarray}
&&\hspace{-1mm}
s{d\over ds}\langle \hat{\cal U}(n,\gamma)\rangle~
\label{lip5}\\
&&\hspace{-1mm}=~-{1\over 2\pi^2}
\cos{\pi n\over 2}{\Gamma(-\gamma+{n\over 2})\over \Gamma(1+\gamma+{n\over 2})}
\!\int\!{d^2 q\over q^2}
{d^2 q'\over {q'}^2}e^{-in\theta}\alpha_s(q)\Big({q^2\over 4\mu^2}\Big)^\gamma
\Phi_B(q')\!\int_{a-i\infty}^{a+i\infty}\!{d\omega\over 2\pi i}\Big({s\over qq'}\Big)^\omega
\int\! d^2p K(q,p)G_\omega(p,q')
\nonumber
\end{eqnarray}
The integration over q can be performed using 
\begin{eqnarray}
&&\hspace{-1mm}
\int\! d^2q ~\alpha_s(q)\Big({q^2\over p^2}\Big)^{\gamma-1}e^{in\phi}K(q,p)~=~
{\alpha_s^2(p)\over \pi}N_c\Big[\chi(n,\gamma)-{b\alpha_s\over 4\pi}\chi'(n,\gamma)+{\alpha_sN_c\over 4\pi}\delta(n,\gamma)
\Big]
\label{lipxz}\\
&&\hspace{-1mm}
\nonumber
\end{eqnarray}
 (recall that $K(q,p)=K(p,q)$ and $\alpha_s(p)=\alpha_s-{b\alpha_s^2\over 4\pi}\ln{p^2\over \mu^2}$
 with our accuracy). The result is
\begin{eqnarray}
&&\hspace{-1mm}
s{d\over ds}\langle \hat{\cal U}(n,\gamma)\rangle~
=~-{\alpha_s\over 2\pi^2}
\cos{\pi n\over 2}{\Gamma(-\gamma+{n\over 2})\over \Gamma(1+\gamma+{n\over 2})}
\!\int\!{d^2 p\over p^2}
{d^2 q'\over {q'}^2}e^{-in\varphi}\Big({p^2\over 4\mu^2}\Big)^\gamma
\Phi_B(q')\label{lip6}\\
&&\hspace{-1mm}
\times~
\!\int_{a-i\infty}^{a+i\infty}\!{d\omega\over 2\pi i}\Big({s\over pq'}\Big)^\omega
G_\omega(p,q'){\alpha_s(p)\over \pi}N_c\Big[\chi(n,\gamma-{\omega\over 2})
-{b\alpha_s\over 4\pi}\chi'(n,\gamma-{\omega\over 2})
+{\alpha_sN_c\over 4\pi}\delta(n,\gamma-{\omega\over 2}))
\Big]
\nonumber
\end{eqnarray}
where the angle $\varphi$ corresponds to $\vec{p}$.
Since $\omega\sim\alpha_s$ we can neglect terms $\sim \omega$ in the argument of $\delta$ and expand $\chi(n,\gamma-{\omega\over 2})\simeq\chi(n,\gamma)
-{\omega\over 2}\chi'(n,\gamma)$. Using again Eq. (\ref{wgw}) in the leading order we
can replace extra $\omega$ by ${\alpha_s\over \pi}N_c\chi(n,\gamma)$ and obtain
\begin{eqnarray}
&&\hspace{-1mm}
s{d\over ds}\langle \hat{\cal U}(n,\gamma)\rangle~=~-{\alpha_s\over 2\pi^2}
\cos{\pi n\over 2}{\Gamma(-\gamma+{n\over 2})\over \Gamma(1+\gamma+{n\over 2})}
\!\int\!{d^2 p\over p^2}
{d^2 q'\over {q'}^2}e^{-in\varphi}\Big({p^2\over 4\mu^2}\Big)^\gamma
\nonumber\\
&&\hspace{-1mm}
\times~
\Phi_B(q')\!\int_{a-i\infty}^{a+i\infty}\!{d\omega\over 2\pi i}\Big({s\over pq'}\Big)^\omega
G_\omega(p,q'){\alpha^2_s(p)\over \pi}N_c\Big[\chi(n,\gamma)
-{b\alpha_s\over 4\pi}\chi'(n,\gamma)+{\alpha_sN_c\over 4\pi}[\delta(n,\gamma)-2\chi(n,\gamma)\chi'(n,\gamma)]
\Big]
\label{lip7}
\end{eqnarray}
Finally, expanding 
 $\alpha_s^2(p)\simeq \alpha_s(p)(\alpha_s-{b\alpha_s^2\over 4\pi}\ln{p^2\over \mu^2})\alpha_s(\mu)$ 
 we obtain

\begin{eqnarray}
&&\hspace{-1mm}
s{d\over ds}\langle \hat{\cal U}(n,\gamma)\rangle~=~-{\alpha_sN_c\over 2\pi^3}
\cos{\pi n\over 2}{\Gamma(-\gamma+{n\over 2})\over \Gamma(1+\gamma+{n\over 2})}
\Big\{\chi(n,\gamma)\Big(1-{b\alpha_s\over 4\pi}{d\over d\gamma}-\ln 4 +2C\Big)
-{b\alpha_s\over 4\pi}\chi'(n,\gamma)
\nonumber\\
&&\hspace{-1mm}
+~{\alpha_sN_c\over 4\pi}[\delta(n,\gamma)-2\chi(n,\gamma)\chi'(n,\gamma)]\Big\}
\!\int\!{d^2 p\over p^2}
{d^2 q'\over {q'}^2}e^{-in\varphi}\alpha_s(p)\Big({p^2\over 4\mu^2}\Big)^\gamma
\Phi_B(q')\!\int_{a-i\infty}^{a+i\infty}\!{d\omega\over 2\pi i}\Big({s\over pq'}\Big)^\omega
G_\omega(p,q')
\label{lip8}
\end{eqnarray}
which can be rewritten as an evolution equation
\begin{eqnarray}
&&\hspace{-3mm}
s{d\over ds}\langle \hat{\cal U}(n,\gamma)\rangle~
\nonumber\\
&&\hspace{-3mm}
=~
{\alpha_sN_c\over \pi}\Big\{\Big(1
+{b\alpha_s\over 4\pi}\Big[\chi(n,\gamma)-{2\gamma\over \gamma^2-{n^2\over 4}}+2C-\ln 4
-{d\over d\gamma}\Big]\Big)\chi(n,\gamma)
+{\alpha_sN_c\over 4\pi}[\delta(n,\gamma)-2\chi(n,\gamma)\chi'(n,\gamma)]
\Big\}
\langle {\cal U}(n,\gamma)\rangle
\nonumber\\
&&\hspace{-3mm}
=~{\alpha_s N_c\over \pi}
\Big\{\Big[1+{b\alpha_s\over 4\pi}\big(2C-\ln 4-{d\over d\gamma}\big)+{\alpha_sN_c\over 4\pi}
\Big({67\over 9}-{\pi^2\over 3}-{10\over 9}{n_f\over N_c^3}\Big)\Big]\chi(n,\gamma)
+{\alpha_s b\over 8\pi}
\Big[\chi^2(n,\gamma)-\chi'(n,\gamma)-{4\gamma\chi(\gamma)\over\gamma^2-{n^2\over 4} }
\Big]
\nonumber\\
&&\hspace{13mm}
+~{\alpha_s N_c\over 4\pi}\Big[-\chi''(n,\gamma)-2\chi(n,\gamma)\chi'(n,\gamma)+6\zeta(3)+F(n,\gamma)
-2\Phi(n,\gamma)-2\Phi(n,1-\gamma)\Big]\Big\}\langle \hat{\cal U}(n,\gamma)\rangle
\label{leigenvalue}
\end{eqnarray}
We see that this eigenvalue coincides with Eq. (\ref{eigenvalue}). 


\end{document}